\documentclass[aps,prc,twocolumn,superscriptaddress]{revtex4-1}

\usepackage{lineno}
\usepackage{graphicx}  
\usepackage{dcolumn}   
\usepackage{bm}        
\usepackage{amssymb}   
\usepackage{amsmath}
\usepackage{longtable}
\usepackage{lipsum}
\usepackage[bottom]{footmisc}

\usepackage{soul} 
\usepackage{color} 

\begin{document}




\title{Using excitation-energy dependent fission yields to identify key fissioning nuclei \newline in $\bm{r}$-process nucleosynthesis}


\author{N. Vassh}
\affiliation{Department of Physics, University of Notre Dame, Notre Dame, Indiana 46556, USA}

\author{R. Vogt}
\affiliation{Nuclear and Chemical Science Division, Lawrence Livermore National Laboratory, Livermore, CA 94551, USA}
\affiliation{Department of Physics, University of California, Davis, CA 95616, USA}

\author{R. Surman}
\affiliation{Department of Physics, University of Notre Dame, Notre Dame, Indiana 46556, USA}

\author{J. Randrup}
\affiliation{Nuclear Science Division, Lawrence Berkeley National Laboratory, Berkeley, CA 94720, USA}

\author{T. M. Sprouse}
\affiliation{Department of Physics, University of Notre Dame, Notre Dame, Indiana 46556, USA}

\author{M. R. Mumpower}
\affiliation{Theoretical Division, Los Alamos National Laboratory, Los Alamos, NM, 87545, USA}

\author{P. Jaffke}
\affiliation{Theoretical Division, Los Alamos National Laboratory, Los Alamos, NM, 87545, USA}

\author{D. Shaw}
\affiliation{Department of Physics, University of Notre Dame, Notre Dame, Indiana 46556, USA}

\author{E. M. Holmbeck}
\affiliation{Department of Physics, University of Notre Dame, Notre Dame, Indiana 46556, USA}

\author{Y. Zhu}
\affiliation{Department of Physics, North Carolina State University, Raleigh, North Carolina 27695 USA}

\author{G. C. McLaughlin}
\affiliation{Department of Physics, North Carolina State University, Raleigh, North Carolina 27695 USA}

\date{\today}

\begin{abstract}
The possibility that nucleosynthesis in neutron star mergers may reach fissioning nuclei introduces significant uncertainties in predicting the relative abundances of $r$-process material from such events. We evaluate the impact of using sets of fission yields given by the 2016 GEF code for spontaneous (sf), neutron-induced ((n,f)), and $\beta$-delayed ($\beta$df) fission processes which take into account the approximate initial excitation energy of the fissioning compound nucleus. We further explore energy-dependent fission dynamics in the $r$ process by considering the sensitivity of our results to the treatment of the energy sharing and de-excitation of the fission fragments using the FREYA code. We show that the asymmetric-to-symmetric yield trends predicted by GEF 2016 can reproduce the high-mass edge of the second $r$-process peak seen in solar data and examine the sensitivity of this result to the mass model and astrophysical conditions applied. We consider the effect of fission yields and barrier heights on the nuclear heating rates used to predict kilonova light curves. We find that fission barriers influence the contribution of $^{254}$Cf spontaneous fission to the heating at $\sim100$ days, such that a light curve observation consistent with such late-time heating would both confirm that actinides were produced in the event and imply the fission barriers are relatively high along the $^{254}$Cf $\beta$-feeding path. We lastly determine the key nuclei responsible for setting the $r$-process abundance pattern by averaging over thirty trajectories from a 1.2--1.4 $M_{\odot}$ neutron star merger simulation. We show it is largely the odd-$N$ nuclei undergoing ($Z$,$N$)(n,f) and ($Z$,$N$)$\beta$df that control the relative abundances near the second peak. We find the ``hot spots" for $\beta$-delayed and neutron-induced fission given all mass models considered and show most of these nuclei lie between the predicted $N=184$ shell closure and the location of currently available experimental decay data.
\end{abstract}


\maketitle

\section{Introduction}\label{sec:intro}
\noindent Over 60 years ago, Burbidge {\it et al.} suggested that nuclear fission was responsible for the behavior of supernova light curves \cite{Burbidge56}. We now know the process that synthesizes fissioning nuclei---rapid neutron capture, or $r$-process, nucleosynthesis---is unrelated to supernova light curves and unlikely to occur robustly, if at all, in ordinary supernovae \cite{Ji+2016,Roederer+16}. The most attractive astrophysical site for an $r$ process that reaches fissioning nuclei is within a neutron star merger \cite{Lattimer+74,Meyer89,Freiburghaus+99,Goriely+11,Korobkin+12,Wanajo+14,Just+15}. The electromagnetic counterpart to the GW170817 neutron star merger \cite{Cowperthwaite2017,AbbottGW170817,Kasen} indicated some thousandths of a solar mass of lanthanides were produced in the event, possibly enough for mergers to account for all of the $r$-process lanthanides in the galaxy if the event was typical \cite{Cote}. If neutron star mergers are indeed the source of all $r$-process elements, including the actinides, we can look to such events as opportunities to probe fission properties.

Fission processes can play an important role in determining $r$-process observables such as abundance patterns and light curves. For example, lanthanide abundances can be influenced by late-time deposition of fission products \cite{Cote} and neutrons from fission can affect the amount of late-time neutron capture that sets the overall abundance pattern \cite{Eichler15}. Nuclear heating by fission can shape kilonova light curves \cite{Barnes+16}, with the late-time heating possibly dominated by the spontaneous fission of $^{254}$Cf \cite{Cfpaper}.  Understanding these effects requires knowledge of fission properties for hundreds of nuclei on the neutron-rich side of stability, about which little is experimentally known. Calculations of the $r$ process instead rely almost entirely on theoretical descriptions that vary widely. Here we examine the influence of two key fission inputs in $r$-process calculations: the fission fragment distributions and fission barrier heights (i.e.\ the maximum energy along the optimum path toward scission).

The importance of the fission fragment treatment in $r$-process calculations is well established \cite{Goriely+13,GorielyGEF,GorielyGMPGEF,Panov,Kodama,Eichler15,Eichler16}. Parameterized, semi-empirical formulae based on systematics, such as those in Refs. \cite{Panov,Kodama,ShibagakiMathews}, are an improvement over symmetric splits which assume the nucleus to divide in half, but are still a simplification of complex fission dynamics. Phenomenological descriptions, such as ABLA \cite{ABLA91,ABLA07}, Wahl \cite{Wahl}, and GEF (a GEneral description of Fission observables) \cite{GEF}, take into account the influence of shell structure, fission barriers, angular momentum, and neutron emission from the excited fragments. These fission yield descriptions have also been applied in $r$-process simulations \cite{Mendoza-Temis+15,Mendoza-Temis+16,RobertsWahl,GorielyGEF,GorielyGMPGEF}. 

An aspect of the phenomenological descriptions of fission that has so far remained relatively unexplored in the $r$ process is the dependence of the fission yields on the excitation energy of the compound nuclei. This effect is generally assumed to be small, as the $r$-process temperatures are low ($\sim 0.1$ MeV), and is often ignored. The possibility for the yields of $r$-process nuclei to explicitly depend on the excitation energy was considered in Ref. \cite{PanovNIF}, but yield distributions were found to vary smoothly with initial energy, thus the yields at vanishing neutron bombarding energy were taken to be appropriate at all energies relevant for the $r$ process. Here we revisit an examination of the role of excitation energy on the fission yields of $r$-process nuclei by treating the three main processes occurring in a fission cycling $r$ process, namely neutron-induced, $\beta$-delayed, and spontaneous fission, with distinct initial excitation energies when determining the fission yields to apply to each process. To examine the impact of such an energy dependence on the $r$ process, we use fission yields from the publicly available GEF code (version GEF-2016-V1-2 \cite{GEFweb}).  

We explore the sensitivity of the $r$ process to the assumptions made for the evolution of excited fission fragments by comparing results with GEF 2016 and results given an alternate treatment for the energy sharing and de-excitation. To do so we modified the published FREYA (Fission Reaction Event Yield Algorithm) code \cite{CPC,CPC_NVA} to use the GEF 2016 fission fragment yields (pre prompt neutron emission) as a function of post-scission fission fragment mass, charge and total kinetic energy, as input. The fission product yields (post prompt neutron emission) and average neutron multiplicities from the two codes are compared. We explore the effect of the additional neutron emission predicted by FREYA in the $r$ process for merger dynamical ejecta conditions.

Perhaps the most important characteristic of a heavy nucleus for predicting fission properties within a given model is the height of the fission barrier. It is well known that the fission barriers assumed for heavy, neutron-rich nuclei can have a great influence on the outcome of $r$-process calculations \cite{Petermann,GorielyGEF,GorielyGMPGEF,Eichler15,ShibagakiMathews,Samuel18}. Fission barriers often determine whether the nuclear flow ($\lambda_i Y_i$ where $\lambda_i$ is the rate of the reaction or decay and $Y_i$ is the abundance) will permit the population of nuclei of interest, such as the predicted superheavy island of stability \cite{Thielemann+83,PanovSF,GorielyGEF,GorielyGMPGEF,Petermann,BDFrp}. The influence of fission barriers can also lead to dramatically different conclusions regarding the origin of the second $r$-process peak as discussed in Ref. \cite{ShibagakiMathews}, where high fission barriers near $N=184$ coupled with very broad fission fragment distributions caused a disappearance of this main $r$-process feature in neutron star merger conditions. 

Here we examine results using the Finite Range Droplet Model (FRDM2012) \cite{FRDM2012}, Thomas-Fermi (TF) \cite{TFmass}, Hartree-Fock-Bogoliubov (HFB-17) \cite{HFB17PRL}, and Extended Thomas-Fermi with Strutinsky Integral (ETFSI) \cite{ETFSI1992,ETFSI1995} model masses and corresponding barriers along with the GEF+FREYA yields described above. We explore how the different termination points for the $r$ process predicted by these models influence the final abundance pattern, and identify fissioning nuclei most accessed under a range of neutron star merger conditions. We also consider the population of $^{254}$Cf and find it to be sensitive to the fission barrier treatment, as suggested recently in Ref. \cite{WuBGMP}.   

The paper is organized as follows: our application of GEF 2016 and implementation of FREYA is described in Section~\ref{sec:GEF}. In Section~\ref{sec:rpimpact} we explore the $r$-process impact of the fission yields and neutron multiplicities that result from this approach. Variations of fission barrier heights and other nuclear physics inputs are considered in Section~\ref{sec:rpMMastro}. In Section~\ref{sec:maxflows} we conclude by identifying the key fissioning nuclei important in a variety of neutron star merger conditions and common to all fission barrier models considered, some of which we find to be potentially within reach of future experimental facilities.   

\section{GEF and FREYA fission treatments}\label{sec:GEF}

Before describing the fission properties of neutron-rich nuclei we obtain from the GEF code, we note this and other codes presently applied in the $r$ process, such as ALBA and Wahl, are phenomenological descriptions which extrapolate into unmeasured regions using systematics. Therefore, reminiscent of some nuclear mass models, predictions for the properties of neutron-rich nuclei are subject to large uncertainties. We look forward to progress in theoretical campaigns to calculate the fission yields of neutron-rich nuclei, such as density functional theory and microscopic-macroscopic approaches, in the future. Here we choose GEF 2016 as a tool to explore the sensitivity of the $r$ process to various yield properties due to its public availability and documented physics inputs.

We now describe the GEF and FREYA fission models, the differences between them, and how those differences affect the resulting fission product yields and average neutron multiplicities for nuclei relevant for the $r$ process.  Both codes use Monte Carlo techniques to produce fission events that provide complete kinematic information, including angular momentum, for all fission fragments; see Refs.~\cite{GEF,CPC_NVA} for more details.  (We note that two versions of GEF are available, the stand-alone Monte Carlo version used here and a deterministic subroutine for use in codes like TALYS \cite{TALYSRef,TALYScode}.)   

While other codes such as FIFRELIN \cite{FIFRELIN} and CGMF \cite{Talou:2017qlc} are also available, GEF and FREYA are the fastest and thus most suitable for studies that require calculations for hundreds of nuclei, as is the case in this work.  Here we first describe the fission fragment yields, as well as their excitation energy dependence, followed by a discussion of particle emission in GEF and FREYA and how the two models can lead to different average neutron multiplicities, relevant for these studies.  All the calculations in this section, for both GEF and FREYA, are based on one million fission events generated for each fissioning isotope.

While GEF and FREYA can achieve similar end results as far as the output of complete fission events, the general approaches are rather different and worth some discussion.  The published version of FREYA is generally more limited in the number of isotopes available because of its approach. Like FIFRELIN and CGMF, FREYA requires the fission fragment yields and total kinetic energies (TKE) of the fragments in some form as inputs.  Thus the current published version of FREYA, FREYA 2.0.2 \cite{CPC_NVA}, is limited to certain isotopes: spontaneous fission of $^{244}$Cm, $^{252}$Cf, $^{238}$U and $^{238,240,242}$Pu as well as neutron-induced fission of $^{233,235,238}$U and $^{239,241}$Pu.

Although FREYA, FIFRELIN and CGMF differ in detail, the basic numerical approaches are similar.  The yields are sampled to choose one of the fragments with the partner chosen to conserve mass number $A$ and charge $Z$, followed by sampling of the TKE for the event.  Given the masses and TKE, the total excitation energy (TXE) is obtained and shared between the two fragments according to a model-dependent prescription.  Once this excitation energy sharing is complete, neutron evaporation follows. The excitation energy dependence of the input yields is modeled based on limited energy-dependent data.  Some progress has recently been made using models based on potential energy surfaces calculated in the macroscopic-microscopic approach \cite{RandrupMoller,PatrickJetal}.  Other yields calculated from many-body approaches based on density functional theory may become available in the future but these are so far quite limited \cite{schunck2016}.

The philosophy of the GEF code is quite different.  The only user inputs required by GEF 2016 are the charge and mass number of the fissioning nucleus, incident neutron energy or initial excitation energy for a compound nucleus, and the number of simulated events. The GEF yields and fragment excitation energies are then based on incident energy, nuclear mass and charge systematics.  This allows a broader range of isotopes to be studied and extrapolations made to regions of the nuclear chart where no data exist.  Thus GEF requires no input data for the yields and total kinetic energies, unlike all other available complete event fission codes. The fission fragment yields in GEF depend on three things: the fission barriers, the fission modes, and the excitation energy sharing.  We will touch upon all three here.

GEF 2016 employs the Thomas-Fermi (TF) macroscopic fission barriers \cite{TFBH} and nuclear masses \cite{TFmass}, along with a microscopic correction to the ground state mass and a pairing correction to the binding energy at the barrier.  Additionally, GEF makes use of experimentally-inferred fission barriers to derive further parameterized corrections to the TF barrier heights (see \cite{GEF}). The mass and charge systematics determined from these nuclear inputs are then extrapolated to unmeasured regions.

  There are four fission modes used in GEF 2016 to describe the fission fragment mass yields immediately after scission (called pre-neutron in GEF).
  Three of these were introduced by Brosa {\it et al.} \cite{Brosa}: standard 1 (S1, related to near-spherical nuclei near the doubly-magic closed proton and neutron shells at $Z = 50$ and $N=82$), standard 2 (S2, associated with the deformed neutron shell closure at $N=88$), and super long (SL, a symmetric mode).  The fourth is a super asymmetric mode (SA or S3) introduced by Mulgin {\it et al.} \cite{Mulgin}.
  
  The proximity to a closed neutron or proton shell determines whether the yields are symmetric or antisymmetric and also governs the widths of the yield distributions.  The effects of nuclear deformation, fragment angular momentum and charge polarization are taken into account.  Empirical $A$ and $Z$ systematics are used to determine the yields where no data are available.
 
To employ GEF 2016 to calculate the fission fragment yields for $r$-process nuclei, we extended the ``range of validity", defined by $A/Z > 172/80$ and $A/Z < 250/90$ for $76<Z<120$, implemented in the default GEF code. We do not, however, go beyond the nuclei included in the default mass and shell correction tables, $1\le Z \le136$ and $1\le N \le203$, to ensure that our output fission data is consistent with standard user outputs. We will later show that, given astrophysical conditions leading to fission, there is minimal $r$-process fission flow beyond $N=203$ for most of the nuclear mass models considered.  Therefore, implementing yields for nuclei with $N\le 203$ is sufficient.

Finally, the energy sharing mechanism in GEF determines how the intrinsic (statistical) excitation energy is divided between the heavy and light fragments.  Before scission, the fragments are coupled so that nucleons unpaired before scission are preferentially transferred to the heavy fragment.  Thus, at higher excitation energies, the transfer of unpaired nucleons to the heavy fragment gives the additional neutron multiplicity to the heavy fragment while the light fragment neutron multiplicity remains relatively constant.  The charge polarization, determining the charge yields, is assumed to remain essentially unchanged by an increase in excitation energy.  While all fission codes assume binary fission, FREYA assumes that neutrons are only emitted from the fully-accelerated fragments.  On the other hand, GEF has some probability for neutron emission prior to scission.  These ``scission neutrons'' are postulated to be emitted even in spontaneous fission.

Because the excitation energy is modeled at scission, no input TKE is required by GEF; instead, the TKE is fixed by energy conservation.  Recall that the TKE is an input in FREYA and other fission models. To then fix the fragment excitation energies at scission, GEF again applies $A$ and $Z$ systematics determined from data which are extrapolated to unmeasured isotopes.

The fission product yields (termed post-neutron emission in GEF) depend on the excitation energy and the identity of the compound nucleus.  In the case of spontaneous and $\beta$-delayed fission, the compound nucleus is simply the mass number $A$ while, for neutron-induced fission, it is $A+1$.  The excitation energy is zero for spontaneous fission. In the case of neutron-induced fission, it is the sum of the incident neutron energy, $E_n$, and the neutron separation energy, $S_n$, with $E_n + S_n$ on the order of a few MeV, except for the most neutron-rich nuclei near the neutron dripline. For $\beta$-delayed fission, the excitation energy (denoted here as $\langle E \rangle_\beta$) can be as large as $9$~MeV.

Prior evaluations of cumulative neutron-induced fission yields (after both prompt neutron emission and delayed emission from the $\beta$-decay of fission products) indicated that the yields become more symmetric with increasing incident neutron energy \cite{england1994endf}. However, a more recent experiment saw the cumulative fission yields for $^{239}$Pu(n,f) in particular to have a non-monotonic energy dependence on the energy of the incident neutron \cite{gooden2016}. Thus the energy dependence of the fission yields may be more complicated than previously assumed.

\begin{figure}
\begin{center}
\includegraphics[scale=0.485]{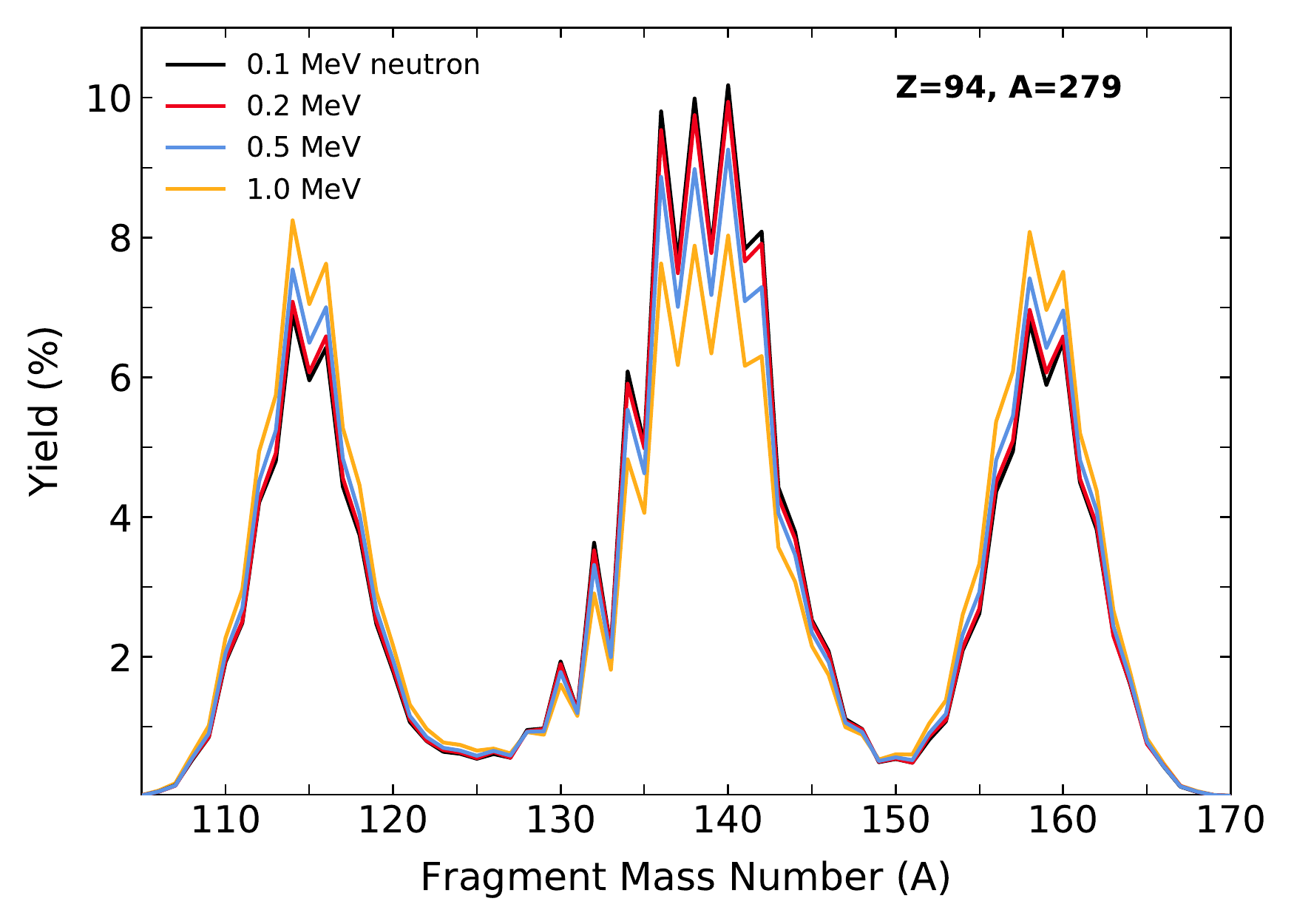}
\end{center}
\caption{(Color online) The fission yield for $^{278}$Pu(n,f) with an incoming neutron energy of 0.1 MeV (black) as compared to 0.2 MeV (red), 0.5 MeV (blue), and 1.0 MeV (orange).}
\label{fig:Yenergy94279}
\end{figure}

Neutron-induced fission in the $r$ process occurs late in time at low temperatures, corresponding to incident neutron energies between $\sim 0.01 - 0.2$~MeV, with most fission taking place at $\sim 0.1$~MeV. Although the $r$-process indeed occurs over a range of temperatures, and the neutrons are characterized by a thermal distribution rather than a single energy, the variation in fission yields over the energy range important for the $r$ process is small. Figure~\ref{fig:Yenergy94279} shows that for the incident neutron energy range of relevance to the $r$ process, even a very neutron-rich nucleus such as $^{279}$Pu with a low separation energy will not exhibit significant differences in its fission fragment yields. Percent-level differences begin to appear when comparing yields at  $\sim 0.1$~MeV and  $\sim 1.0$~MeV, however $1.0$~MeV ($\approx 10$~GK) is not a temperature at which fission participates in the $r$-process since here the environment is governed by nuclear statistical equilibrium (NSE) and has yet to synthesize fissioning nuclei.
The differences in the yields of a less neutron-rich nucleus with a separation energy on the order of MeV, such as $^{258}$Pu, are even less relevant. In this case, even a neutron incident energy of $\sim1.0$~MeV produces differences on the sub-percent level relative to $E_n = 0.1$~MeV. Therefore, unless the $r$-process proceeds through conditions in which nuclear reheating produces a rise in late time temperatures which reaches NSE values, the variance in the fission yields over the temperature evolution of the $r$ process can be safely ignored. Thus we apply a constant incident neutron energy of  $\sim 0.1$~MeV.

\begin{figure*}
\begin{center}
\includegraphics[scale=0.43]{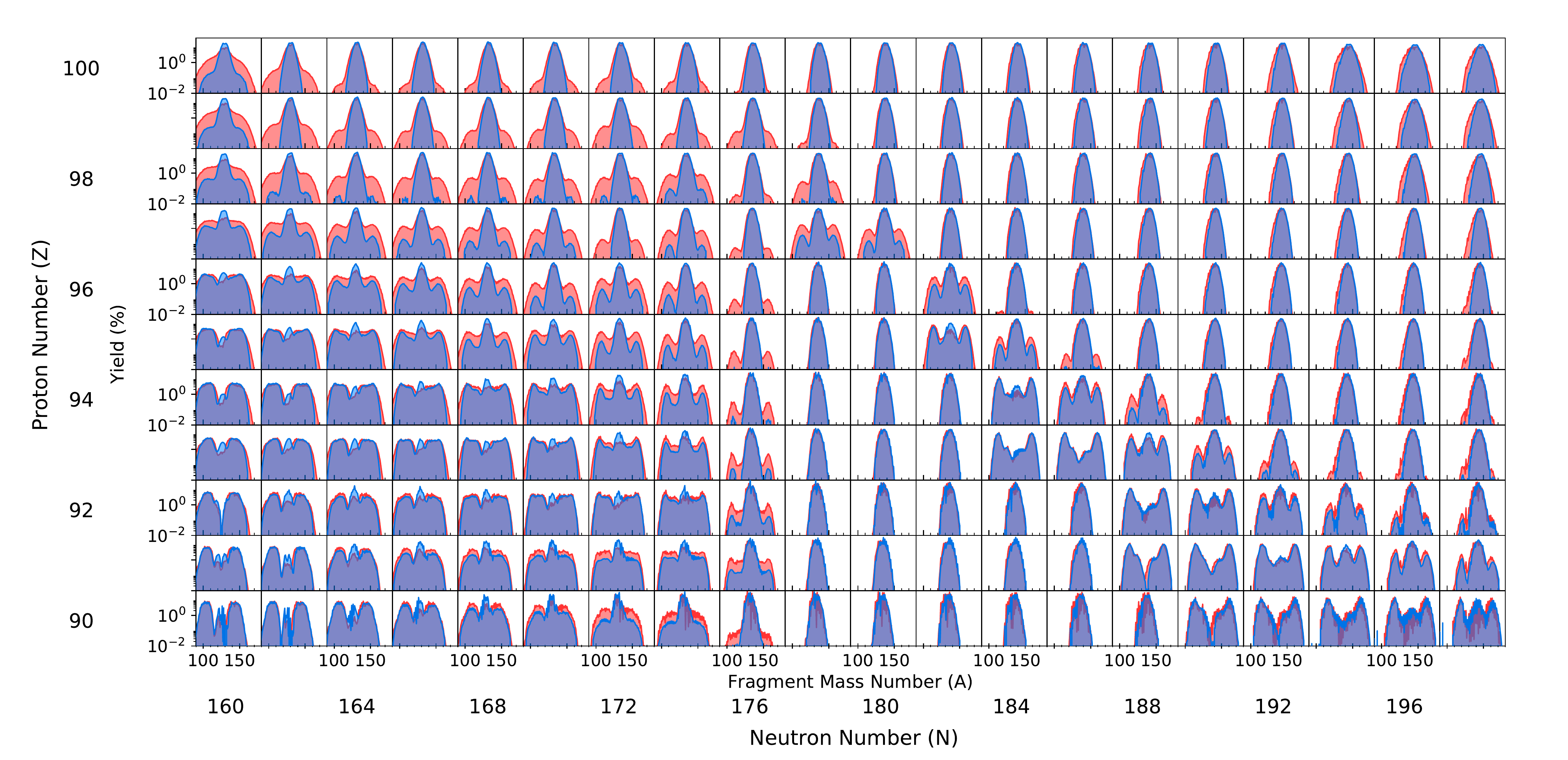}
\end{center}
\caption{(Color online) The GEF 2016 product yields for neutron-induced fission from a $0.1$~MeV neutron (red) compared to the product yields from spontaneous fission of the same compound nucleus (blue).}
\label{fig:yieldsNIF}
\end{figure*}

\begin{figure}
\begin{center}
\includegraphics[scale=0.5]{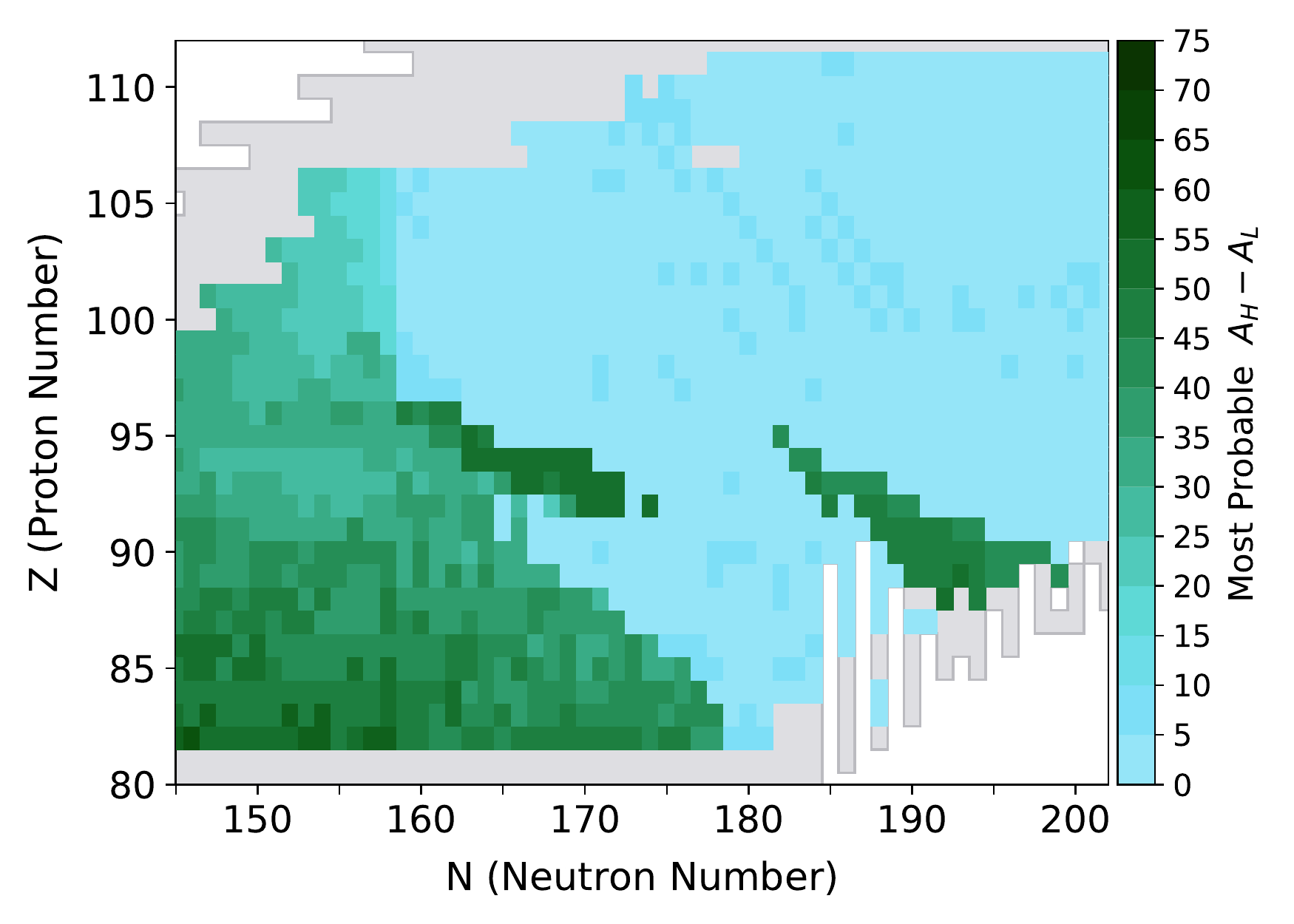}
\end{center}
\caption{(Color online) The most probable mass difference between the light ($A_L$) and heavy ($A_H$) fragments given GEF 2016 product yields for neutron-induced fission from a $0.1$~MeV neutron. The grey region shows the TF dripline.}
\label{fig:yieldwidthNIF}
\end{figure}

Even though the incident neutron energy of relevance to the $r$ process is low, the excitation energy of neutron-induced fission is typically a few MeV, considerably higher than the zero excitation energy of spontaneous fission. One could thus expect a difference in the spontaneous and neutron-induced fission yields. The resultant fission product yields are shown in Fig.~\ref{fig:yieldsNIF}. Some increased asymmetry is seen in the tails of the neutron-induced yields.
For the neutron-rich nuclei of interest here, the GEF 2016 systematics suggest a global trend of transition from asymmetric toward symmetric yields along most isotopic chains, with a region of primarily symmetric yields near the shell closure at $N=184$. Along an isotonic chain in this neutron-rich region, on average the yields become increasingly symmetric with increasing $Z$. This behavior can be further demonstrated by examining the trend in the width of the yields, represented in Fig.~\ref{fig:yieldwidthNIF} by the most probable mass number difference between the light and heavy fragment. To obtain this value, we find the maximum, $A'$, of the yield distributions shown in Fig.~\ref{fig:yieldsNIF}, which gives a most probable width of $|A' - (A_f - A' -\bar{\nu})|$ where $A_f$ and $\bar{\nu}$ are the mass number and average neutron multiplicity of the fissioning nucleus, respectively. Since, as can be seen in Fig.~\ref{fig:yieldsNIF}, many of the GEF 2016 yields in the neutron-rich regions contain both a symmetric and asymmetric component, a metric based on the maximum of the yields will not fully capture their complex behavior. However Fig.~\ref{fig:yieldwidthNIF} is still representative of the dominant yield trends predicted for $r$-process isotopes. In Section III, we will examine the impact that such asymmetric-to-symmetric yield trends have on the $r$-process abundance pattern and will show that the enhanced asymmetric yield contributions for finite excitation energies appear in key regions for a fission cycling $r$ process. We note that for some nuclei the fission yields predicted by other versions of the GEF code can be significantly different from those of GEF 2016. Nevertheless, the general arguments we lay out in this work remain the same.

\begin{figure*}
\begin{center}
\includegraphics[scale=0.53]{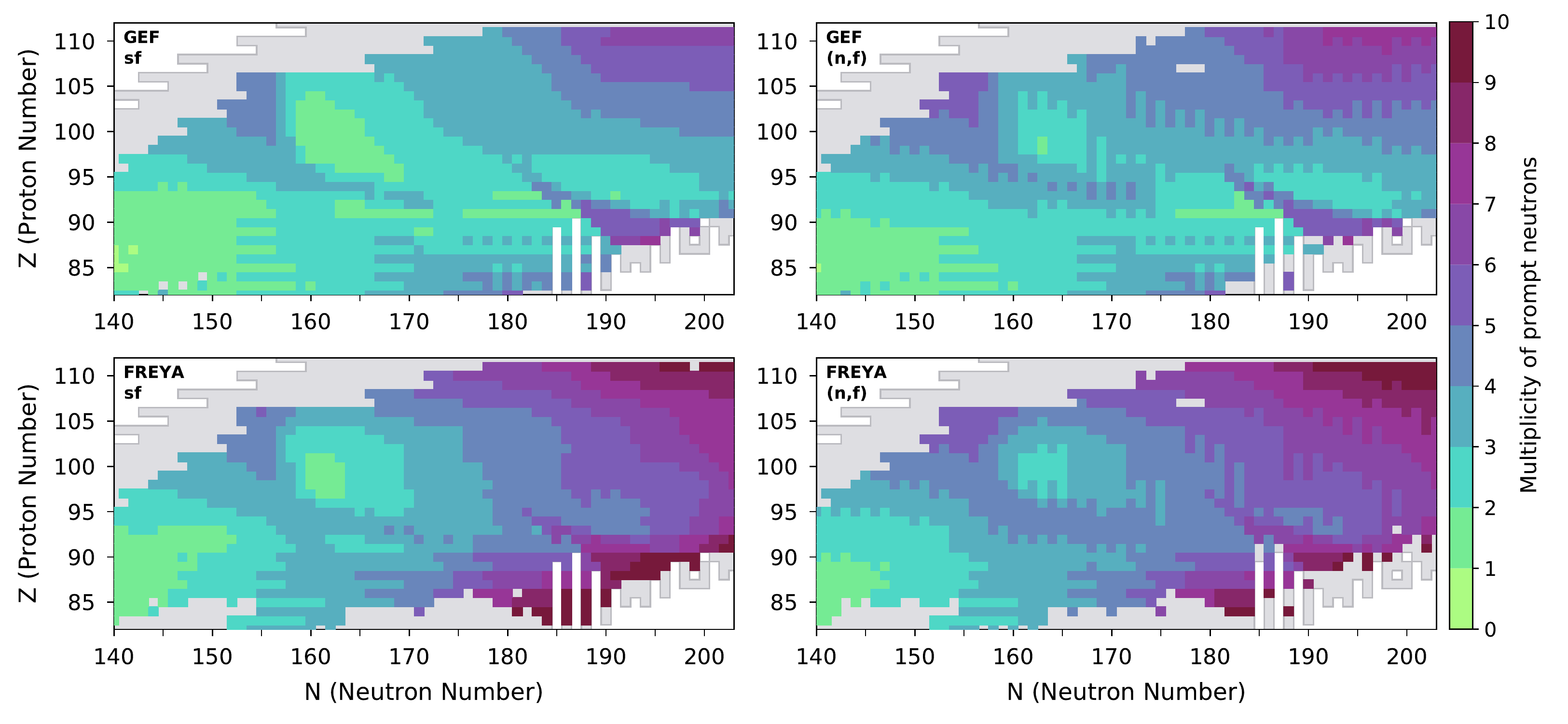}
\end{center}
\caption{(Color online) The average prompt neutron multiplicity for fission induced by the capture of a 0.1 MeV neutron (right panels) compared to the multiplicity from spontaneous fission (left panels) using the energy sharing and de-excitation treatments of GEF 2016 (top panels) and FREYA (bottom panels).}
\label{fig:FREYAnu}
\end{figure*}  

  We now turn to neutron emission from the fragments and the resulting fission product yields.  First we discuss how we modify FREYA to make use of the GEF 2016 fission fragment yields.  We then discuss how neutron emission differs in the two codes and how, even though we may start with identical yields in both GEF and FREYA, we may end up with different fission product yields and average neutron multiplicities, $\overline \nu$.  The difference in neutron emission is important for the $r$ process because prompt fission neutrons can be a substantial fraction of the late-time neutrons available for capture \cite{GorielyGEF}, thereby influencing the movement of lighter nuclei near the second and third $r$-process abundance peaks \cite{GorielyGMPGEF}.

For FREYA to be used in calculating the fission product yields and neutron emission relevant to the $r$ process, it needed to be adapted to use the GEF 2016 fission fragment yields as input.  FREYA was thus modified to take the fission fragment yields as function of mass and charge, $Y(A,Z)$, as well as the yields as a function of mass and TKE, $Y(A)$ and $Y({\rm TKE})$, given by GEF.  Thus, in the calculations that follow, the primary fragment yields employed in the two codes are identical and the differences are due to how neutron evaporation is treated.

Recall that, because GEF models the excitation energy partition between the fragments at scission, the energy available for neutron emission is set with no additional parameters required.  Neutron emission from the fully accelerated fragments in GEF then proceeds through a statistical model, using the relative neutron emission width from \cite{GEFGammaNref}.  The width depends on fragment mass, excitation energy, angular momentum, fragment temperature, and neutron separation energy, which are all modeled internally in GEF.

On the other hand, FREYA, like CGMF and FIFRELIN, starts with the TKE as input instead of the total excitation energy.  Thus the excitation energy partition between the fragments in these codes is done empirically, which requires a number of parameters.  FREYA 2.0.2 has five physics-based parameters: $d$TKE, which shifts TKE as a function of heavy fragment mass by some amount to ensure agreement with the average neutron multiplicity; $x$, representing additional excitation energy given to the light fragment, $x > 1$; $c$, setting the level of thermal fluctuations in the fragments; $e_0$, the value of the asymptotic level density parameter and $c_S$, governing the fragment spin magnitude, see Ref.~\cite{RVJR_gamma2}.  
Neutron emission in FREYA proceeds assuming a Weisskopf-Ewing energy spectrum with a maximum temperature in the daughter fragment decreased appropriately \cite{CPC}.

Because FREYA does not have any systematic way of setting the parameters for unknown isotopes, we adjust $d$TKE in FREYA to match the known $\overline \nu$ in two specific cases, $^{252}$Cf(sf), for spontaneous fission, and $^{239}$Pu(n,f), for neutron-induced fission, while leaving the other FREYA parameters ($x$, $c$, $c_S$, and $e_0$) fixed at their default values for these isotopes.  In these two cases, it was found that the $d$TKE required in FREYA when using the GEF 2016 yields was negligibly small, and the calculated average neutron multiplicity in both models agreed with each other and with the data.  We use the $^{252}$Cf(sf) parameters for all spontaneous fission and the $^{239}$Pu(n,f) parameters for all neutron-induced and $\beta$-delayed fissions.

Because FREYA and GEF employ different methods of fragment de-excitation and the FREYA parameters are not fixed for each isotope, it is clear that we will not obtain the same neutron multiplicity from FREYA and GEF aside from our two matching points.  Indeed, there are no data for us to test the parameter values in either model.  However, we can take any differences between the fission product yields and neutron multiplicities in FREYA and GEF as an indication of the fission uncertainties affecting the $r$ process.
    
\begin{figure*}
\begin{center}
\includegraphics[scale=0.42]{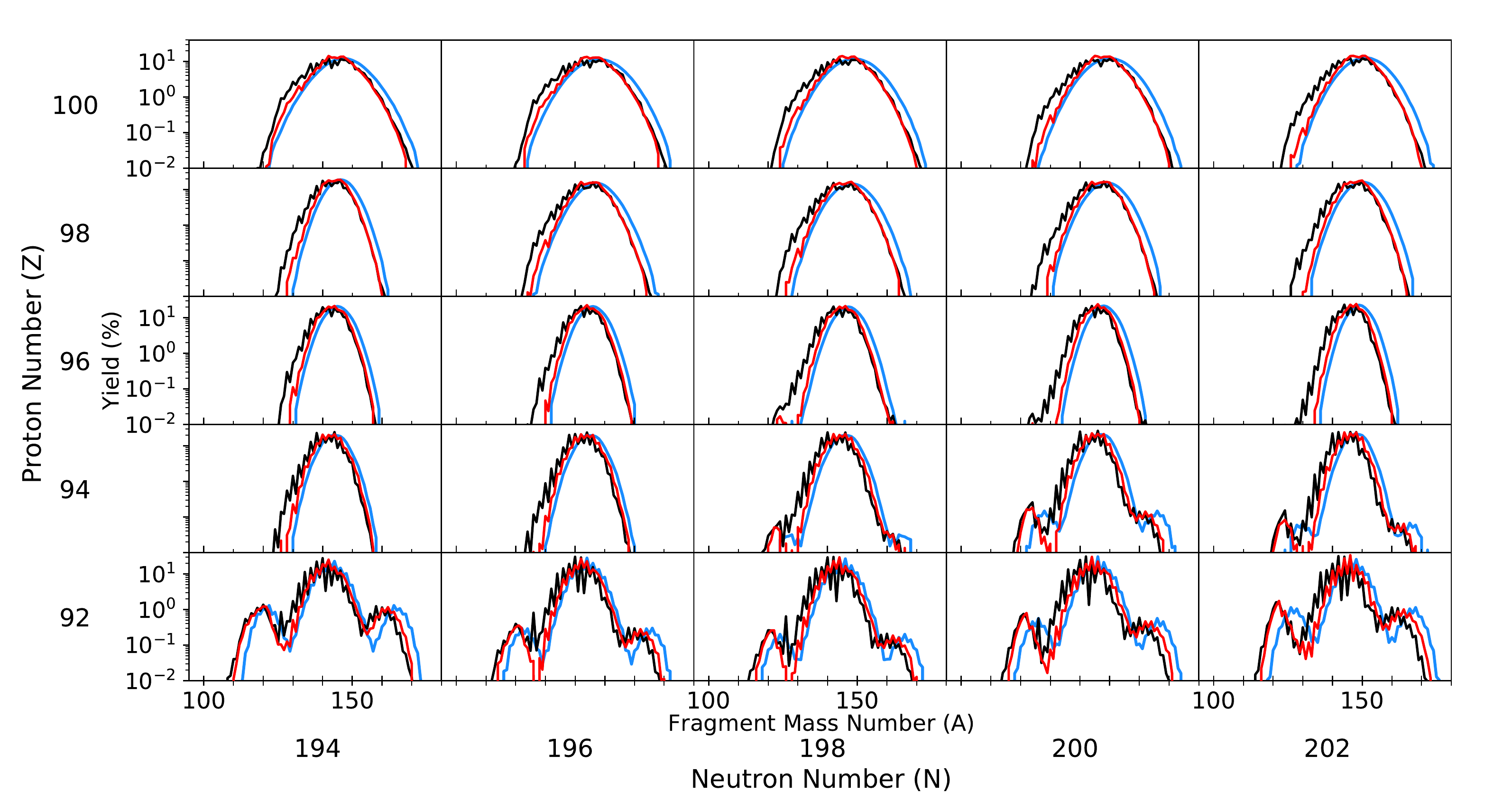}
\end{center}
\caption{(Color online) The GEF 2016 fragment yields for fission induced by a 0.1 MeV neutron (blue) along with the resultant product yields after applying the de-excitation treatments of GEF (red) and FREYA (black). The increase in the predicted prompt neutron emission shown in Fig.~\ref{fig:FREYAnu} shifts the FREYA yields to lower mass number.}
\label{fig:FREYAy}
\end{figure*}

  Figure~\ref{fig:FREYAnu} shows the resulting neutron multiplicities for spontaneous (left panels) and neutron-induced (right panels) fission from GEF 2016 (top panels) and FREYA (bottom panels).  In both cases the multiplicities are generally higher for neutron-induced fission, with $1-2$ more neutrons emitted than for spontaneous fission.  In the very neutron-rich region, in many cases the neutron multiplicities can be quite high, with some nuclei emitting $8-10$ neutrons, presumably because the outer neutrons are not strongly bound and thus emission is more probable.  It is also clear that more neutrons are emitted through the de-excitation process in FREYA than in GEF, even starting from the same initial fission fragment yields.  Both GEF and FREYA see a diagonal region of systematically higher neutron emission near $N=184$ which comes from the ability of the asymmetric yields seen in this region (recall Fig.~\ref{fig:yieldwidthNIF}) to emit more neutrons from a neutron-rich light fragment. We now describe where the differences in multiplicity may come from and discuss possible consequences of this difference later.
  
Recall that the model of excitation energy sharing in GEF is replaced in FREYA by the constant parameter $x$.  We note that in FIFRELIN and CGMF, the excitation energy sharing is also parameterized but, in those cases, mass dependent ratios are derived from data on the measured neutron multiplicity for specific isotopes as a function of fragment mass, $\overline \nu(A)$. Both FREYA's $x$ and the parameterized ratios in FIFRELIN and CGMF can reproduce the ``sawtooth" shape of $\overline \nu(A)$ in gross or fine detail, depending on which approach is employed.  The sawtooth shape is thought to arise from closed nuclear shells, in particular at $A = 132$ where there is a doubly closed shell and $\overline \nu(A)$ is at a minimum because closed-shell nuclei are harder to excite than highly deformed nuclei, resulting in fewer neutrons emitted. The $x$ parameter in FREYA gives more excitation energy to fragments with $A < 132$, resulting in a higher neutron multiplicity for the light fragment in spontaneous and thermal-neutron-induced fission, as suggested by data on well-studied nuclei such as $^{252}$Cf(sf), $^{235}$U(n,f) and $^{239}$Pu(n,f) \cite{Lemaire}.  

In the case of very neutron rich nuclei, the structure of $\overline \nu(A)$ is wholly unknown.
While FREYA will minimize $\overline \nu(A)$ in the proximity of a closed shell, forcing $x > 1$ for very neutron-rich nuclei, where shell effects may be less important, would artificially increase neutron emission, as seen in Fig.~\ref{fig:FREYAnu}.  As suggested earlier, one can take this difference in neutron emission as a theoretical uncertainty.

The enhanced neutron multiplicities in FREYA relative to GEF 2016, due to the fixed energy sharing in FREYA, as seen in Fig.~\ref{fig:FREYAnu}, also has an effect on the shape of the fission product yields, as shown in Fig.~\ref{fig:FREYAy}.  The fission product yields are shifted toward lower mass numbers for FREYA.  The larger product yields on the low mass side of symmetry, especially for fragment masses $A<132$, is the result of larger neutron emission from the light fragments in FREYA.  We will discuss the sensitivity of the $r$ process to the energy sharing and de-excitation treatment in Sec. III. The sensitivity of the $r$ process depends on how much isotopic material reaches the most neutron-rich region with $N>184$ where FREYA predicts as many as $\sim 3$ more neutrons emitted per fission than GEF. 

\begin{figure}
\begin{center}
\includegraphics[scale=0.48]{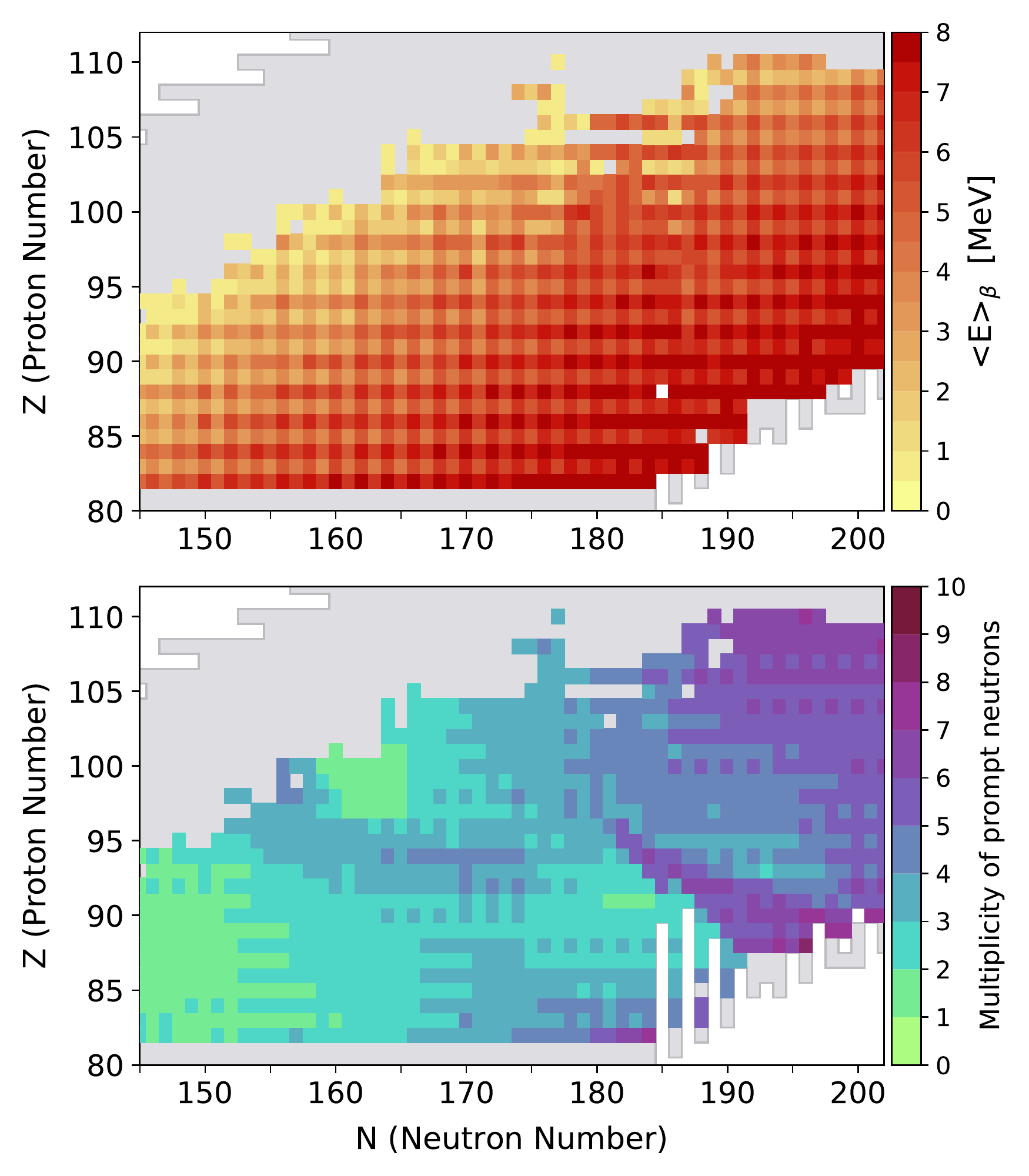}
\end{center}
\caption{(Color online) The average excitation energy for the daughter nucleus populated by $\beta$-decay calculated as in \cite{BDFrp} with $\beta$-strength functions from \cite{MollerSd0} (upper panel) and the corresponding average prompt neutron emission as predicted by GEF 2016 when this excited daughter fissions (lower panel).}
\label{fig:enubdaught}
\end{figure}

$\beta$-delayed fission is a good test of the excitation energy dependence, as seen in Fig.~\ref{fig:enubdaught}. The excitation energy of this process can range from near zero, as is the case for spontaneous fission, up to $8-9$~MeV, as seen in the upper panel of Fig.~\ref{fig:enubdaught}, typically somewhat higher than the few MeV excitation energies of neutron-induced fission relevant for the $r$ process (the average excitation energies used to calculate $\beta$-delayed fission yields from GEF 2016, shown in Fig.~\ref{fig:enubdaught}, are tabulated in Supplemental Materials).  The average neutron multiplicity in $\beta$-delayed fission, shown in the lower panel of Fig.~\ref{fig:enubdaught}, correspondingly tends to be higher than that of neutron-induced fission shown in the upper right panel of Fig.~\ref{fig:FREYAnu}.

\begin{figure*}
\begin{center}
\includegraphics[scale=0.44]{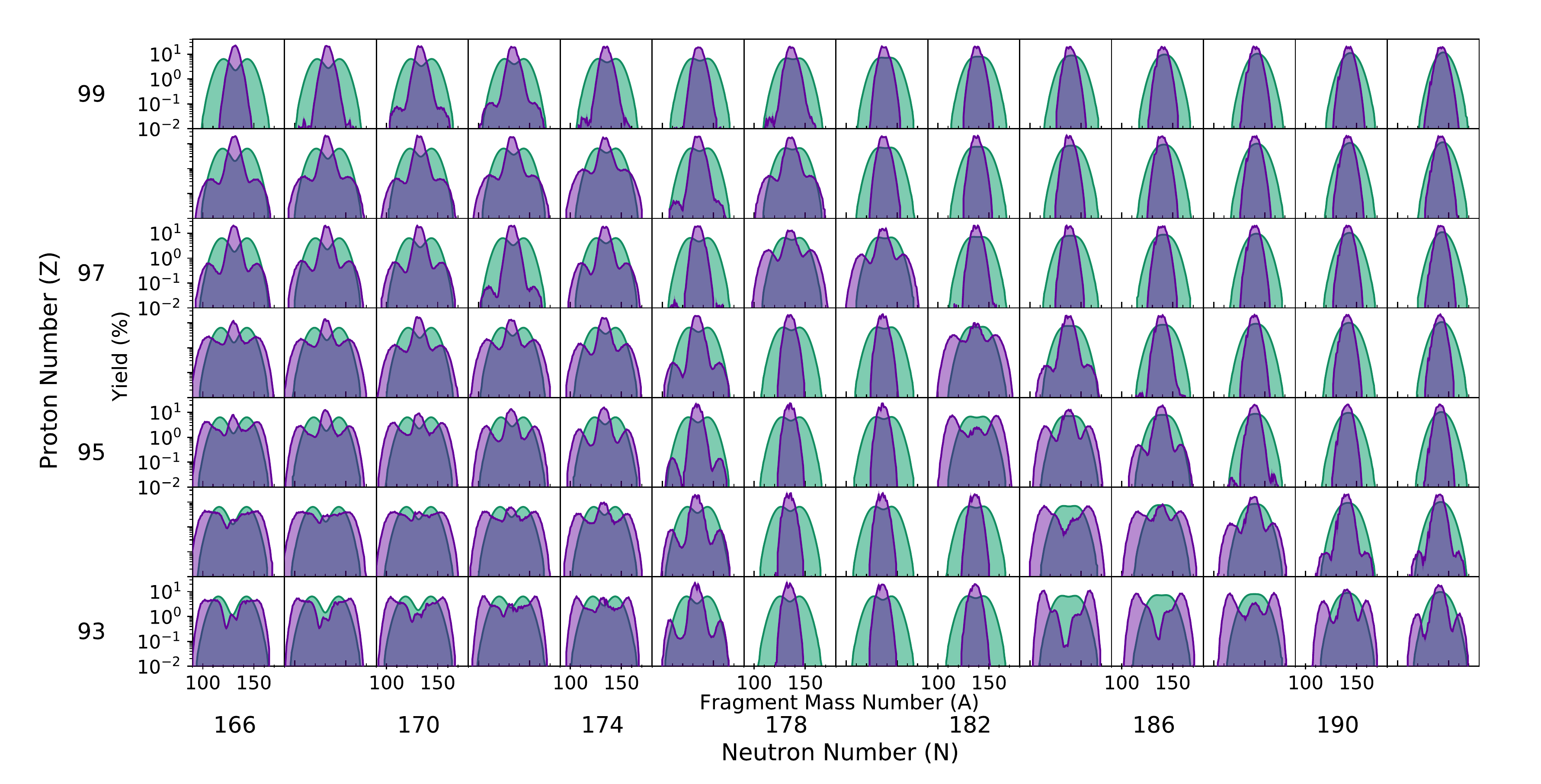}
\end{center}
\caption{(Color online) The product yields predicted for the fission of a daughter nucleus populated by $\beta$-decay after using the excitation energies shown in Fig.~\ref{fig:enubdaught} in GEF 2016 (purple) as compared to the yields from Kodama and Takahasi \cite{Kodama} (green).}
\label{fig:yieldcomp}
\end{figure*}

The $\beta$-delayed fission product yields predicted by GEF 2016, given the average excitation energy for the $\beta$ daughter shown in the upper panel of Fig.~\ref{fig:enubdaught}, also suggest that higher excitation energies will increase the asymmetry of the yields of neutron-rich nuclei.  
Compared to the case of neutron-induced fission shown in Fig.~\ref{fig:yieldsNIF}, even larger deviations from symmetry can be observed in the GEF 2016 $\beta$-delayed yields shown in purple in Fig.~\ref{fig:yieldcomp}.  For reference, the parameterized double-Gaussian fission yields of Kodama and Takahashi \cite{Kodama}, widely used in $r$-process calculations, are also shown in Fig.~\ref{fig:yieldcomp} in green.  These yields also show a transition from asymmetric to symmetric fission near neutron number $N=184$, similar to GEF 2016.  However, in regions of the nuclear chart of interest for the $r$ process, the broad distributions predicted by Kodama and Takahashi are centered near $A\sim144$ for both asymmetric and symmetric fission while the asymmetric GEF 2016 yields tend to prefer $A\sim150$ daughters. Although both models predict asymmetric yield contributions for similar nuclei, note that the GEF 2016 yields for such nuclei often contain a symmetric component as well. 

\section{Impact of Fission Yield Trends on $\bm{r}$-process abundances}\label{sec:rpimpact}

For nucleosynthesis calculations, we use the network Portable Routines for Integrated nucleoSynthesis Modeling (PRISM) developed jointly at the University of Notre Dame and Los Alamos National Laboratory \cite{BDFrp,Cote,Cfpaper}. PRISM permits a straightforward implementation of mass model-dependent nucleosynthesis rates due to its flexibility with nuclear data inputs. For the masses of neutron-rich nuclei, we first apply the Thomas-Fermi (TF) model in order to employ masses and fission barriers consistent with the GEF 2016 inputs used to determine the yields. For this, we explicitly use the barriers assumed in GEF 2016 which include corrections to TF barriers, as discussed in Sec. II. Where available we use experimental masses \cite{AME2016} as well as experimentally established half-lives and branching ratios from NUBASE \cite{NUBASE2016}. For theoretical $\alpha$-decay rates we use the well-established Viola-Seaborg formula \cite{VS1966}:

\begin{equation}
\text{log}_{10}T_{1/2}^{\alpha}(s) = \frac{aZ+b}{\sqrt{Q_{\alpha}(\text{MeV})}}+cZ+d+h_{\text{log}}
\end{equation}

\noindent where we apply a least-squares fit to NUBASE2016 half-life data that takes into account the reported experimental uncertainties when optimizing coefficients. Using this procedure we find values of $a=1.6606 \pm 0.0007$, $b=-9.2990 \pm 0.0656$, $c=-0.2121 \pm 0.0003$, and $d=-32.5432 \pm 0.0267$. Since the fit is performed using even-even nuclei \cite{VS1966}, species with unpaired nucleons are accounted for by finding the average difference between the fitted and experimental half-life values. These hindrance factors were found to be:
\begin{equation}
  h_{\text{log}}=
  \begin{cases}
    0.5325, & \text{$Z$ odd, $N$ even} \\
    0.5253, & \text{$Z$ even, $N$ odd} \\
    0.9222, & \text{$Z$ odd, $N$ odd.}
  \end{cases}
\end{equation}

\noindent We note, however, that the uncertainties in the treatment of theoretical $\alpha$-decay rates are relatively unimportant in the $r$ process since this decay is most influential at late times when the nuclei populated are within the experimentally probed regions. We use neutron capture, $\beta$-decay, neutron-induced fission and $\beta$-delayed fission rates as in Refs. \cite{Kawano+17,Mumpower+14,Mumpower+16,BDFrp,MollerQRPA}, with all rates determined from the same model masses as in Ref. \cite{Mumpower+15} and updated to be self-consistent with the fission barrier heights of a given model. For spontaneous fission we apply a parameterized prescription with a simple dependence on barrier height as in Refs. \cite{Karpov,Zagrebaev}. Therefore with the same fission barriers used to determine the fission yields and rates of all fission reaction and decay channels, our calculations which apply TF inputs represent the most fully self-consistent fission cycling $r$-process calculations in this work. Note that in this section, we focus on the influence of the fission yields in the $r$ process and therefore keep the fission rates identical in all comparisons shown.

\begin{figure*}
\begin{center}
\includegraphics[scale=0.48]{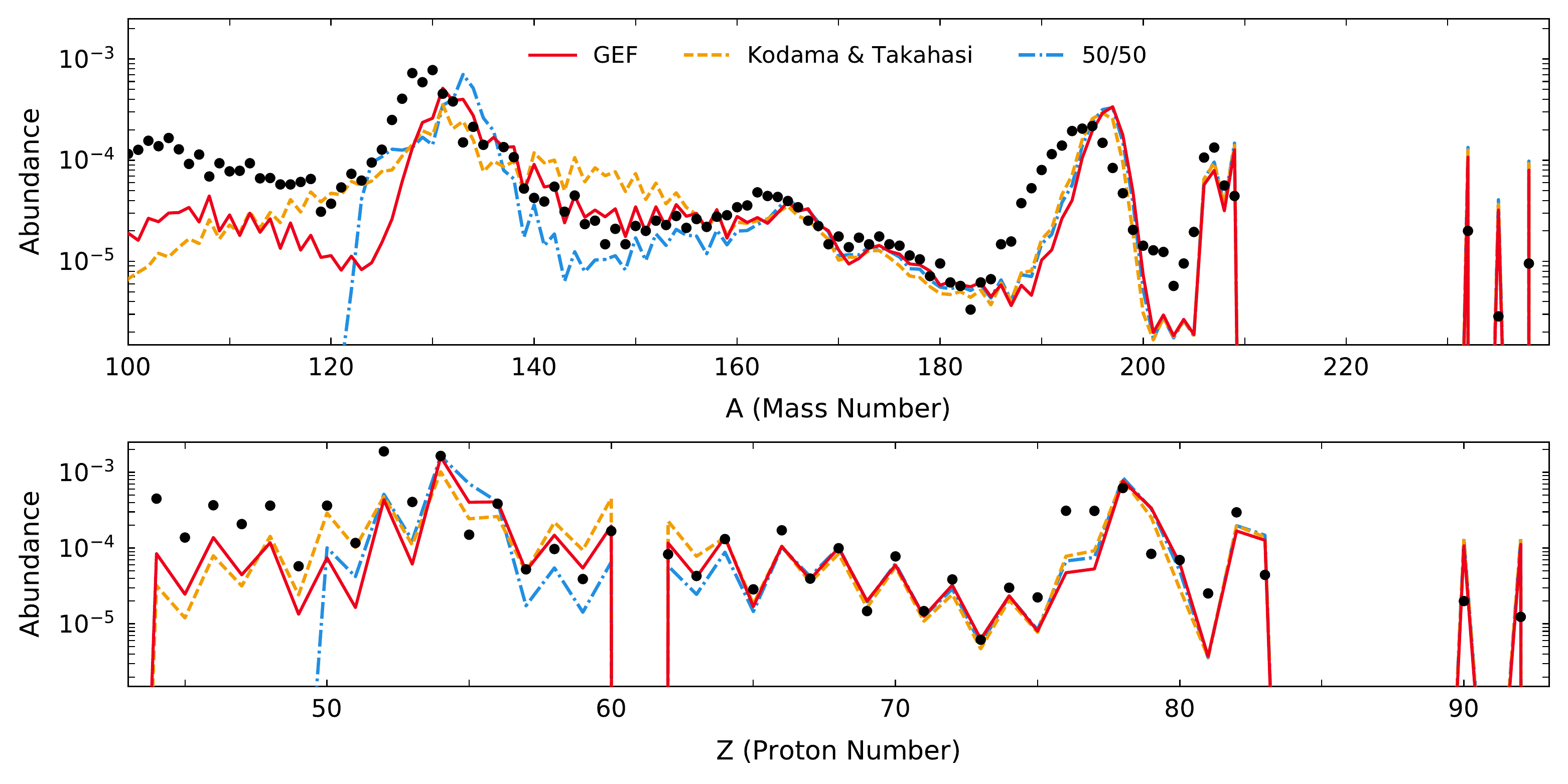}
\end{center}
\caption{(Color online) The $r$-process abundances at 1 Gyr as a function of mass number (upper panel) and charge number (lower panel) using GEF 2016 yields for spontaneous fission, neutron-induced fission and $\beta$-delayed fission (red) as compared to using the fission yields of Kodama and Takahasi (orange) and simple symmetric splits (blue). The solar data is that of Sneden {\it et al.} \cite{Sneden}.}
\label{fig:abFFD}
\end{figure*}

\begin{figure*}
\begin{center}
\includegraphics[scale=0.585]{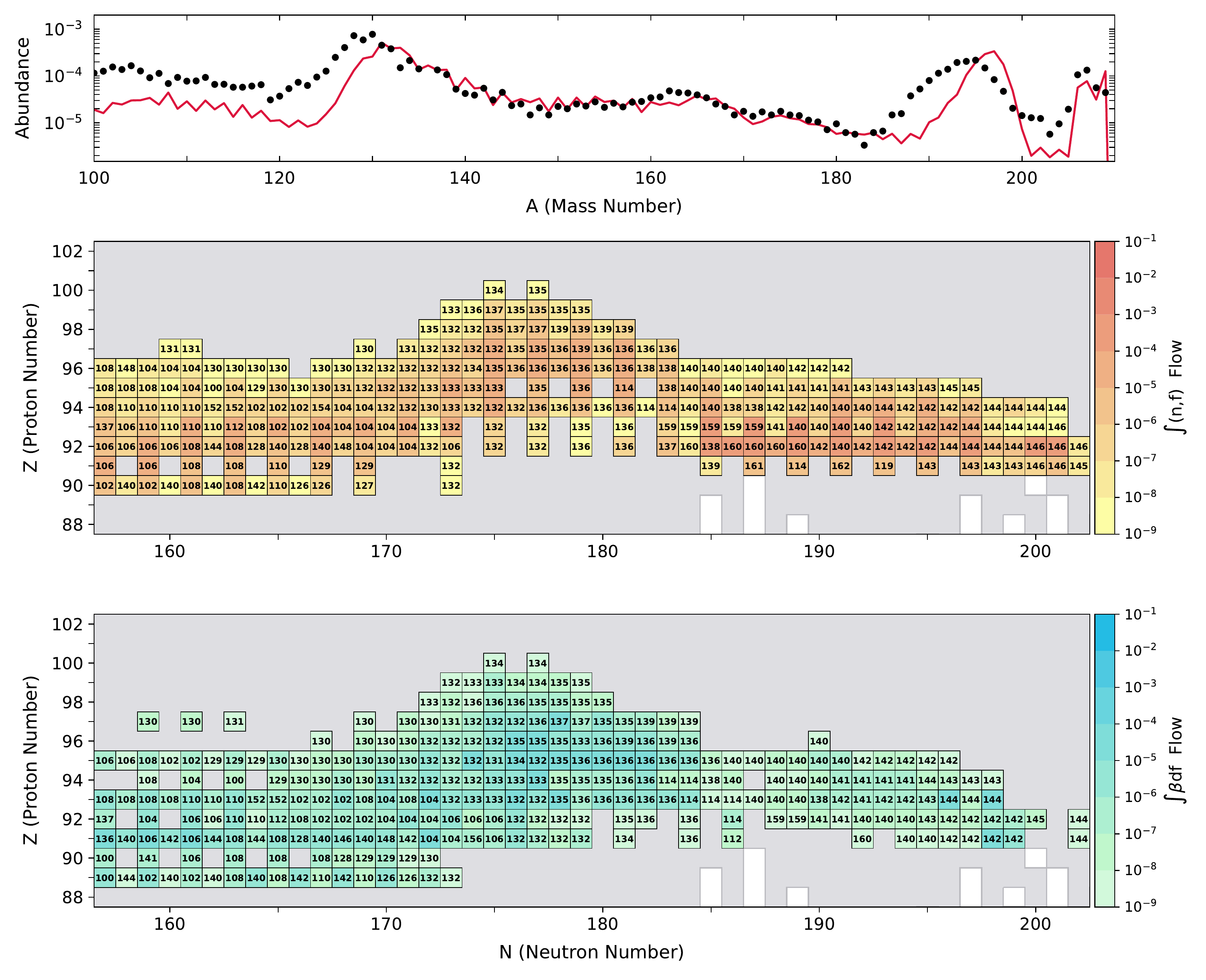}
\end{center}
\caption{(Color online) The fission flow ($\lambda_i Y_i$ where $\lambda_i$ is the rate of the reaction or decay and $Y_i$ is the abundance) for nuclei undergoing neutron-induced (middle panel) and $\beta$-delayed (lower panel) fission integrated over time along with the resultant $r$-process abundance pattern (upper panel). The numbers labeling nuclei in middle and lower panels denote the mass number at the location of the maximum of the daughter fission yield distribution (here from GEF 2016).}
\label{fig:rtraj1}
\end{figure*}

For astrophysical conditions, we consider dynamical ejecta from a 1.2--1.4 $M_{\odot}$ neutron star merger simulation \cite{Rosswog,PiranRoss} for which we calculate nuclear reheating self-consistently with the chosen nuclear inputs when extrapolating beyond the reported simulation trajectory (with an assumed $50\%$ heating efficiency as in \cite{Oleg}; note heating efficiencies have been argued to lie between $25-100\%$ \cite{Metzger2010} and often a variety of efficiency values are considered \cite{Jonas}). We extrapolate the density by assuming free expansion as in Korobkin {\it et al.} \cite{Oleg} and assume NSE at 10 GK to obtain seed nuclei abundances using the SFHo equation of state \cite{SFHoSteiner}. Since these simulation trajectories are publicly available, we refer to the original number labeling (1--30) in order to permit direct comparisons with the results presented here. We first consider a ``cold", very neutron-rich ($Y_e=0.01957$) tidal tail trajectory in this set (traj.\ 1) which permits significant fission cycling due to its extreme neutron richness. Here the term ``cold" when used in reference to an astrophysical trajectory implies photodissociation drops out of equilibrium early, leaving $\beta$-decay to compete with neutron capture. In contrast, the term ``hot" when applied to trajectories implies conditions which support an extended (n,$\gamma$)$\rightleftharpoons$($\gamma$,n) equilibrium.

We show the impact of the asymmetric-to-symmetric yield trends predicted by GEF 2016 as compared to more simplistic descriptions in such fission rich environments in Fig.~\ref{fig:abFFD}. The underproduction at $A\sim144$ seen with a simple symmetric split (50/50) is a consequence of their narrow distribution which exclusively deposits material near $A\sim130$ when $r$-process material encounters the region with $Z\gtrsim90$ at $N<184$. In contrast, the fission yields of Kodama and Takahashi, whose $r$-process impact has been previously explored in Refs. \cite{Eichler15,Eichler16,BDFrp}, are exclusively asymmetric at $N<184$ and transition to symmetric distributions at higher neutron number (see Fig. \ref{fig:yieldcomp}). However the width of these fission yields places material over a fairly broad range around $A\sim144$ at both early and late times. The overproduction of the light lanthanides with the Kodama and Takahashi yields is a direct consequence of its global preference for $A\sim144$ fission daughters. The results obtained with the GEF 2016 yields here predict a more gradual fall off for the right edge of the second $r$-process peak, and the lanthanide abundances in this region, as well as near $A\sim150$, follow solar data trends. This lanthanide abundance behavior given the GEF 2016 yields is due to the transition from asymmetric to symmetric yields discussed in Sec. II.

\begin{figure*}
\begin{center}
\includegraphics[scale=0.51]{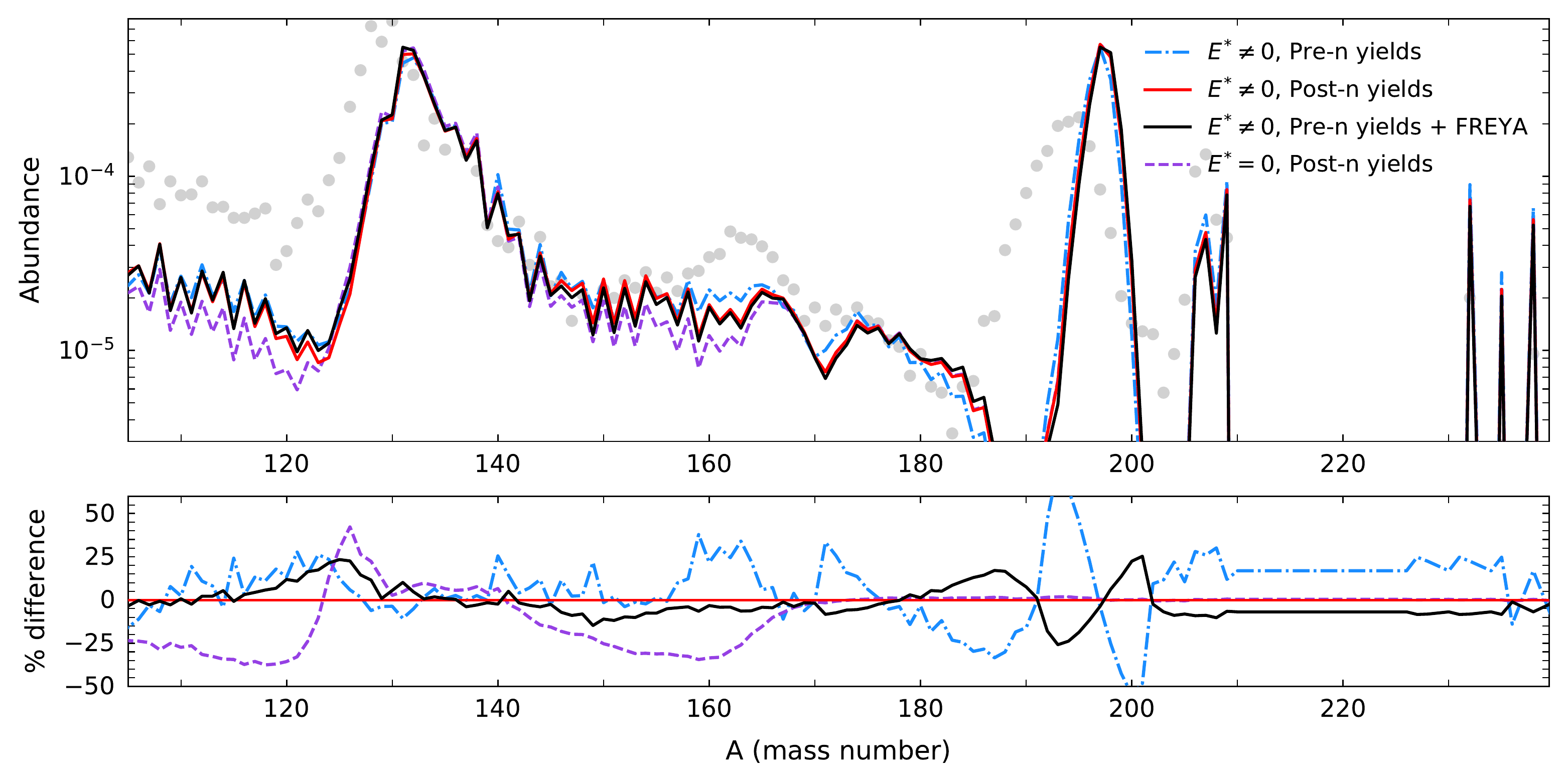}
\end{center}
\caption{(Color online) An example of the $r$-process abundance pattern sensitivity to the treatment of excited fission fragments from examining results using GEF 2016 fragment yields (pre-neutron emission, blue), GEF product yields (post-neutron emission, red), FREYA product yields (black), and GEF product yields with zero excitation energy ($E^*=0$) (purple). The upper panel shows abundances at 1 Gyr while the lower panel shows the percent difference relative to results with the GEF 2016 product yields which account for excitation energy ($E^*\neq0$).}
\label{fig:abFREYA}
\end{figure*}

To explicitly demonstrate the role that GEF 2016 yield trends play in determining the $r$-process abundance pattern, the integrated fission flow can be cross checked with the fissioning daughter yields for these nuclei as shown in Fig.~\ref{fig:rtraj1}.  The fission flows are calculated with raw $r$-process abundances which satisfy $\sum Y_iA_i =1$ and no other scalings are applied so that the flows of different fission processes, such as neutron-induced and $\beta$-delayed fission, are directly comparable. In Fig.~\ref{fig:rtraj1}, we label each reactant species at $Z$, $N$ with the mass number which locates the maximum of the fission yield for the fissioning nucleus $Z_f$, $N_f$ (where $Z_f = Z+1$, $N_f = N-1$ for $\beta$df and $Z_f = Z$, $N_f = N+1$ for (n,f)). In the case of symmetric yields, the mass number, $A$, in the boxes of Fig. \ref{fig:rtraj1} will be $A\sim (N_f+Z_f)/2$, where $N_f$ and $Z_f$ correspond to the fissioning nucleus populated from the reactant species at $Z,N$. If the fission is asymmetric, the mass number shown is for one fragment peak so material will be deposited at $A'\sim (N_f+Z_f)-A$ as well. Of course the yields can contain both symmetric and asymmetric contributions that are similarly probable. However, it is still instructive to see explicitly where each fissioning nucleus will preferably deposit material. With the GEF 2016 yields, the $r$-process material in ``cold" dynamical ejecta trajectories, as in the very neutron-rich tidal tail ejecta conditions of Fig.~\ref{fig:abFFD}, first encounters symmetric yields centered at $A\sim144$ beyond the $N=184$ shell closure. Significant fission flow passes through this region as neutron-induced fission acts to terminate the $r$ process. As material decays back to stability, nuclei at neutron numbers below $N=184$ with substantial asymmetric yield contributions are encountered, placing material primarily near $A\sim110,130,$ and $150$. The contributions from fission products at early times near $A\sim144$, followed by late-time contributions near $A\sim130,150$, work to smooth the right edge of the second $r$-process peak. It is interesting to note that with the GEF 2016 yields there are only a few nuclei with $A\sim160$ fission daughters which place material directly into the rare-earth peak region. Therefore, with these yields it is not possible to explain rare-earth peak formation by a late-time deposition of fission material at $A\sim164$, as can occur with the four hump fragment distributions predicted by the SPY yield model \cite{GorielyGEF}. 

We next turn to examine the influence of prompt fission neutrons in the $r$-process. Here we consider merger dynamical ejecta trajectory 22 with $Y_e\sim0.054$ which starts similarly cold to traj.\ 1 but later reaches higher temperatures. The most straight-forward approach to evaluate the impact of fission neutrons is to compare the $r$-process abundances using the GEF 2016 fragment yields (pre-neutron emission) and GEF 2016 product yields (post-neutron emission) as is shown in Fig.~\ref{fig:abFREYA}. The effect of prompt neutron emission on the $r$-process abundance pattern is two fold. The widening of the yield distributions toward lower mass number that accompanies neutron emission places more material to the left of the $N=82$ shell closure. This can lead to an increase on the order of $10\%$ in the height of the second peak. The increase of material held at the $N=82$ shell closure effectively decreases the number of available isotopes that can neutron capture back up past the $N=126$ shell closure which therefore reduces overall actinide abundances. This tendency to decrease the ability of material to access the heaviest nuclei also implies less fission activity in the region of mostly asymmetric yields between $N=126$ and $N=184$ which leads to less lanthanide material at $A\sim150$. The influence of the extra neutrons alone can be seen from the overall shift in the pattern to the right of the rare-earth peak toward higher mass number. Such a narrowing of the third $r$-process peak due to late-time neutron capture from $\beta$-delayed neutron emission and prompt neutron emission from excited fission fragments has been noted in previous work \cite{Goriely+13,Oleg,Eichler15,GorielyGMPGEF}. 

The influence of the fission fragment energy sharing and de-excitation treatment, which determines prompt neutron emission, is also demonstrated in Fig.~\ref{fig:abFREYA} by the $r$-process abundances using the FREYA fission product yields as compared to those from GEF 2016. The widening of the yields toward lower mass number (recall Fig.~\ref{fig:FREYAy}) is primarily responsible for the differences seen between GEF and FREYA, as can be seen from the increase in height of the second peak produced by FREYA yields which place more material below $N=82$ relative to GEF 2016 yields. 
Although the treatment of prompt neutron emission can lead to differences in the main $r$-process peaks, these effects are modest relative to the influence of the global yield trend demonstrated in Fig.~\ref{fig:abFFD}, which points to the fragment yields (prior to neutron emission) as being of primary importance for $r$-process calculations. 

The finding that the location of fission fragment deposition is of dominant importance over prompt neutron emission is further demonstrated in Fig.~\ref{fig:abFREYA} by the effect of using excitation-energy dependent sets of fission yields for neutron-induced and $\beta$-delayed fission as compared to applying the zero excitation energy yields of spontaneous fission for all fission processes. The enhanced asymmetries in the fission yields discussed in Sec. II (and shown in Fig. \ref{fig:yieldsNIF}) due to an excited compound parent nucleus places more material at $A\sim110$ and $150$ as compared to the case when all fission processes make use of spontaneous fission yields. These enhanced asymmetric contributions produce a clear signature in the final abundance pattern with the fission yields which account for excitation energy giving a result more consistent with observed solar data. Meanwhile, the differing prompt neutron emission between these cases (recall from Figs.~\ref{fig:FREYAnu} and \ref{fig:enubdaught} that neutron-induced and $\beta$-delayed yields will contribute slightly more prompt neutrons than spontaneous fission) shows no clear effect on the abundances in the third peak region where late-time neutrons shape the final results.

\begin{figure}
\begin{center}
\includegraphics[scale=0.49]{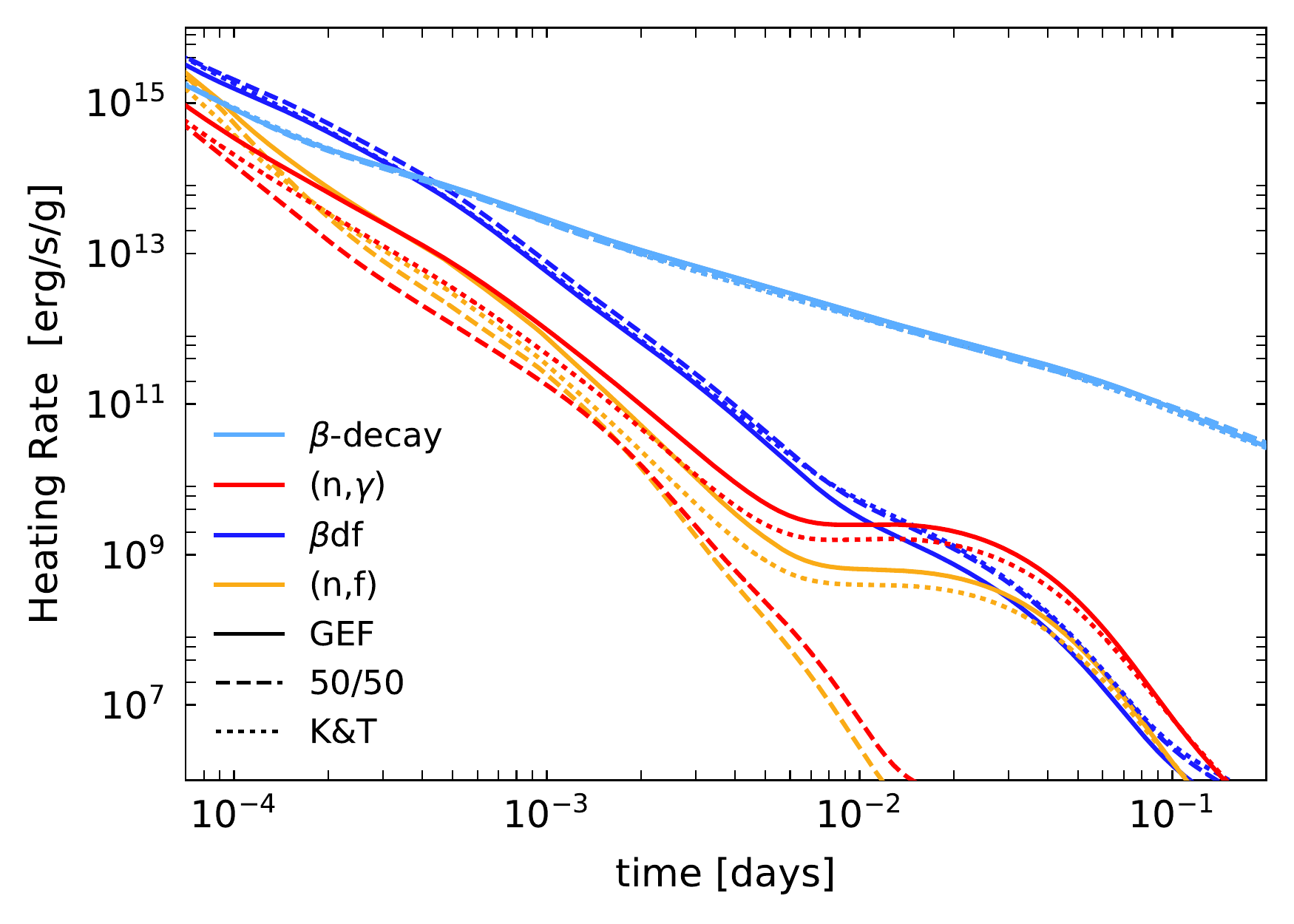}
\end{center}
\caption{(Color online) The nuclear heating rates for $\beta$-decay (blue), $\beta$-delayed fission (light blue), neutron capture (red) and neutron-induced fission (orange) as a function of time using GEF 2016 fission yields (solid lines) as compared to results using simple symmetric splits (dashed lines) and the yields of Kodama and Takahashi (dotted lines).  The masses and fission barriers applied are those of TF and GEF 2016 respectively.}
\label{fig:heatyield}
\end{figure}

We next consider the effect of more realistic fission yields and prompt neutron emission, as calculated by GEF 2016 and FREYA, on the nuclear heating rates needed to calculate kilonova light curves. For such heating rate calculations, we use the ``cold" dynamical ejecta conditions applied in Figs.~\ref{fig:abFFD} and \ref{fig:rtraj1}. We calculate the heating contributions from $\beta$-decay, neutron capture, $\beta$-delayed fission, and neutron-induced fission as flow $\times$ $Q$-value for each channel and compare the results using GEF and FREYA to simpler treatments in Fig.~\ref{fig:heatyield}.  
We first focus on the heating before $\sim 1$ day when $\beta$-delayed and neutron-induced fission are most active.  Both the $\beta$-decay and $\beta$-delayed fission heating rates show some dependence on the yield distribution, however it is the neutron capture reaction channels that show a rather pronounced sensitivity to the fission yields. Applying simple symmetric splits produces early time heating contributions from neutron capture and neutron-induced fission which are about three orders of magnitude lower than those predicted using the GEF 2016 yields. This difference comes from the tendency of symmetric splits to deposit material in a concentrated region near the $N=82$ shell closure as compared to the wider distribution of daughter products seen with GEF 2016. The late-time addition of nuclei to the right of the $N=82$ shell closure permits further neutron capture and therefore more material is driven up in mass number toward the fissioning regions near $N=184$, correspondingly increasing the heating contribution from neutron capture and neutron-induced fission. A comparison of heating rates when using the yield model of Kodama and Takahashi, which does not contain a prescription for neutron emission from excited daughter fragments, confirms that it is the narrow placement of daughter nuclei near $N=82$ that is most responsible for the lower heating rate for neutron-induced processes in the symmetric yield case. The inclusion of prompt neutron emission from the excited fragments as in GEF can further increase the heating from neutron capture reaction channels by a factor of around three as compared to Kodama and Takahashi. We note that applying the yields obtained using the FREYA energy sharing and de-excitation treatment, while not shown in Fig.~\ref{fig:heatyield}, can further increase the heating from neutron capture processes by roughly $10\%$ at these early times.

\begin{figure}
\begin{center}
\includegraphics[scale=0.49]{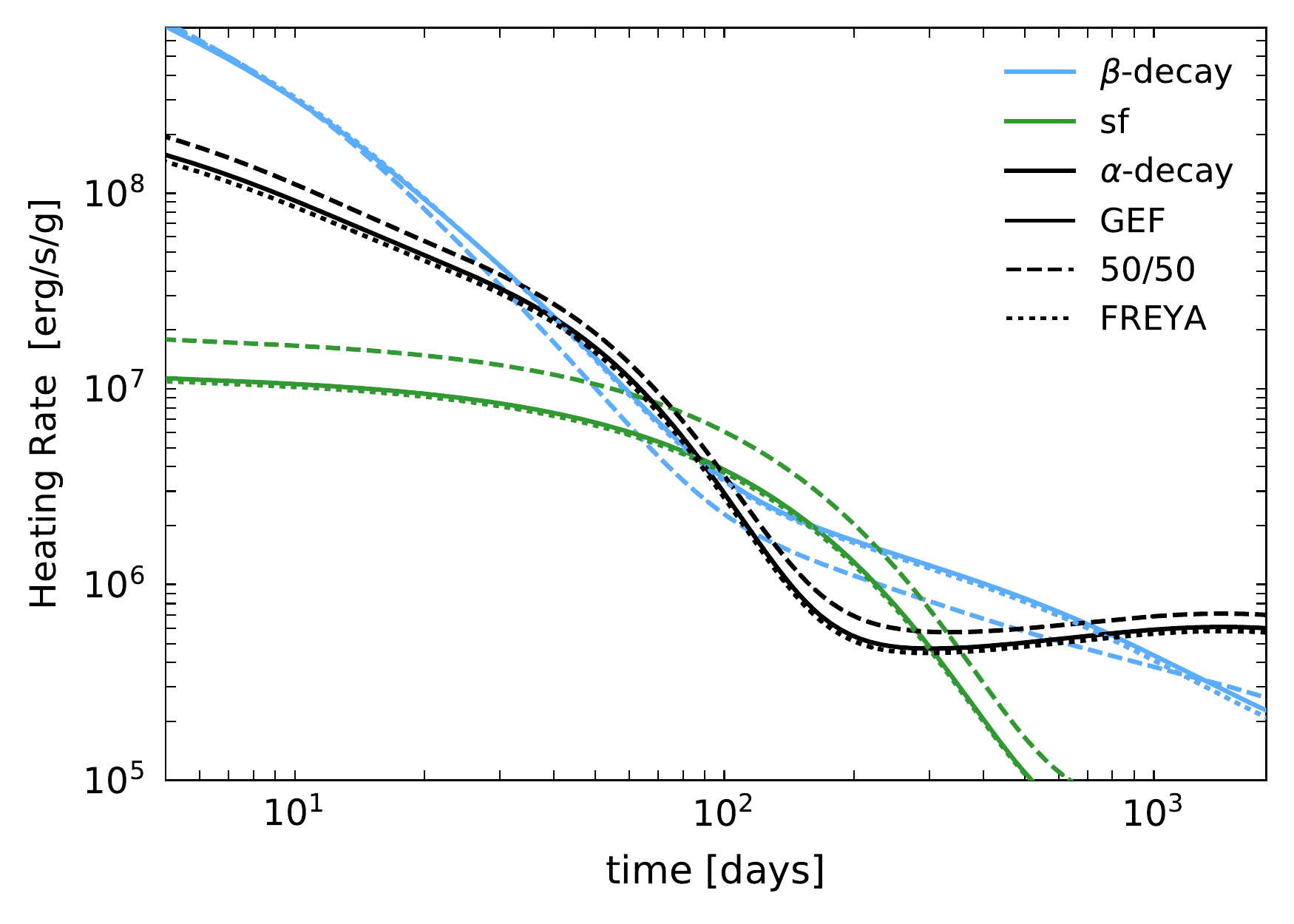}
\end{center}
\caption{(Color online) The nuclear heating rate for $\beta$-decay (blue), alpha decay (black) and spontaneous fission (green) as a function of time using GEF 2016 fission yields (solid lines) as compared to results using simple symmetric splits (dashed lines) and the yields with the FREYA energy sharing and de-excitation treatment (dotted lines). The masses and fission barriers applied are those of TF and GEF 2016 respectively.}
\label{fig:spfheat}
\end{figure}

When timescales on the order of days or longer are considered, the influence of the fission treatment on nuclear heating becomes more pronounced due to the late-time dominance of fission, specifically the spontaneous fission of $^{254}$Cf \cite{Cfpaper}. Fig.~\ref{fig:spfheat} shows the heating rate for $\beta$-decay, spontaneous fission, and $\alpha$-decay when the yield distributions from GEF 2016 and FREYA are compared with those obtained using simple symmetric splits. The 50/50 splits result in the largest predicted late-time dominance of the spontaneous fission heating with the greatest deviation between $\beta$-decay and spontaneous fission heating curves. The spread of fission recycled material produced by the GEF yields can produce nuclei closer to the neutron dripline where $\beta$-decay rates are faster, thereby increasing the total effective $\beta$ heating. When considering the extra late-time neutrons and increased yield widths predicted by FREYA, Fig.~\ref{fig:spfheat} shows that $\alpha$-decay and spontaneous fission heatings are decreased relative to the GEF case with slightly more material getting stuck near $N=82$ and therefore less material populating the highest mass number regions (as seen in Fig.~\ref{fig:abFREYA}). Figures~\ref{fig:heatyield} and \ref{fig:spfheat} explicitly demonstrate that in fission cycling conditions the exact details of the heating rates applied to calculations of kilonova light curves depend on the nuclear physics assumptions for the heaviest, fissioning $r$-process nuclei. 

\begin{figure}
\begin{center}
\includegraphics[scale=0.545]{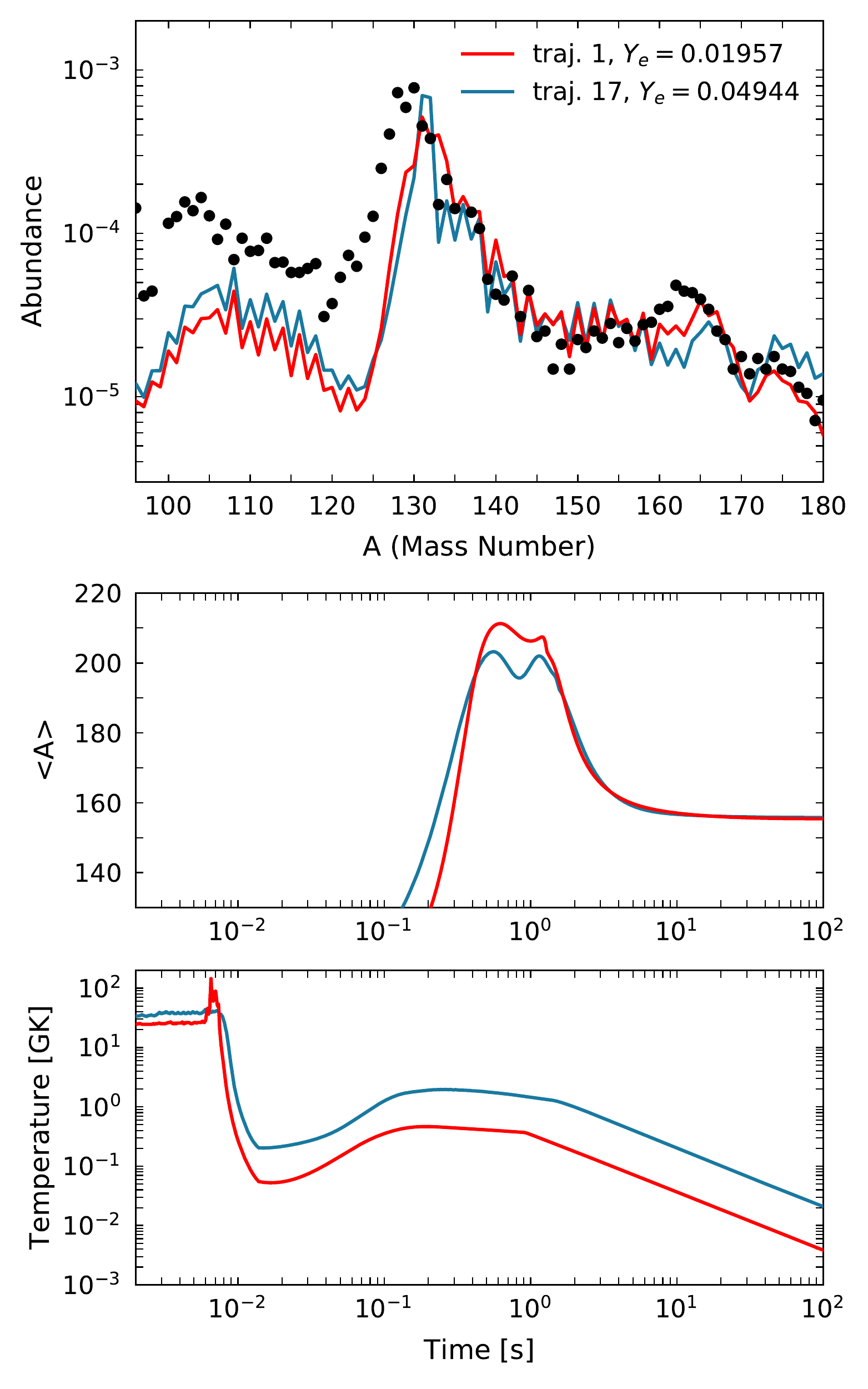}
\end{center}
\caption{(Color online) The $r$-process abundance pattern (upper panel) given two dynamical ejecta trajectories from a 1.2--1.4 M$_{\odot}$ neutron star merger simulation \cite{Rosswog} (lower panel) along with the abundance weighted mass number as a function of time for each set of conditions (middle panel).}
\label{fig:rtraj117}
\end{figure}

As described above, exactly how fission yields shape abundance patterns and heating curves depends sensitively on which nuclei are fissioning. This is in part set by the $r$-process astrophysical conditions. To examine this sensitivity, we consider astrophysical conditions from the same 1.2--1.4 M$_{\odot}$ neutron star merger simulation \cite{Rosswog} used throughout this section. While all thirty of these dynamical ejecta trajectories are similarly neutron-rich, with $Y_e$ ranging from $\sim0.015-0.055$, they exhibit a variety of density and temperature profiles. A comparison of the final abundances (upper panel), abundance wighted mass number (middle panel), and temperature profile (lower panel) for the two trajectories found to represent the extremes in temperature evolution can be seen in Fig.~\ref{fig:rtraj117}. In contrast with the ``cold" conditions of trajectory 1, in the ``hot" dynamical ejecta conditions of trajectory 17, the $r$-process path does not significantly populate the mostly symmetric yield region past $N=184$ and therefore underproduces near $A\sim140$ relative to results with cold conditions. This is partially due to the ability of photodissociation to prevent material from reaching the most neutron-rich nuclei past $N=184$ but also due to the slightly higher $Y_e$ ($\sim0.049$) producing somewhat lower fission flow (total integrated fission flow for traj.\ 17 of 0.00488 as compared to 0.00567 for traj.\ 1). The region below $N=184$ with mostly asymmetric yields, however, is still accessed in such hot conditions resulting in increases to the abundances near $A\sim100,150$ (to show this behavior explicitly the version of Fig. \ref{fig:rtraj1} given the astrophysical conditions of traj.\ 17 used in Fig. \ref{fig:rtraj117} has been included in Supplemental Materials). Therefore which fissioning nuclei are accessed is influenced by the impact of the astrophysical conditions on the location and termination point of the $r$-process path.

\section{Variations in other nuclear inputs influencing fission deposition}\label{sec:rpMMastro}

\begin{figure*}
\begin{center}
\includegraphics[scale=0.53]{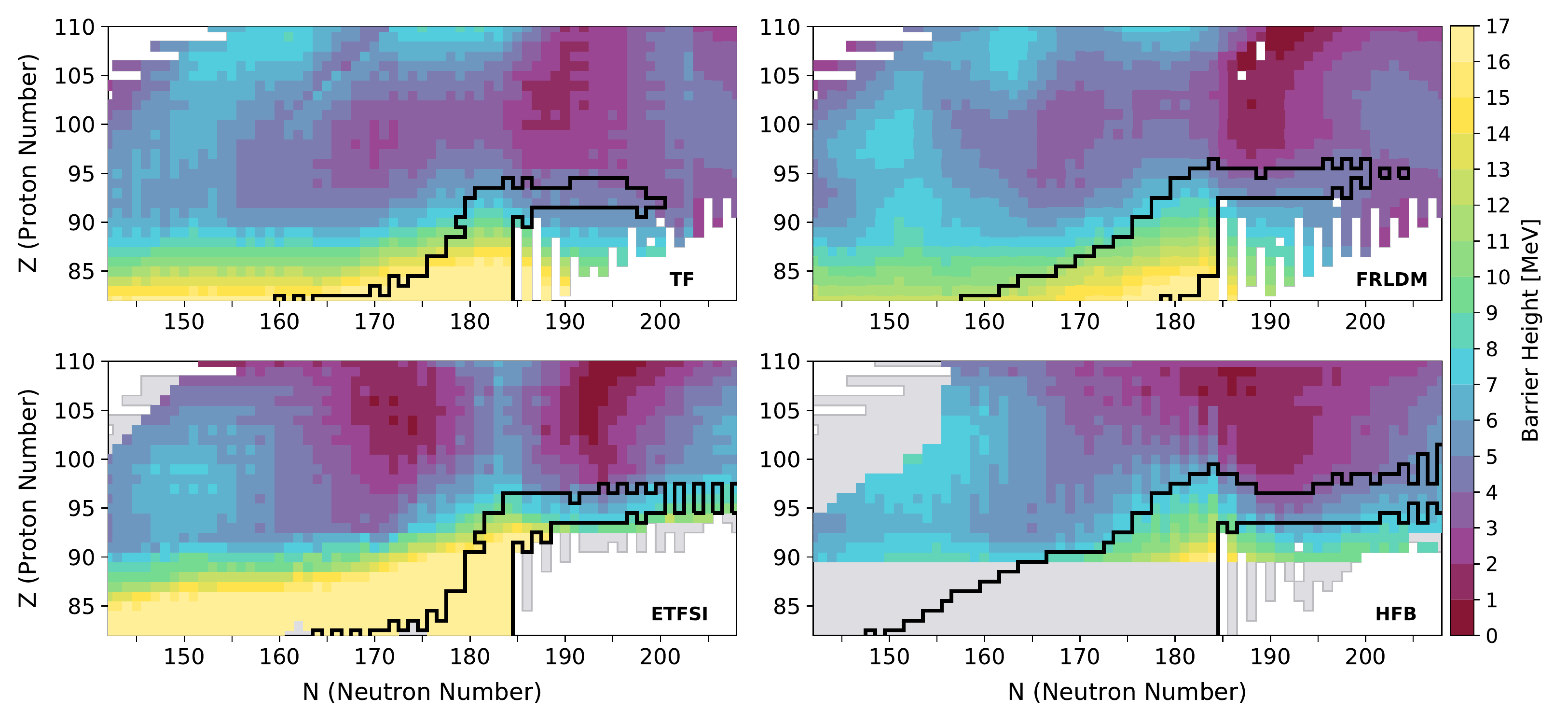}
\end{center}
\caption{(Color online) Fission barrier heights for GEF 2016 (TF) (upper left), FRLDM (upper right), ETFSI (lower left) and HFB-14 (lower right) models along with a snapshot of the $r$-process material with an abundance $\ge 10^{-10}$ (black outline) just before the $r$-process path begins to move back toward stability in the cold, dynamical ejecta conditions of trajectory 1.}
\label{fig:bhrp}
\end{figure*}

Having established that the location and termination point of the $r$-process path influences which fissioning nuclei will most impact the $r$ process, we next consider how other nuclear inputs, such as the nuclear masses, fission barriers, and $\beta$-decay rates, can affect the nuclear flow. The connection between the path termination point and the fission yields most influencing $r$-process abundances has been made previously. For example, in Ref. \cite{ShibagakiMathews} the authors find that termination near $N=184$ along with symmetric yields deposits daughter nuclei near the second $r$-process peak while higher mass number termination points see more influence from their broad yield distributions for high mass nuclei. Here we explore such considerations by examining the termination behavior predicted by different mass models. We show that it is not only how far the $r$-process path proceeds, but also the structure of predicted fission barriers near $N=184$, which determine the fission yields of most relevance.

To study these dependencies, we compare results when employing the Finite Range Droplet Model (FRDM2012), Thomas-Fermi (TF), Hartree-Fock-Bogoliubov (HFB-17), and Extended Thomas-Fermi with Strutinsky Integral (ETFSI) mass models. We chose these four models since the data for both the masses and fission barriers consistent with model masses are publicly available and commonly used in $r$-process calculations. All reaction and decay rates are consistently calculated with model masses as in Ref. \cite{Mumpower+15} and all fission rates are updated based on the fission barrier heights of a given model. The fission barriers are illustrated in Fig.~\ref{fig:bhrp}. The barriers for the Finite Range Liquid Droplet Model (FRLDM) are from \cite{MollerBH1,MollerBH2}, Thomas-Fermi barriers are again considered to be those applied in GEF 2016, and ETFSI barriers taken from \cite{Mamdouh98,Mamdouh01}. The HFB-14 fission paths used to determine the barriers are available for $Z\geq 90$ in the BRUSLIB database \cite{BRUSLIB} as well as TALYS. When the data truncates at a mass or charge number, as for HFB barriers at $Z<90$ and ETFSI masses at $A>300$, we apply rates based on FRDM2012 masses and FRLDM barriers for the absent nuclei. For the fission yields in this section, we apply the default GEF 2016 distributions presented in Sec. II and do not update these to reflect the barriers of each model. With a fixed fission yield model, we can study how the abundance pattern is shaped by these yields when different sets of fissioning nuclei, as determined by the fission barriers, are accessed.

\begin{figure*}
\begin{center}
\includegraphics[scale=0.53]{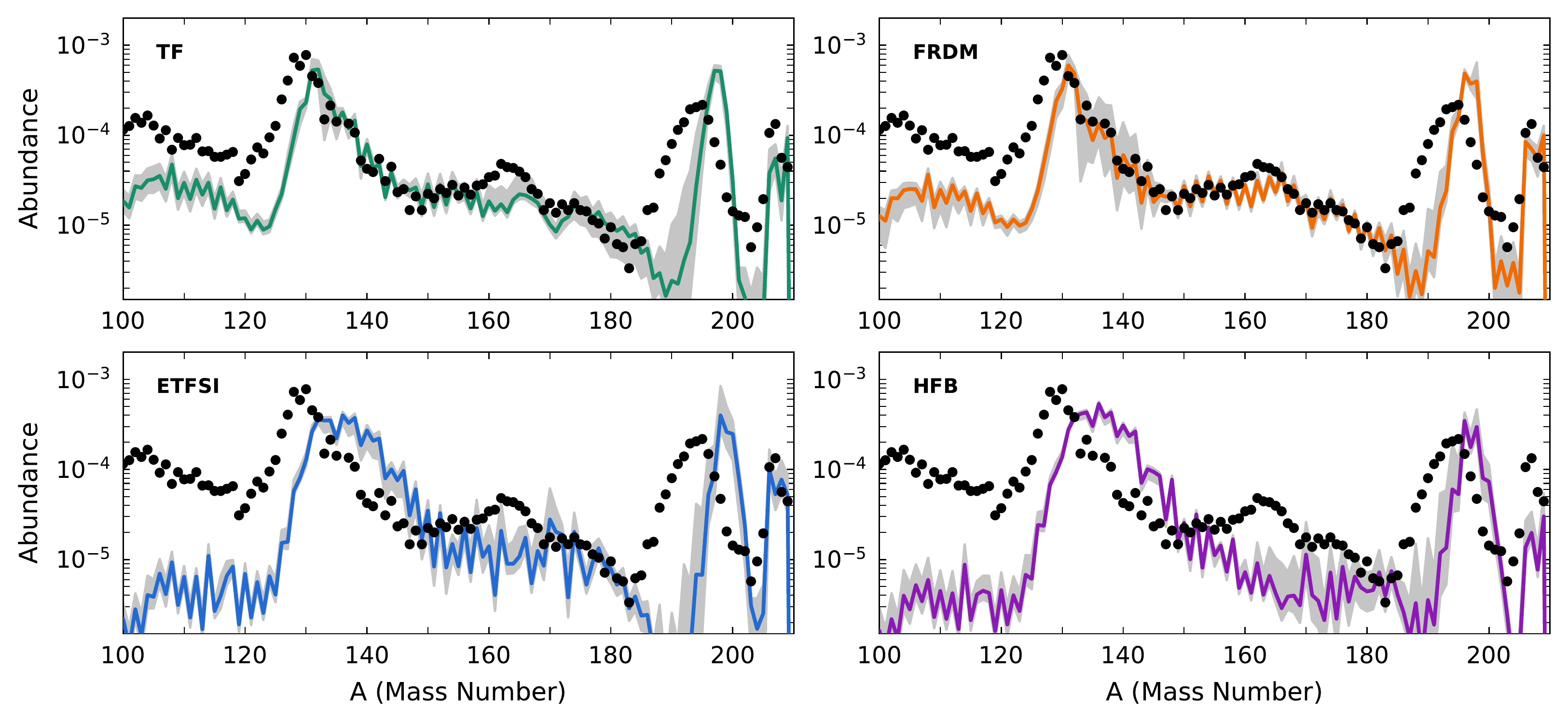}
\end{center}
\caption{(Color online) Range (grey band) and average (colored line) for the final $r$-process abundances with TF (upper left), FRDM (upper right), ETFSI (lower left) and HFB (lower right) models given thirty 1.2--1.4 M$_{\odot}$ neutron star merger simulation trajectories \cite{Rosswog}.}
\label{fig:rpRossAll}
\end{figure*}

We first consider ``cold" dynamical ejecta conditions of trajectory 1 \cite{Rosswog} in order to examine the case in which the nuclear flow can reach the highest possible mass numbers because it is less impeded by photodissociation. The black outline in Fig. \ref{fig:bhrp} shows the location of the most populated nuclei (abundance $\ge 10^{-10}$) just before the $r$-process path begins to move back toward stability (taken to be the time when the abundance weighted mass number reaches its last maximum). With FRDM and TF, material encounters relatively low barrier heights $\sim 4-5$ MeV as it pushes past the $N=184$ shell closure, permitting fission to occur quickly and thus preventing the nuclear flow from continuing much higher in mass number. As noted in Ref. \cite{Eichler15}, we find HFB permits the synthesis of heavier nuclei than FRDM and TF models due to higher barriers near the predicted $N=184$ shell closure. The same is true for the ETFSI model \cite{Petermann,Eichler16}, which was previously found to have its nuclear flow terminated by $\beta$-delayed fission \cite{Petermann} instead of neutron-induced fission which terminates the path using TF and FRLDM barriers. We also find $\beta$-delayed fission to be more active in the ETFSI case, with a $\sim60\%$ enhancement in the total integrated $\beta$-delayed fission flow relative to the flow found with TF or FRDM. We note that the ability to synthesize nuclei with higher mass number makes $r$-process calculations using ETFSI and HFB models more sensitive than results with FRDM and TF to the increase in prompt neutrons and widening of yields predicted by FREYA which become more significant at higher neutron and proton numbers (as shown in Figs.~\ref{fig:FREYAnu} and \ref{fig:FREYAy}) (to show this explicitly the version of Fig. \ref{fig:abFREYA} given the astrophysical conditions of traj.\ 22 with HFB model inputs has been included in Supplemental Materials).

We next turn to recalculate the thirty trajectories of the 1.2--1.4 M$_{\odot}$ neutron star merger simulation \cite{Rosswog} considered in Sec.\ III with the four sets of mass and barrier models described above. The results appear in Fig.~\ref{fig:rpRossAll}. Although these dynamical ejecta trajectories are all very neutron-rich, the variations in their density and temperature profiles cause a spread in the range of predicted abundances along the right edge of the second peak. As previously discussed in the context of Fig.~\ref{fig:rtraj117}, this is largely due to whether material reaches the region past $N=184$ where GEF 2016 yields are mostly symmetric and deposit material near $A\sim144$. 
The ability of ETFSI and HFB-17 to reach nuclei higher in mass number due to higher fission barriers near $N=184$ means these models access more of this region found by GEF 2016 to have mostly symmetric yields as compared to the nuclei accessed by FRDM and TF.  We find that the higher nuclear flow through the symmetric GEF yield region with ETFSI and HFB, seen in Fig.~\ref{fig:bhrp}, contributes to their overproduction of the right edge of the second peak as compared to solar data. The model dependent shell closure predictions also play a role with FRDM having a stronger $N=82$ shell closure than ETFSI and HFB which keeps fission daughter products closer to the $A\sim130$ region at late times. It is therefore the interplay between the barrier height landscape around $N=184$ and the structure of the $N=82$ shell closure which determines the shape of the second $r$-process peak in fission cycling conditions. 
It was previously found in Ref. \cite{Eichler15} that given HFB-14 barriers, most of the fissioning nuclei accessed had fission fragments lying between $A = 125$ and $A = 155$, similar to their results with the FRDM case. 
We find that the HFB-14 barriers produce fission flow which dominantly accesses fragment distributions centered at $132\leq A\leq 150$, even after material is primarily located at $N<184$. This contributes to the overall shift in the second $r$-process peak with HFB in Fig.~\ref{fig:rpRossAll}  (to show this explicitly the version of Fig. \ref{fig:rtraj1} given the astrophysical conditions of traj.\ 1 and HFB model inputs has been included in Supplemental Materials). 

\begin{figure}
\begin{center}
\includegraphics[scale=0.527]{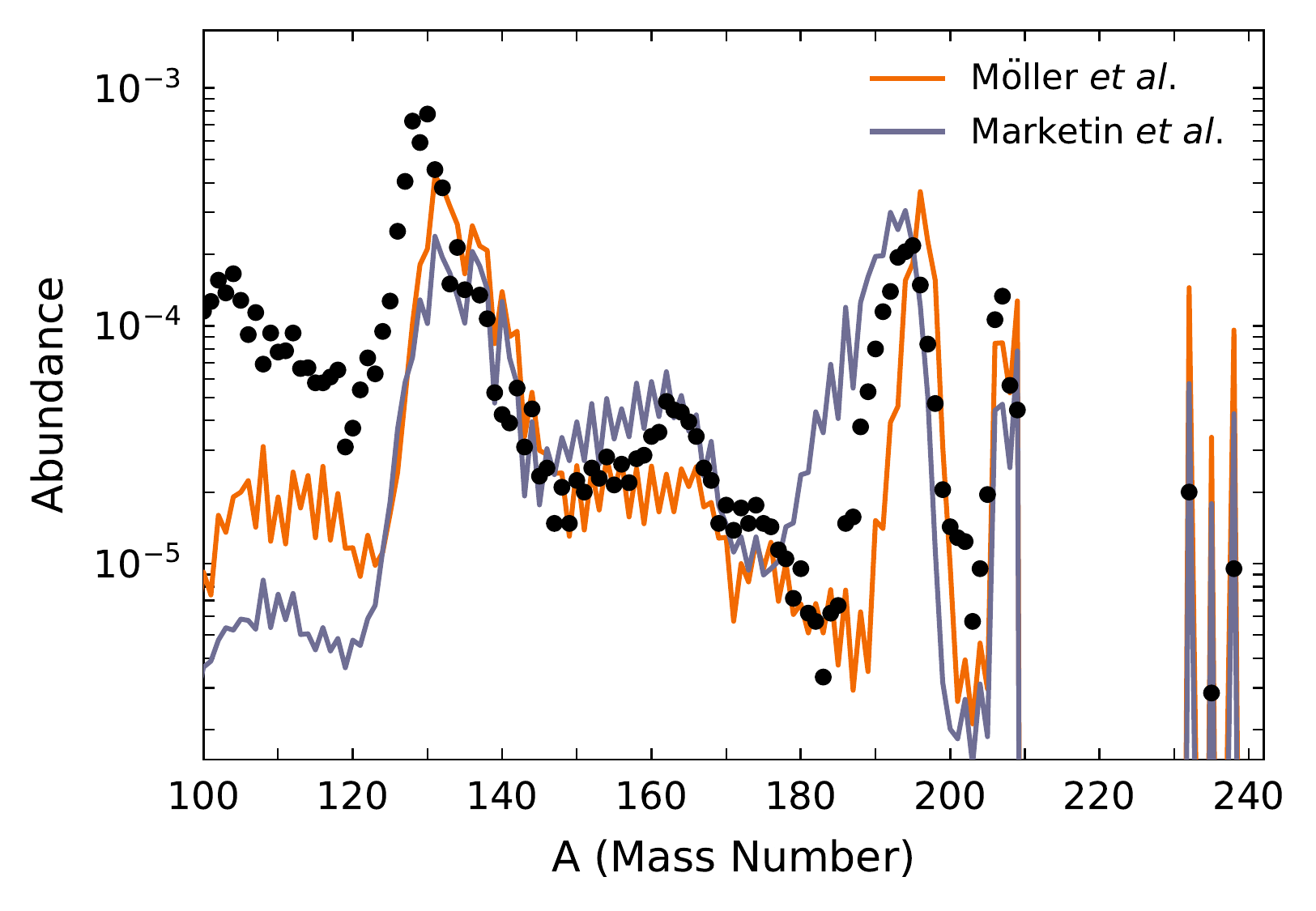}
\end{center}
\caption{(Color online) The $r$-process abundances at 1 Gyr using the cold tidal tail conditions of trajectory 1 \cite{Rosswog} and the FRDM2012 mass model with $\beta$-decay rates determined from M$\ddot{\mathrm{o}}$ller {\it et al.} QRPA calculations \cite{MollerQRPA} (orange) as compared to those from Marketin {\it et al.} \cite{Marketin} (grey).}
\label{fig:rpMark}
\end{figure}

Having considered the role of masses and fission barriers, we next turn to how $\beta$-decay rates influence which fissioning nuclei are accessed in an $r$-process calculation. We repeat the FRLDM/FRDM simulations described above with the M$\ddot{\mathrm{o}}$ller {\it et al.} \cite{MollerQRPA} $\beta$-decay rates replaced by those from Marketin {\it et al.} \cite{Marketin} and show an example abundance pattern comparison in Fig.~\ref{fig:rpMark}. The two sets of rates are generally similar except for nuclei above the $N=126$ shell closure, where the Marketin {\it et al.} rates are faster. Simulations with the faster Marketin {\it et al.} rates tend to show less material hung up in the higher mass regions than with M$\ddot{\mathrm{o}}$ller {\it et al.} rates which reduces the extra post-freeze-out neutrons produced via fission and $\beta$-delayed neutron emission. This in turn thwarts the shifting and narrowing of the third peak from late-time neutron capture (as noted in Refs. \cite{Eichler15,Eichler16}). This reduction of nuclei present near $N=184$ hinders the opportunity for fission to build and shape the second $r$-process peak at late times. Specifically, with nuclei near $N=184$ less populated at freeze-out, the region of the nuclear chart where GEF 2016 predicts fission yields to be centered mostly near $A\sim130$ is not accessed very heavily. This is responsible for the differences in the height of the second $r$-process peak when results using the M$\ddot{\mathrm{o}}$ller {\it et al.} \cite{MollerQRPA} and Marketin {\it et al.} rates are compared (to show this explicitly the versions of Fig. \ref{fig:rtraj1} given the Marketin {\it et al.} and M$\ddot{\mathrm{o}}$ller {\it et al.} conditions applied in Fig. \ref{fig:rpMark} have been included in Supplemental Materials).

 \begin{figure}
\begin{center}
\includegraphics[scale=0.49]{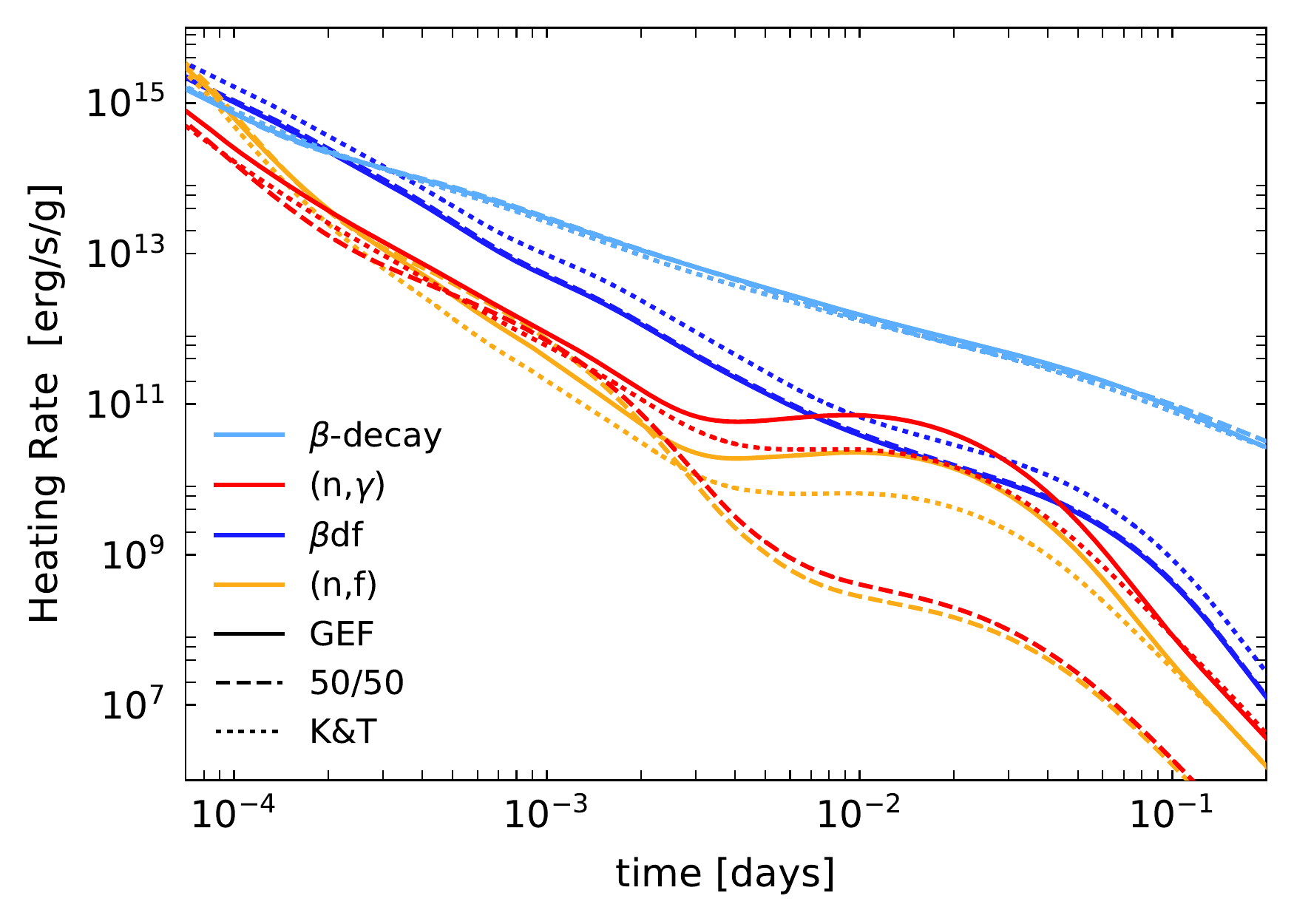}
\end{center}
\caption{(Color online) The nuclear heating rates for $\beta$-decay (blue), $\beta$-delayed fission (light blue), neutron capture (red) and neutron-induced fission (orange) as a function of time using GEF 2016 fission yields (solid lines) as compared to results using simple symmetric splits (dashed lines) and the yields of Kodama and Takahashi (dotted lines).  The masses and fission barriers applied are those of FRDM2012 and FRLDM respectively.}
\label{fig:heatyield2}
\end{figure}

 \begin{figure}
\begin{center}
\includegraphics[scale=0.49]{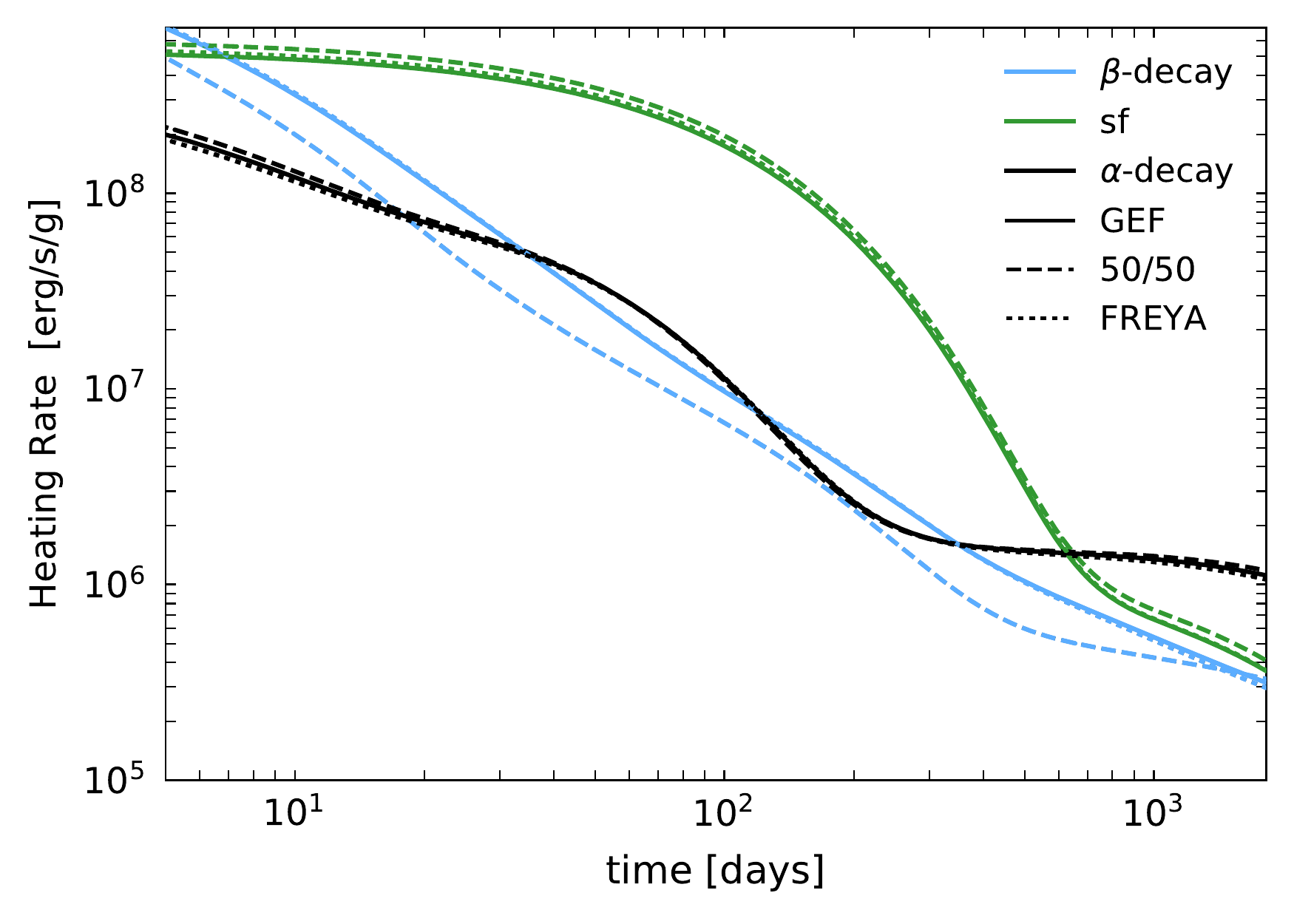}
\end{center}
\caption{(Color online) The nuclear heating rate for $\beta$-decay (blue), alpha decay (black) and spontaneous fission (green) as a function of time using GEF 2016 fission yields (solid lines) as compared to results using simple symmetric splits (dashed lines) and the yields with the FREYA energy sharing and de-excitation treatment (dotted lines). The masses and fission barriers applied are those of FRDM2012 and FRLDM respectively.}
\label{fig:spfheat2}
\end{figure}

We lastly consider how the masses and fission barriers influence the sensitivity of the nuclear heating rates to fission yield treatments. To do so we repeat the calculations presented in Sec. III with nuclear rates determined from FRDM2012 masses and FRLDM fission barriers, shown in Figs.~\ref{fig:heatyield2} and  \ref{fig:spfheat2}. Figure~\ref{fig:heatyield2} confirms that the three order of magnitude discrepancy in the heating for neutron capture channels when 50/50 and GEF 2016 yield results are compared is not isolated to the TF model. This increase in neutron-induced fission in the GEF case is even stronger given FRDM inputs (compare Figs.~\ref{fig:heatyield} and  \ref{fig:heatyield2}) since here the rise in heating rate of neutron capture channels reaches a value $\sim$ one order of magnitude lower than the dominant $\beta$-decay heating channel.

\begin{figure}
\begin{center}
\includegraphics[scale=0.56]{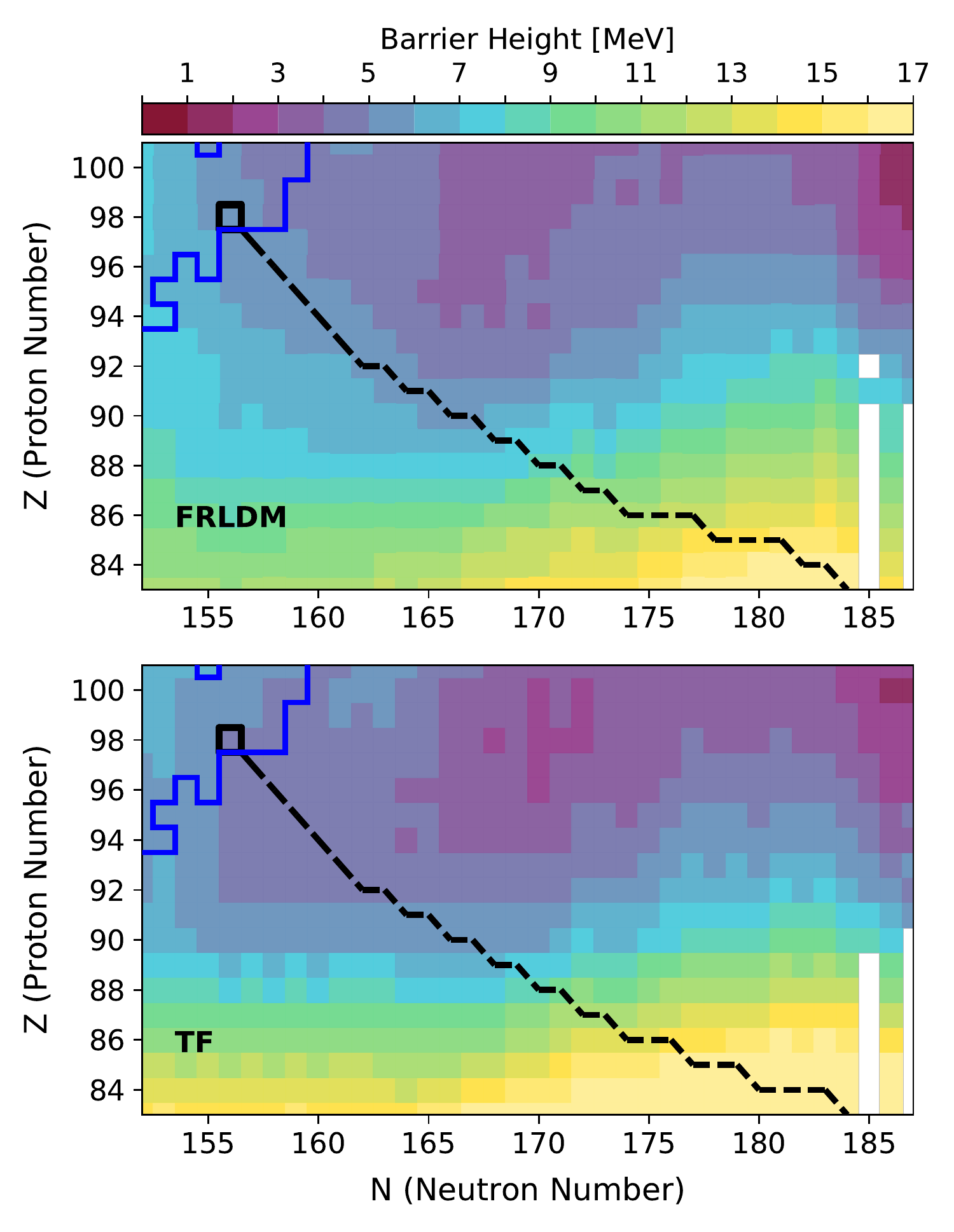}
\end{center}
\caption{(Color online) Fission barrier heights for FRLDM (upper panel) and GEF 2016 (TF) (lower panel) along with the $\beta$-decay path for $^{254}$Cf in each case (black dashed line). The blue outline shows the region of experimentally established decay rates.}
\label{fig:Cfbar}
\end{figure}

When late time heating is considered, a comparison of Figs.~\ref{fig:spfheat} and  \ref{fig:spfheat2} shows that the spontaneous fission contribution to the late-time nuclear heating is two orders of magnitude larger with the FRDM masses and FRLDM barriers as compared to the result using TF masses and GEF 2016 barriers. In Fig.~\ref{fig:Cfbar} we show that it is the difference between the barriers of these two models along the $A=254$ isobaric chain that is primarily responsible for the difference in the late-time contributions from the long-lived californium nucleus $^{254}$Cf. We find that in this region of the nuclear chart, fission flows are highest in the presence of barrier heights around $4-5$ MeV. In the TF model, this is precisely the height of the barrier which $A=254$ nuclei must overcome in order to eventually populate californium, and we find fission of these nuclei transfers material out of this isobaric chain. In contrast, with the FRLDM model, the nuclei set to populate $^{254}$Cf pass just to the left of the region with barrier heights of $4-5$ MeV and the fission encountered along the path to $^{254}$Cf is insignificant. Note from the discussion in Sec. II that the GEF 2016 barriers include some systematic corrections in this region of the nuclear chart. However, along the $A=254$ isobaric line, the TF barriers without GEF corrections are also $4-5$ MeV or lower. Specifically we find that the lower barriers of TF/GEF roughly translate into an order of magnitude higher neutron-induced fission rate for two key nuclei ($^{254}$Np and $^{255}$U) along the path feeding $^{254}$Cf. Though these two key nuclei are populated late in the $r$ process (on the order of $\sim 1-10$ seconds) when the neutron abundance has dropped significantly, their neutron-induced fission rates can be sufficiently high to interrupt the $\beta$ feeding of $^{254}$Cf. We do not find the differences in $\beta$-delayed fission with TF and FRDM models to significantly affect the population of $^{254}$Cf. 

\begin{figure}
\begin{center}
\includegraphics[scale=0.53]{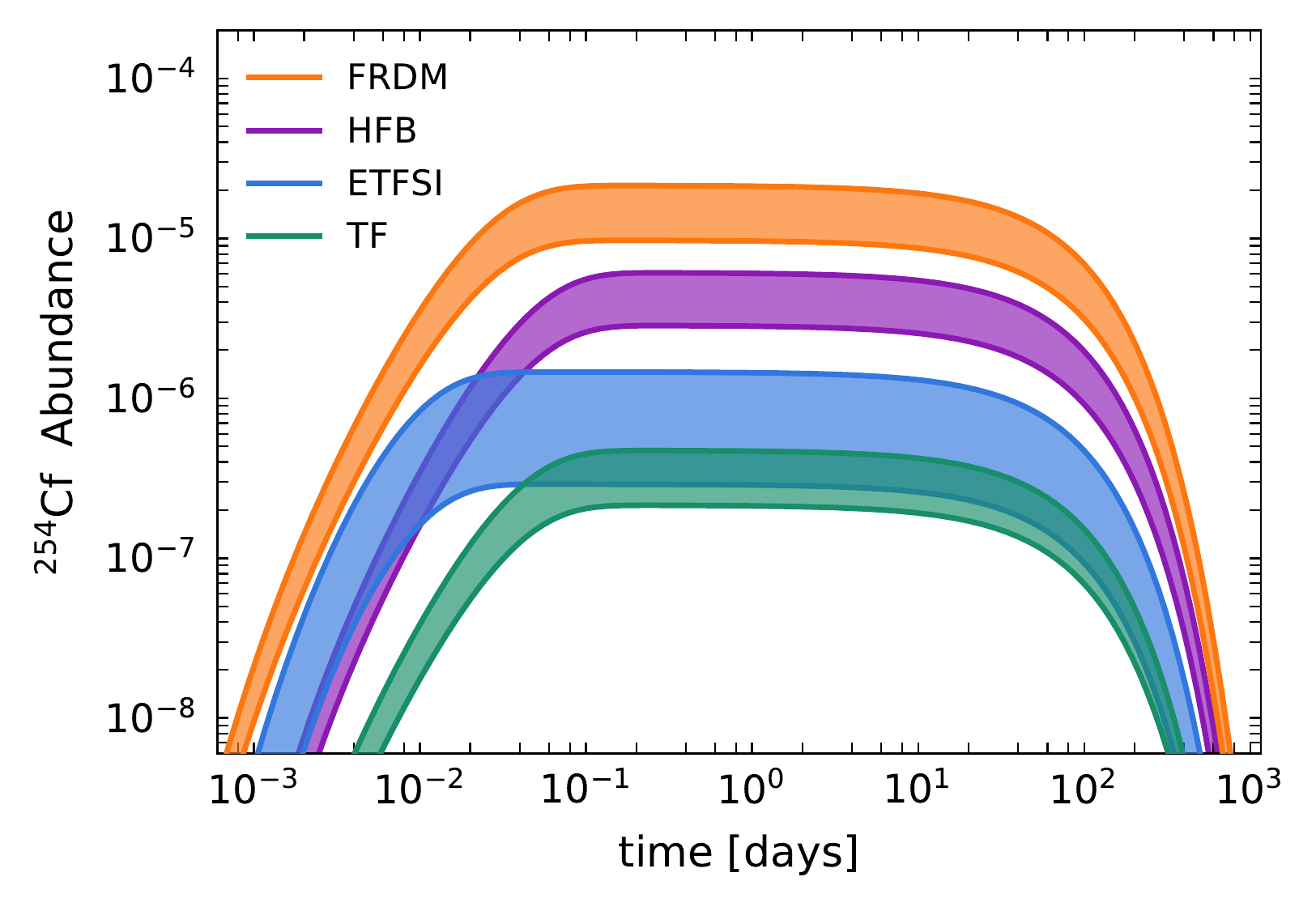}
\end{center}
\caption{(Color online) The $^{254}$Cf abundance as a function of time with TF (green), FRDM (orange), ETFSI (blue) and HFB (purple) models. The bands represent the range in $^{254}$Cf production given thirty 1.2--1.4 M$_{\odot}$ neutron star merger simulation trajectories \cite{Rosswog}.}
\label{fig:254Cfallmod}
\end{figure}

Finally, we comment on the population of $^{254}$Cf given all the nuclear mass/fission barrier models considered in this section. The $^{254}$Cf abundance as a function of time for all cases is shown in Fig. \ref{fig:254Cfallmod}. The two order of magnitude reduction in the abundance of this nucleus with the TF model relative to FRDM is expected given the differences in barrier heights and fission rates previously discussed. The ETFSI model also populates $^{254}$Cf less than FRDM. In this case, the barrier height landscape near $N=184$ leaves less material available to populate  $^{254}$Cf when it pushes significantly past this predicted shell closure. After fission terminates the flow of material near $N=210$, the significant amount of material piled up here encounters the right edge of a large region with low barriers and fissions quickly. In addition, the ETFSI model predicts the $^{254}$Cf path to encounter a small region of barriers with heights $4-5$ MeV which reduce the population $\beta$ feeding this nucleus. Much like ETFSI, the HFB model barriers allow a significant flow of material past $N=184$ which will then not be available to populate the $A=254$ isobaric chain. However the material that remains piled up near $N=184$ at freeze-out which is capable of populating $^{254}$Cf is not inhibited by low $4-5$ MeV barrier heights during its decay back to stability. This produces a higher predicted $^{254}$Cf abundance with HFB as compared to ETFSI and TF models. We find that it is the FRDM model which populates $^{254}$Cf most strongly through the coupled effects of low barriers just past $N=184$ preventing material from moving to higher mass number as well as sufficiently high barriers along the $^{254}$Cf path that prevent a significant depopulation from fission. It is clear that the fission barrier heights of heavy unstable nuclei are key to assessing the influence of $^{254}$Cf on kilonova light curves. Therefore, potential observations of the increased light curve luminosity associated with the heating ``bump" from $^{254}$Cf at late times would not only be able to confirm the synthesis of long-lived actinides, but actually inform nuclear physics calculations of fission properties in this heavy, unstable region. 

\section{Fission Hot Spots}\label{sec:maxflows}

With fission in the $r$ process occurring over a large range of mass numbers, and with much of the initial fission concentrated as far out as the neutron drip line, it is not obvious how practical it is for experimental and theoretical efforts to refine our knowledge of fissioning $r$-process nuclei. To push experiments even a few neutron numbers out from presently-studied nuclei could require tremendous efforts (for a brief review of experimental campaigns to study the neutron-rich isotopes of heavy $r$-process nuclei see \cite{HorowitzRIB2018} and references therein). Theoretical campaigns to calculate fission yields starting from the nuclear properties assumed for a given model, such as density functional theory approaches, can assist with predictions of fission yields for nuclei far from experimental reach. However such methods can be computationally expensive. Therefore given the impracticality for experimental, and even theoretical, studies to provide information on the fission yields of all nuclei of interest to the $r$ process, a guide as to which nuclei participate most during fission in the $r$ process is needed. Here we provide such information by finding the ``hot spots" of nuclei with the highest fission flow in our $r$-process calculations. To do so, we average over $r$-process results given thirty trajectories from a 1.2--1.4 M$_{\odot}$ neutron star merger simulation \cite{Rosswog} (as in Fig.~\ref{fig:rpRossAll}). For each mass model considered, we report the average integrated fission flow for the neutron-induced and $\beta$-delayed fission processes which we find determine the final $r$-process abundances. 

We confine our discussion to neutron-induced and $\beta$-delayed fission since here spontaneous fission flows are comparatively much lower. In this work we apply spontaneous fission rates determined by the phenomenological equation of Karpov {\it et al.} \cite{Karpov} which depends on fissility ($Z^2/A$) and barrier height. These spontaneous fission rates are very low until $Z>100$. Therefore, for all of the simulations considered here, $r$-process material tends to only encounter spontaneous fission at late times when the main abundance pattern is nearly finalized. This is consistent with previous studies which surveyed the influence of several descriptions of spontaneous fission rates \cite{PanovSF} and observed this process to weakly influence the abundances of the second to third $r$-process peaks. However, given the sparsity of experimental fission data, there exist phenomenological fits to spontaneous fission half-lives which, when extrapolated into neutron-rich regions, give very high spontaneous fission rates starting at $Z>94$ \cite{XuRen}. Such rates effectively cut-off the influence of neutron-induced and $\beta$-delayed fission at higher proton numbers. Therefore, the ``hotspots" reported here are meant to represent a case where neutron-induced and $\beta$-delayed fission alone shape the final $r$-process abundance pattern.

\begin{figure*}
\begin{center}
\includegraphics[scale=0.54]{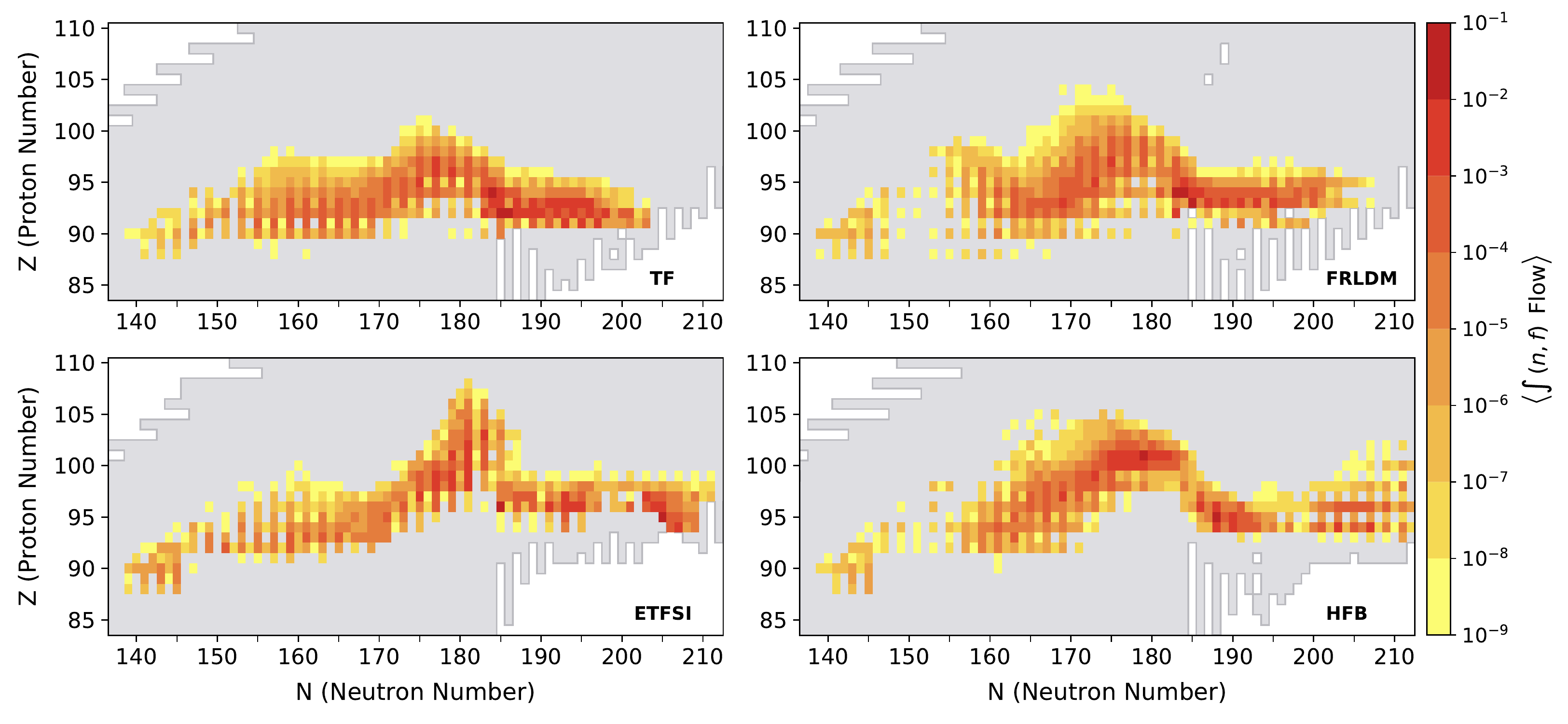}
\end{center}
\caption{(Color online) The integrated neutron-induced fission flow averaged over thirty astrophysical trajectories from a 1.2--1.4 M$_{\odot}$ neutron star merger simulation \cite{Rosswog} assuming GEF (TF) (top left), FRLDM (top right), ETFSI (bottom left), and HFB-14 (bottom right) barrier heights.}
\label{fig:hsnif}
\end{figure*}

\begin{figure*}
\begin{center}
\includegraphics[scale=0.54]{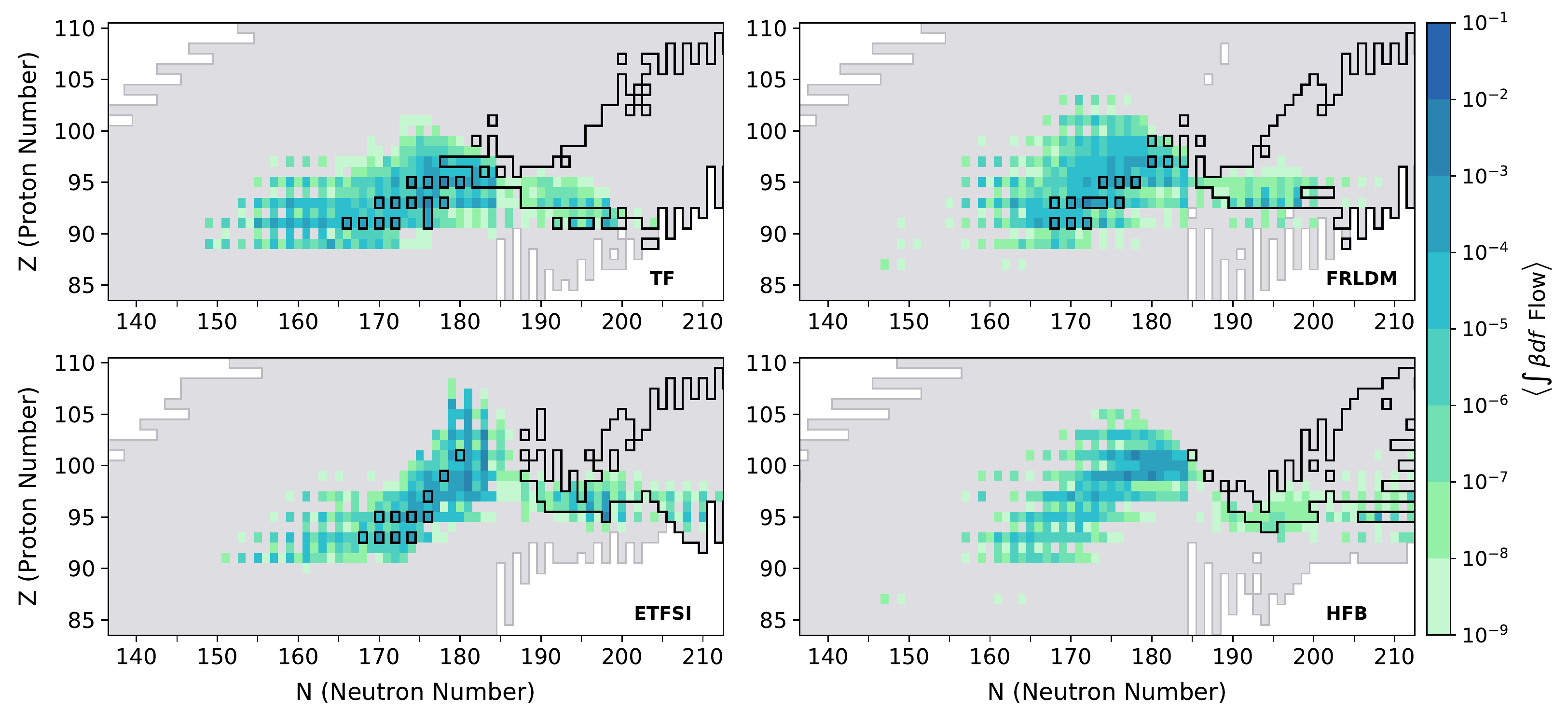}
\end{center}
\caption{(Color online) The integrated $\beta$-delayed fission flow averaged over thirty astrophysical trajectories from a 1.2--1.4 M$_{\odot}$ neutron star merger simulation \cite{Rosswog} assuming GEF (TF) (top left), FRLDM (top right), ETFSI (bottom left), and HFB-14 (bottom right) barrier heights. The black outline here shows where the probability for nuclei to undergo multi-chance $\beta$-delayed fission (relative to ordinary $\beta$-decay) exceeds $10\%$.}
\label{fig:hsbdf}
\end{figure*}

The averaged integrated neutron-induced and $\beta$-delayed fission flows are shown in Figs.~\ref{fig:hsnif} and \ref{fig:hsbdf}. Each panel shows high fission flows are obtained in two general regions of the nuclear chart: along isotopic chains $90<Z<95$ where fission terminates the $r$-process path in $A$, and along the $\beta$-decay pathways of nuclei initially hung up near the $N=184$ shell closure. The exact nuclei in each general region which will have high fission flow is strongly dependent on nuclear masses and barrier heights.  The Thomas-Fermi model used in GEF 2016 as well as the FRLDM model predict a large range of nuclei with relatively low barrier heights $\sim4-6$ MeV near the $N=184$ shell closure. These models therefore show significant fission flow to the left of $N=184$ for the neutron-rich actinide region of $89<Z<95$ which is where GEF 2016 yields show asymmetric yield contributions. Less of this asymmetric region is accessed by HFB-17 and ETFSI models which have higher barrier heights $\sim6-8$ MeV near $N=184$ which hinders fission of material until it moves higher in mass number during the decay back to stability.   

We next identify the nuclei whose fission yields are primarily responsible for the shape of the second $r$-process peak. To do so we consider the threshold of integrated fission fission flow that contributes to finalizing the $r$-process abundance pattern. For each mass model, we apply GEF 2016 neutron-induced fission yields for the nuclei with an average integrated neutron-induced fission above a set threshold with symmetric splits assumed for all remaining nuclei. We apply the same threshold criterion for $\beta$-delayed fission. We find that applying GEF yields to only nuclei with an average integrated fission flow larger than $10^{-5}$ reproduces the final abundance trend almost exactly for all mass models. In the case of TF, HFB, and ETFSI models, implementing the GEF yields for only nuclei having an average integrated fission flow larger than $10^{-4}$ was found to be sufficient to reproduce the relative abundances.

\begin{figure}
\begin{center}
\includegraphics[scale=0.53]{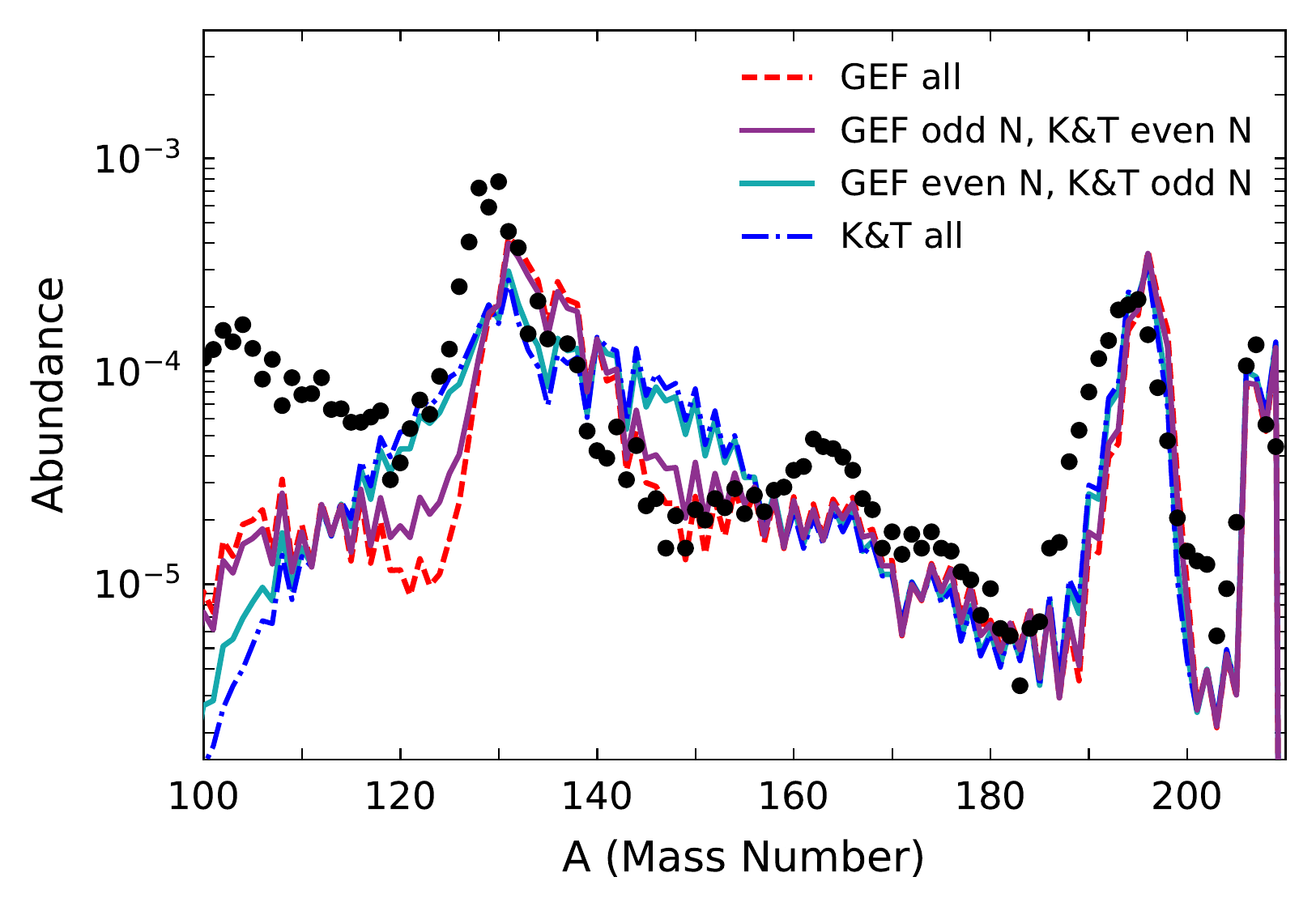}
\end{center}
\caption{(Color online) The $r$-process abundance pattern using the cold tidal tail conditions of trajectory 1 \cite{Rosswog} and the FRDM2012 mass model when GEF 2016 yields are applied to all nuclei (dashed red) as compared to GEF yields for only the odd-$N$ fission reactions of ($Z$,$N$)(n,f) and ($Z$,$N$)$\beta$df (solid purple) with all other fissioning nuclei assuming the fission yields of \cite{Kodama} (K$\&$T). For comparison the abundances with GEF yields applied to only the even-$N$ fission reactions of ($Z$,$N$)(n,f) and ($Z$,$N$)$\beta$df (solid blue) as well as K$\&$T applied to all nuclei (dot-dashed dark blue) are also shown.}
\label{fig:abthres}
\end{figure}

We note Figs.~\ref{fig:hsnif} and \ref{fig:hsbdf} show many of the nuclei with fission flows higher than the $10^{-5}$ threshold have odd neutron number. Are these flows primarily responsible for shaping the final abundance pattern, or do even-$N$ nuclei play a larger role? It is not immediately obvious that high odd-$N$ flow implies a great influence on the abundance pattern since, roughly speaking, the even-$N$ abundances are greater than those of odd-$N$ nuclei throughout the $r$ process. However fission, particularly in the case of neutron-induced, can have rates for an odd-$N$ initial species which are $\sim8-10$ orders of magnitude larger than their even-$N$ isotopic neighbor. We find this disparity in the rates to dominate over the abundance preference for even-$N$ nuclei. Previous work noted the expectation for neutron-induced fission to mainly occur with an even-$N$ fissioning species due to lower values in the difference between the fission barrier and neutron separation energy (since fission barriers do not suffer from strong odd-even effects as is the case for the neutron separation energies) \cite{Petermann}. Here our conclusions are not drawn solely from the odd-even dependence of the fission rates, but also account for the population of fissioning nuclei in the $r$ process by considering flow. We also find that the dominance of odd-$N$ flow, which suggests the importance of the fission yields of even-$N$ species, generally extends to $\beta$-delayed fission as well. In fact, we find fission reactions which have an odd-$N$ initial species are so influential that the $r$-process abundance pattern can nearly be reproduced when GEF 2016 yields are applied to the daughters of just odd-$N$ nuclei, as shown in Fig.~\ref{fig:abthres}. Therefore it is the fission rates of odd-$N$ target nuclei and the fission yields of the corresponding even-$N$ compound nuclei which most impact our $r$-process calculations.

\begin{figure*}
\begin{center}
\includegraphics[scale=0.54]{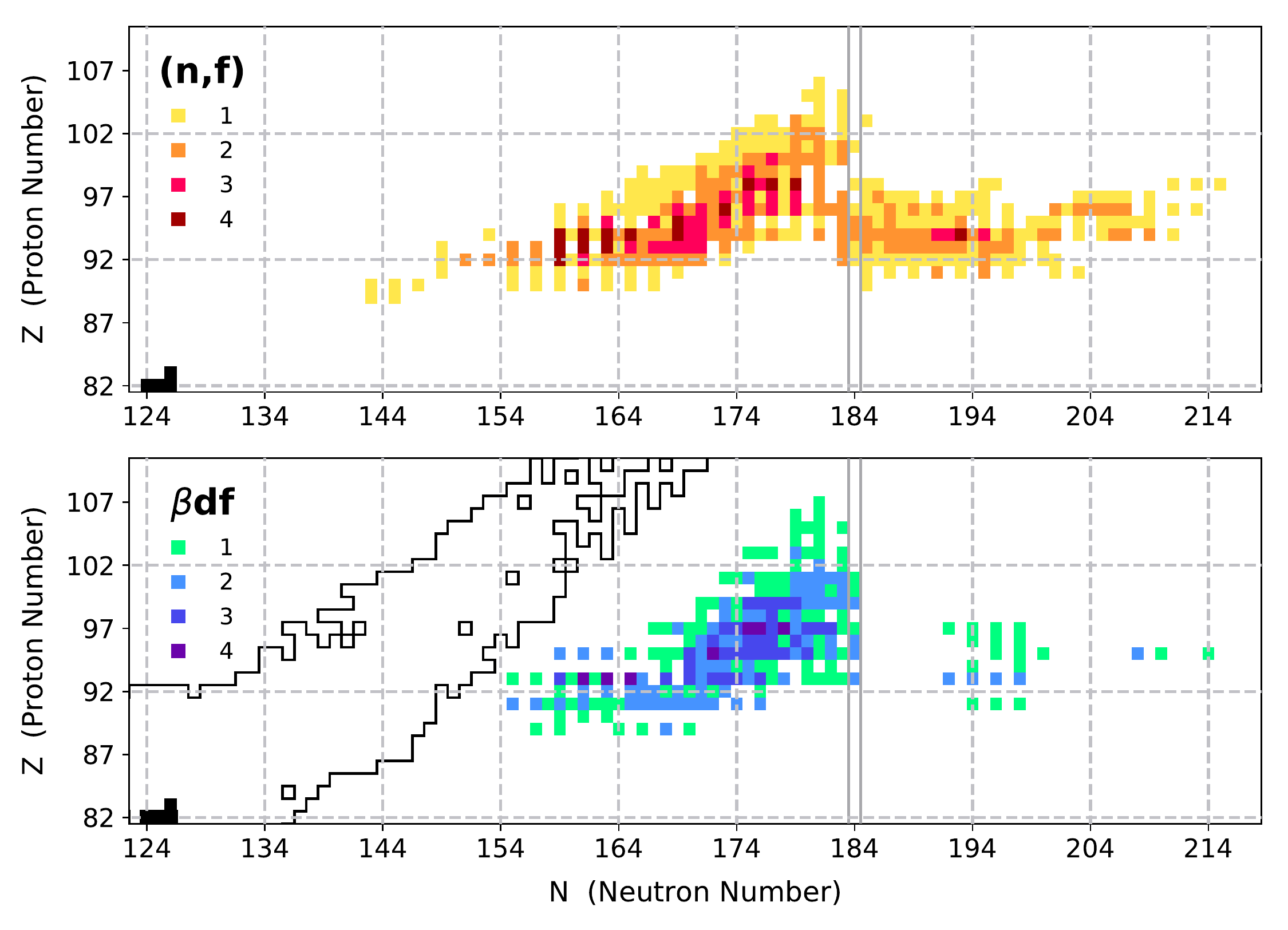}
\end{center}
\caption{(Color online) The nuclei found to have an average integrated neutron-induced (upper) /$\beta$-delayed (lower) fission flow above the $10^{-5}$ threshold found to control the $r$-process abundance pattern in one (yellow/green), two (orange/light blue), three (red/dark blue), or four (dark red/purple) of the mass models considered in Figure \ref{fig:hsnif}/\ref{fig:hsbdf}. The black boxes show stable nuclei while the black outline shows the location of known experimental decay rates.}
\label{fig:hotspots}
\end{figure*}

To report the exact nuclei whose daughter yields are of most consequence when assuming a particular mass model, in Supplemental Materials we tabulate the nuclei that satisfy the $10^{-5}$ threshold criterion for each of the mass models considered. We find that all mass models predict the fission outcomes of $200$ nuclei or less to be of relevance in setting the $r$-process abundance pattern. If only the odd-$N$ nuclei are considered, then all mass models predict $120$ nuclei or less to be relevant.  Although the nuclei which most impact the shape of the abundance pattern are dependent on the mass model and fission barriers, in Fig.~\ref{fig:hotspots} we highlight nuclei commonly found to have a high average integrated fission flow ($\ge10^{-5}$) given the range of nuclear inputs considered. The earlier onset of neutron-induced fission, as compared to $\beta$-delayed fission, is reflected in the reach of this fission channel beyond $N=184$ where abundances are high only at early times. Note that Fig.~\ref{fig:hotspots} essentially reports the overlap of nuclei found to have high fission flow in Figs.~\ref{fig:hsnif} and \ref{fig:hsbdf}, which was mostly determined by a given model's fission barriers. Therefore, the shape of these hotspots is highly influenced by where models agree nuclei have low $4-6$ MeV barrier heights.

Our calculations with the nuclear inputs applied here see all four models agree upon 15 nuclei to be of importance for the neutron-induced fission channel, with over half of these found in the $Z=93$ and $94$ isotopic chains (all of which have odd-$N$). The $\beta$-delayed fission channel sees all models agree upon only 7 nuclei of importance, with nearly all of these nuclei in $Z=93$ and $Z=97$ isotopic chains and most (but not all) having odd-$N$. We note that since the fission yields do affect the flow of material in a fission cycling $r$-process, the ``hotspots" found using 50/50 splits differ for a small handful of nuclei, but the overall region of importance remains the same. In previous work, the mass region with $93 \leq Z \leq 95$ and $180 \leq N \leq 186$ was identified as the dominant region for neutron-induced and $\beta$-delayed fission flow for the FRDM case considered in Ref.~\cite{Eichler15}. In Refs.~\cite{GorielyGEF,GorielyGMPGEF}, nuclei with $A\simeq278$ were singled out as the isobars whose fission products determine the abundance of nuclei in the $110\lesssim A \lesssim 170$ region. Although we also see high fission flow in this region near the $N=184$ shell closure, we see much of the high flow concentrated at $N<184$ since fission is still very active during the decay back to stability.

\section{Conclusions}

The influence of the astrophysical conditions on whether a fission cycling $r$ process is achieved, as well as the dependence on unknown nuclear physics properties for the heaviest neutron-rich nuclei, imply astrophysical environments which host fission to involve the greatest number of $r$-process uncertainties. If ``red'" kilonova components originate from the very neutron-rich conditions seen in neutron star merger dynamical ejecta, fission most certainly plays a role in setting the lanthanide mass fraction which determines wavelength and peak luminosity. If the ``red" kilonova is instead generated primarily from an accretion disk wind, astrophysical uncertainties in the exact neutron richness of such conditions make it difficult to know how much the treatment of fission can influence observation. Since neutron star mergers permit many possible routes to a fission cycling $r$ process, it is important to understand the potential impact on $r$-process observables from variations in the fission treatment. 

We have shown that taking into account the fission yield dependence on the initial excitation energy can influence the final $r$-process abundances by using the publicly available 2016 GEF code to obtain sets of fission yields for neutron-induced, $\beta$-delayed, and spontaneous fission with an appropriate energy applied for each respective fission process. We considered the sensitivity of our results to the treatment of the excitation-energy sharing and de-excitation of the fission fragments using FREYA and showed that such considerations can change the average prompt neutron multiplicity by $\sim1-3$ neutrons in the most neutron-rich regions. However, the sensitivity of the $r$ process to the energy sharing and de-excitation treatment was found to be secondary compared to the effect of the fission yield dependence on initial excitation energy, demonstrating the treatment of the primary fragment yields to be of dominant influence in the $r$ process. Thus careful theoretical calculations of fission fragment yields (prior to neutron emission) that are consistently derived within the framework of a given mass model are crucial to understanding lanthanide production in a fission cycling $r$ process. 

We showed that the trend in GEF 2016 yields, which transition from asymmetric to mostly symmetric yields along an isotopic chain, can reproduce the trend of the right edge of the second $r$-process peak seen in solar data given fission cycling conditions that reach the most neutron-rich regions beyond $N=184$. We considered the influence of nuclear mass models by applying fission rates that self-consistently account for the dependence on nuclear masses and fission barriers. We found that the fission flow explicitly follows the regions of low $4-6$ MeV fission barrier heights making the fissioning nuclei which most impact $r$-process calculations model dependent. The population of key fissioning actinides, such as $^{254}$Cf, was also shown to sensitively depend on the fission barrier assumptions of a given model.

For each of the four sets of masses and barrier heights considered, we reported the integrated fission flow averaged over 30 simulation trajectories for a 1.2--1.4 M$_{\odot}$ neutron star merger to find the fissioning nuclei of most importance. The odd-even behavior observed in these fission flows lead us to identify odd-$N$ species undergoing ($Z$,$N$)(n,f) and ($Z$,$N$)$\beta$df to substantially dominate over even-$N$ species in setting the $r$-process abundance pattern. The ``hot spots" of key fissioning nuclei given all models considered show that nuclei of importance are often found to have $N<184$ due to pile-up of material at this predicted shell closure. The proximity of these ``hot spots" to the region of experimentally established decay data shows the potential for experiments to access some of the fissioning nuclei found to play a key role in setting the $r$-process abundance pattern. Efforts by experimental and theoretical nuclear physics to further the knowledge of fission properties in neutron-rich regions are necessary to develop a more complete picture of heavy element production in neutron star mergers.

\begin{acknowledgments}
N.V. would like to thank Nicolas Schunck, Karl-Heinz Schmidt, Stephan Goriely, Sean Liddick, Samuel Giuliani and Marius Eichler for useful discussions. The work of N.V., R.S., M.M., P.J. and R.V. was partly supported by the Fission In R-process Elements (FIRE) topical collaboration in nuclear theory, funded by the U.S. Department of Energy.  The work of R.V. was performed under the auspices of the U.S. Department of Energy by Lawrence Livermore National Laboratory under Contract No. DE-AC52-07NA27344.  The work of J.R. was performed under the auspices of the U.S. Department of Energy by Lawrence Berkeley National Laboratory under Contract DE-AC02-05CH11231.  The work of M.M. and P.J. was supported in part under the auspices of the National Nuclear Security Administration of the U.S. Department of Energy at Los Alamos National Laboratory under Contract No. DE-AC52-06NA25396.  The work of R.S., T.M.S., and E.M.H. was also supported in part by the U.S. Department of Energy under grant numbers DE-SC0013039 and was enabled by the National Science Foundation under Grant No. PHY-1430152 (JINA Center for the Evolution of the Elements). This manuscript is locatable via Los Alamos and Lawerence Livermore National Laboratory report numbers LA-UR-19-22109 and LLNL-JRNL-760148, respectively.
\end{acknowledgments}

$\,\,$\newline
$\,\,$\newline
$\,\,$\newline
$\,\,$\newline
$\,\,$\newline

\appendix
\section{List of Supplemental Materials}

Sec. II: Ascii table of the average excitation energies for the daughter nuclei populated by $\beta$-decay calculated as in \cite{BDFrp} with $\beta$-strength functions from \cite{MollerSd0} as shown in the upper panel of Fig.~\ref{fig:enubdaught}. Columns are: $Z$, $A$, and $\left<E\right>_{\beta}$ in MeV.

Sec. III: Integrated fission flows cross referenced with fission yields as in Fig.~\ref{fig:rtraj1} for trajectory 17 with the TF model.

Sec. IV: Integrated fission flows cross referenced with fission yields as in Fig.~\ref{fig:rtraj1} for trajectory 1 with the HFB model.

Sec. IV: Comparison of the $r$-process abundance pattern with GEF versus FREYA as in Fig.~\ref{fig:abFREYA} for trajectory 22 with the HFB model.

Sec. IV: Integrated fission flows cross referenced with fission yields as in Fig.~\ref{fig:rtraj1} for trajectory 1 with the FRDM model.

Sec. IV: Integrated fission flows cross referenced with fission yields as in Fig.~\ref{fig:rtraj1} for trajectory 1 with the FRDM model and Marketin $et\,al.$ $\beta$-decay rates.

Sec. VI: Ascii tables of nuclei with an average integrated fission flow higher than $10^{-5}$ for neutron-induced and $\beta$-delayed fission processes for each mass model considered (as in Figs.~\ref{fig:hsnif} and \ref{fig:hsbdf}). For the neutron-induced case, columns are: $Z$, $A$, average integrated flow$\times 10^{5}$, relative percent flow, barrier height of ($Z$,$A$+1) in MeV, and the neutron separation energy of ($Z$,$A$+1) in MeV. For the $\beta$-delayed case columns are: $Z$, $A$, average integrated flow$\times 10^{5}$, relative percent flow, $\beta$-delayed fission probability relative to ordinary $\beta$-decay, barrier height of ($Z$+1,$A$) in MeV, $\beta$-decay Q-value of ($Z$,$A$) in MeV, and the neutron separation energy of ($Z$+1,$A$) in MeV. When a value is marked with $*$, it implies that mass or fission barrier height information was unavailable for the given model, so the FRDM2012 mass value and FRLDM barrier height are instead used to determine the reaction and decay rates.

\bibliography{gefYrefs}

\begin{thebibliography}{87}%
\makeatletter
\providecommand \@ifxundefined [1]{%
 \@ifx{#1\undefined}
}%
\providecommand \@ifnum [1]{%
 \ifnum #1\expandafter \@firstoftwo
 \else \expandafter \@secondoftwo
 \fi
}%
\providecommand \@ifx [1]{%
 \ifx #1\expandafter \@firstoftwo
 \else \expandafter \@secondoftwo
 \fi
}%
\providecommand \natexlab [1]{#1}%
\providecommand \enquote  [1]{``#1''}%
\providecommand \bibnamefont  [1]{#1}%
\providecommand \bibfnamefont [1]{#1}%
\providecommand \citenamefont [1]{#1}%
\providecommand \href@noop [0]{\@secondoftwo}%
\providecommand \href [0]{\begingroup \@sanitize@url \@href}%
\providecommand \@href[1]{\@@startlink{#1}\@@href}%
\providecommand \@@href[1]{\endgroup#1\@@endlink}%
\providecommand \@sanitize@url [0]{\catcode `\\12\catcode `\$12\catcode
  `\&12\catcode `\#12\catcode `\^12\catcode `\_12\catcode `\%12\relax}%
\providecommand \@@startlink[1]{}%
\providecommand \@@endlink[0]{}%
\providecommand \url  [0]{\begingroup\@sanitize@url \@url }%
\providecommand \@url [1]{\endgroup\@href {#1}{\urlprefix }}%
\providecommand \urlprefix  [0]{URL }%
\providecommand \Eprint [0]{\href }%
\providecommand \doibase [0]{http://dx.doi.org/}%
\providecommand \selectlanguage [0]{\@gobble}%
\providecommand \bibinfo  [0]{\@secondoftwo}%
\providecommand \bibfield  [0]{\@secondoftwo}%
\providecommand \translation [1]{[#1]}%
\providecommand \BibitemOpen [0]{}%
\providecommand \bibitemStop [0]{}%
\providecommand \bibitemNoStop [0]{.\EOS\space}%
\providecommand \EOS [0]{\spacefactor3000\relax}%
\providecommand \BibitemShut  [1]{\csname bibitem#1\endcsname}%
\let\auto@bib@innerbib\@empty
\bibitem [{\citenamefont {{Burbidge}}\ \emph {et~al.}(1956)\citenamefont
  {{Burbidge}}, \citenamefont {{Hoyle}}, \citenamefont {{Burbidge}},
  \citenamefont {{Christy}},\ and\ \citenamefont {{Fowler}}}]{Burbidge56}%
  \BibitemOpen
  \bibfield  {author} {\bibinfo {author} {\bibfnamefont {G.~R.}\ \bibnamefont
  {{Burbidge}}}, \bibinfo {author} {\bibfnamefont {F.}~\bibnamefont {{Hoyle}}},
  \bibinfo {author} {\bibfnamefont {E.~M.}\ \bibnamefont {{Burbidge}}},
  \bibinfo {author} {\bibfnamefont {R.~F.}\ \bibnamefont {{Christy}}}, \ and\
  \bibinfo {author} {\bibfnamefont {W.~A.}\ \bibnamefont {{Fowler}}},\ }\href
  {\doibase 10.1103/PhysRev.103.1145} {\bibfield  {journal} {\bibinfo
  {journal} {Phys.\ Rev.}\ }\textbf {\bibinfo {volume} {103}},\ \bibinfo
  {pages} {1145} (\bibinfo {year} {1956})}\BibitemShut {NoStop}%
\bibitem [{\citenamefont {{Ji}}\ \emph {et~al.}(2016)\citenamefont {{Ji}},
  \citenamefont {{Frebel}}, \citenamefont {{Chiti}},\ and\ \citenamefont
  {{Simon}}}]{Ji+2016}%
  \BibitemOpen
  \bibfield  {author} {\bibinfo {author} {\bibfnamefont {A.~P.}\ \bibnamefont
  {{Ji}}}, \bibinfo {author} {\bibfnamefont {A.}~\bibnamefont {{Frebel}}},
  \bibinfo {author} {\bibfnamefont {A.}~\bibnamefont {{Chiti}}}, \ and\
  \bibinfo {author} {\bibfnamefont {J.~D.}\ \bibnamefont {{Simon}}},\ }\href
  {\doibase 10.1038/nature17425} {\bibfield  {journal} {\bibinfo  {journal}
  {Nature}\ }\textbf {\bibinfo {volume} {531}},\ \bibinfo {pages} {610}
  (\bibinfo {year} {2016})},\ \Eprint {http://arxiv.org/abs/1512.01558}
  {arXiv:1512.01558} \BibitemShut {NoStop}%
\bibitem [{\citenamefont {{Roederer}}\ \emph {et~al.}(2016)\citenamefont
  {{Roederer}}, \citenamefont {{Mateo}}, \citenamefont {{Bailey}},
  \citenamefont {{Song}}, \citenamefont {{Bell}}, \citenamefont {{Crane}},
  \citenamefont {{Loebman}}, \citenamefont {{Nidever}}, \citenamefont
  {{Olszewski}}, \citenamefont {{Shectman}}, \citenamefont {{Thompson}},
  \citenamefont {{Valluri}},\ and\ \citenamefont {{Walker}}}]{Roederer+16}%
  \BibitemOpen
  \bibfield  {author} {\bibinfo {author} {\bibfnamefont {I.~U.}\ \bibnamefont
  {{Roederer}}}, \bibinfo {author} {\bibfnamefont {M.}~\bibnamefont {{Mateo}}},
  \bibinfo {author} {\bibfnamefont {J.~I.}\ \bibnamefont {{Bailey}},
  \bibfnamefont {III}}, \bibinfo {author} {\bibfnamefont {Y.}~\bibnamefont
  {{Song}}}, \bibinfo {author} {\bibfnamefont {E.~F.}\ \bibnamefont {{Bell}}},
  \bibinfo {author} {\bibfnamefont {J.~D.}\ \bibnamefont {{Crane}}}, \bibinfo
  {author} {\bibfnamefont {S.}~\bibnamefont {{Loebman}}}, \bibinfo {author}
  {\bibfnamefont {D.~L.}\ \bibnamefont {{Nidever}}}, \bibinfo {author}
  {\bibfnamefont {E.~W.}\ \bibnamefont {{Olszewski}}}, \bibinfo {author}
  {\bibfnamefont {S.~A.}\ \bibnamefont {{Shectman}}}, \bibinfo {author}
  {\bibfnamefont {I.~B.}\ \bibnamefont {{Thompson}}}, \bibinfo {author}
  {\bibfnamefont {M.}~\bibnamefont {{Valluri}}}, \ and\ \bibinfo {author}
  {\bibfnamefont {M.~G.}\ \bibnamefont {{Walker}}},\ }\href {\doibase
  10.3847/0004-6256/151/3/82} {\bibfield  {journal} {\bibinfo  {journal} {AJ}\
  }\textbf {\bibinfo {volume} {151}},\ \bibinfo {eid} {82} (\bibinfo {year}
  {2016})},\ \Eprint {http://arxiv.org/abs/1601.04070} {arXiv:1601.04070
  [astro-ph.SR]} \BibitemShut {NoStop}%
\bibitem [{\citenamefont {{Lattimer}}\ and\ \citenamefont
  {{Schramm}}(1974)}]{Lattimer+74}%
  \BibitemOpen
  \bibfield  {author} {\bibinfo {author} {\bibfnamefont {J.~M.}\ \bibnamefont
  {{Lattimer}}}\ and\ \bibinfo {author} {\bibfnamefont {D.~N.}\ \bibnamefont
  {{Schramm}}},\ }\href {\doibase 10.1086/181612} {\bibfield  {journal}
  {\bibinfo  {journal} {ApJ}\ }\textbf {\bibinfo {volume} {192}},\ \bibinfo
  {pages} {L145} (\bibinfo {year} {1974})}\BibitemShut {NoStop}%
\bibitem [{\citenamefont {{Meyer}}(1989)}]{Meyer89}%
  \BibitemOpen
  \bibfield  {author} {\bibinfo {author} {\bibfnamefont {B.~S.}\ \bibnamefont
  {{Meyer}}},\ }\href {\doibase 10.1086/167702} {\bibfield  {journal} {\bibinfo
   {journal} {ApJ}\ }\textbf {\bibinfo {volume} {343}},\ \bibinfo {pages} {254}
  (\bibinfo {year} {1989})}\BibitemShut {NoStop}%
\bibitem [{\citenamefont {{Freiburghaus}}\ \emph {et~al.}(1999)\citenamefont
  {{Freiburghaus}}, \citenamefont {{Rosswog}},\ and\ \citenamefont
  {{Thielemann}}}]{Freiburghaus+99}%
  \BibitemOpen
  \bibfield  {author} {\bibinfo {author} {\bibfnamefont {C.}~\bibnamefont
  {{Freiburghaus}}}, \bibinfo {author} {\bibfnamefont {S.}~\bibnamefont
  {{Rosswog}}}, \ and\ \bibinfo {author} {\bibfnamefont {F.-K.}\ \bibnamefont
  {{Thielemann}}},\ }\href {\doibase 10.1086/312343} {\bibfield  {journal}
  {\bibinfo  {journal} {ApJ}\ }\textbf {\bibinfo {volume} {525}},\ \bibinfo
  {pages} {L121} (\bibinfo {year} {1999})}\BibitemShut {NoStop}%
\bibitem [{\citenamefont {{Goriely}}\ \emph {et~al.}(2011)\citenamefont
  {{Goriely}}, \citenamefont {{Bauswein}},\ and\ \citenamefont
  {{Janka}}}]{Goriely+11}%
  \BibitemOpen
  \bibfield  {author} {\bibinfo {author} {\bibfnamefont {S.}~\bibnamefont
  {{Goriely}}}, \bibinfo {author} {\bibfnamefont {A.}~\bibnamefont
  {{Bauswein}}}, \ and\ \bibinfo {author} {\bibfnamefont {H.-T.}\ \bibnamefont
  {{Janka}}},\ }\href {\doibase 10.1088/2041-8205/738/2/L32} {\bibfield
  {journal} {\bibinfo  {journal} {ApJ}\ }\textbf {\bibinfo {volume} {738}},\
  \bibinfo {eid} {L32} (\bibinfo {year} {2011})},\ \Eprint
  {http://arxiv.org/abs/1107.0899} {arXiv:1107.0899 [astro-ph.SR]} \BibitemShut
  {NoStop}%
\bibitem [{\citenamefont {{Korobkin}}\ \emph
  {et~al.}(2012{\natexlab{a}})\citenamefont {{Korobkin}}, \citenamefont
  {{Rosswog}}, \citenamefont {{Arcones}},\ and\ \citenamefont
  {{Winteler}}}]{Korobkin+12}%
  \BibitemOpen
  \bibfield  {author} {\bibinfo {author} {\bibfnamefont {O.}~\bibnamefont
  {{Korobkin}}}, \bibinfo {author} {\bibfnamefont {S.}~\bibnamefont
  {{Rosswog}}}, \bibinfo {author} {\bibfnamefont {A.}~\bibnamefont
  {{Arcones}}}, \ and\ \bibinfo {author} {\bibfnamefont {C.}~\bibnamefont
  {{Winteler}}},\ }\href {\doibase 10.1111/j.1365-2966.2012.21859.x} {\bibfield
   {journal} {\bibinfo  {journal} {MNRAS}\ }\textbf {\bibinfo {volume} {426}},\
  \bibinfo {pages} {1940} (\bibinfo {year} {2012}{\natexlab{a}})},\ \Eprint
  {http://arxiv.org/abs/1206.2379} {arXiv:1206.2379 [astro-ph.SR]} \BibitemShut
  {NoStop}%
\bibitem [{\citenamefont {{Wanajo}}\ \emph {et~al.}(2014)\citenamefont
  {{Wanajo}}, \citenamefont {{Sekiguchi}}, \citenamefont {{Nishimura}},
  \citenamefont {{Kiuchi}}, \citenamefont {{Kyutoku}},\ and\ \citenamefont
  {{Shibata}}}]{Wanajo+14}%
  \BibitemOpen
  \bibfield  {author} {\bibinfo {author} {\bibfnamefont {S.}~\bibnamefont
  {{Wanajo}}}, \bibinfo {author} {\bibfnamefont {Y.}~\bibnamefont
  {{Sekiguchi}}}, \bibinfo {author} {\bibfnamefont {N.}~\bibnamefont
  {{Nishimura}}}, \bibinfo {author} {\bibfnamefont {K.}~\bibnamefont
  {{Kiuchi}}}, \bibinfo {author} {\bibfnamefont {K.}~\bibnamefont {{Kyutoku}}},
  \ and\ \bibinfo {author} {\bibfnamefont {M.}~\bibnamefont {{Shibata}}},\
  }\href {\doibase 10.1088/2041-8205/789/2/L39} {\bibfield  {journal} {\bibinfo
   {journal} {ApJ}\ }\textbf {\bibinfo {volume} {789}},\ \bibinfo {eid} {L39}
  (\bibinfo {year} {2014})},\ \Eprint {http://arxiv.org/abs/1402.7317}
  {arXiv:1402.7317 [astro-ph.SR]} \BibitemShut {NoStop}%
\bibitem [{\citenamefont {{Just}}\ \emph {et~al.}(2015)\citenamefont {{Just}},
  \citenamefont {{Bauswein}}, \citenamefont {{Pulpillo}}, \citenamefont
  {{Goriely}},\ and\ \citenamefont {{Janka}}}]{Just+15}%
  \BibitemOpen
  \bibfield  {author} {\bibinfo {author} {\bibfnamefont {O.}~\bibnamefont
  {{Just}}}, \bibinfo {author} {\bibfnamefont {A.}~\bibnamefont {{Bauswein}}},
  \bibinfo {author} {\bibfnamefont {R.~A.}\ \bibnamefont {{Pulpillo}}},
  \bibinfo {author} {\bibfnamefont {S.}~\bibnamefont {{Goriely}}}, \ and\
  \bibinfo {author} {\bibfnamefont {H.-T.}\ \bibnamefont {{Janka}}},\ }\href
  {\doibase 10.1093/mnras/stv009} {\bibfield  {journal} {\bibinfo  {journal}
  {MNRAS}\ }\textbf {\bibinfo {volume} {448}},\ \bibinfo {pages} {541}
  (\bibinfo {year} {2015})},\ \Eprint {http://arxiv.org/abs/1406.2687}
  {arXiv:1406.2687 [astro-ph.SR]} \BibitemShut {NoStop}%
\bibitem [{\citenamefont {{Cowperthwaite et al.}}(2017)}]{Cowperthwaite2017}%
  \BibitemOpen
  \bibfield  {author} {\bibinfo {author} {\bibfnamefont {P.~S.}\ \bibnamefont
  {{Cowperthwaite et al.}}},\ }\href
  {http://stacks.iop.org/2041-8205/848/i=2/a=L17} {\bibfield  {journal}
  {\bibinfo  {journal} {ApJL}\ }\textbf {\bibinfo {volume} {848}},\ \bibinfo
  {pages} {L17} (\bibinfo {year} {2017})}\BibitemShut {NoStop}%
\bibitem [{\citenamefont {{Abbott et al.}}(2017)}]{AbbottGW170817}%
  \BibitemOpen
  \bibfield  {author} {\bibinfo {author} {\bibfnamefont {B.~P.}\ \bibnamefont
  {{Abbott et al.}}} (\bibinfo {collaboration} {LIGO Scientific Collaboration
  and Virgo Collaboration}),\ }\href {\doibase 10.1103/PhysRevLett.119.161101}
  {\bibfield  {journal} {\bibinfo  {journal} {\prl}\ }\textbf {\bibinfo
  {volume} {119}},\ \bibinfo {pages} {161101} (\bibinfo {year}
  {2017})}\BibitemShut {NoStop}%
\bibitem [{\citenamefont {{Kasen}}\ \emph {et~al.}(2017)\citenamefont
  {{Kasen}}, \citenamefont {{Metzger}}, \citenamefont {{Barnes}}, \citenamefont
  {{Quataert}},\ and\ \citenamefont {{Ramirez-Ruiz}}}]{Kasen}%
  \BibitemOpen
  \bibfield  {author} {\bibinfo {author} {\bibfnamefont {D.}~\bibnamefont
  {{Kasen}}}, \bibinfo {author} {\bibfnamefont {B.}~\bibnamefont {{Metzger}}},
  \bibinfo {author} {\bibfnamefont {J.}~\bibnamefont {{Barnes}}}, \bibinfo
  {author} {\bibfnamefont {E.}~\bibnamefont {{Quataert}}}, \ and\ \bibinfo
  {author} {\bibfnamefont {E.}~\bibnamefont {{Ramirez-Ruiz}}},\ }\href
  {\doibase 10.1038/nature24453} {\bibfield  {journal} {\bibinfo  {journal}
  {\nat}\ }\textbf {\bibinfo {volume} {551}},\ \bibinfo {pages} {80} (\bibinfo
  {year} {2017})},\ \Eprint {http://arxiv.org/abs/1710.05463} {arXiv:1710.05463
  [astro-ph.HE]} \BibitemShut {NoStop}%
\bibitem [{\citenamefont {{C{\^o}t{\'e}}}\ \emph {et~al.}(2018)\citenamefont
  {{C{\^o}t{\'e}}}, \citenamefont {{Fryer}}, \citenamefont {{Belczynski}},
  \citenamefont {{Korobkin}}, \citenamefont {{Chru{\'s}li{\'n}ska}},
  \citenamefont {{Vassh}}, \citenamefont {{Mumpower}}, \citenamefont
  {{Lippuner}}, \citenamefont {{Sprouse}}, \citenamefont {{Surman}},\ and\
  \citenamefont {{Wollaeger}}}]{Cote}%
  \BibitemOpen
  \bibfield  {author} {\bibinfo {author} {\bibfnamefont {B.}~\bibnamefont
  {{C{\^o}t{\'e}}}}, \bibinfo {author} {\bibfnamefont {C.~L.}\ \bibnamefont
  {{Fryer}}}, \bibinfo {author} {\bibfnamefont {K.}~\bibnamefont
  {{Belczynski}}}, \bibinfo {author} {\bibfnamefont {O.}~\bibnamefont
  {{Korobkin}}}, \bibinfo {author} {\bibfnamefont {M.}~\bibnamefont
  {{Chru{\'s}li{\'n}ska}}}, \bibinfo {author} {\bibfnamefont {N.}~\bibnamefont
  {{Vassh}}}, \bibinfo {author} {\bibfnamefont {M.~R.}\ \bibnamefont
  {{Mumpower}}}, \bibinfo {author} {\bibfnamefont {J.}~\bibnamefont
  {{Lippuner}}}, \bibinfo {author} {\bibfnamefont {T.~M.}\ \bibnamefont
  {{Sprouse}}}, \bibinfo {author} {\bibfnamefont {R.}~\bibnamefont {{Surman}}},
  \ and\ \bibinfo {author} {\bibfnamefont {R.}~\bibnamefont {{Wollaeger}}},\
  }\href {\doibase 10.3847/1538-4357/aaad67} {\bibfield  {journal} {\bibinfo
  {journal} {ApJ}\ }\textbf {\bibinfo {volume} {855}},\ \bibinfo {eid} {99}
  (\bibinfo {year} {2018})},\ \Eprint {http://arxiv.org/abs/1710.05875}
  {arXiv:1710.05875} \BibitemShut {NoStop}%
\bibitem [{\citenamefont {{Eichler}}\ \emph {et~al.}(2015)\citenamefont
  {{Eichler}}, \citenamefont {{Arcones}}, \citenamefont {{Kelic}},
  \citenamefont {{Korobkin}}, \citenamefont {{Langanke}}, \citenamefont
  {{Marketin}}, \citenamefont {{Martinez-Pinedo}}, \citenamefont {{Panov}},
  \citenamefont {{Rauscher}}, \citenamefont {{Rosswog}}, \citenamefont
  {{Winteler}}, \citenamefont {{Zinner}},\ and\ \citenamefont
  {{Thielemann}}}]{Eichler15}%
  \BibitemOpen
  \bibfield  {author} {\bibinfo {author} {\bibfnamefont {M.}~\bibnamefont
  {{Eichler}}}, \bibinfo {author} {\bibfnamefont {A.}~\bibnamefont
  {{Arcones}}}, \bibinfo {author} {\bibfnamefont {A.}~\bibnamefont {{Kelic}}},
  \bibinfo {author} {\bibfnamefont {O.}~\bibnamefont {{Korobkin}}}, \bibinfo
  {author} {\bibfnamefont {K.}~\bibnamefont {{Langanke}}}, \bibinfo {author}
  {\bibfnamefont {T.}~\bibnamefont {{Marketin}}}, \bibinfo {author}
  {\bibfnamefont {G.}~\bibnamefont {{Martinez-Pinedo}}}, \bibinfo {author}
  {\bibfnamefont {I.}~\bibnamefont {{Panov}}}, \bibinfo {author} {\bibfnamefont
  {T.}~\bibnamefont {{Rauscher}}}, \bibinfo {author} {\bibfnamefont
  {S.}~\bibnamefont {{Rosswog}}}, \bibinfo {author} {\bibfnamefont
  {C.}~\bibnamefont {{Winteler}}}, \bibinfo {author} {\bibfnamefont {N.~T.}\
  \bibnamefont {{Zinner}}}, \ and\ \bibinfo {author} {\bibfnamefont {F.-K.}\
  \bibnamefont {{Thielemann}}},\ }\href {\doibase 10.1088/0004-637X/808/1/30}
  {\bibfield  {journal} {\bibinfo  {journal} {ApJ}\ }\textbf {\bibinfo {volume}
  {808}},\ \bibinfo {eid} {30} (\bibinfo {year} {2015})},\ \Eprint
  {http://arxiv.org/abs/1411.0974} {arXiv:1411.0974 [astro-ph.HE]} \BibitemShut
  {NoStop}%
\bibitem [{\citenamefont {{Barnes}}\ \emph {et~al.}(2016)\citenamefont
  {{Barnes}}, \citenamefont {{Kasen}}, \citenamefont {{Wu}},\ and\
  \citenamefont {{Mart{\'{\i}}nez-Pinedo}}}]{Barnes+16}%
  \BibitemOpen
  \bibfield  {author} {\bibinfo {author} {\bibfnamefont {J.}~\bibnamefont
  {{Barnes}}}, \bibinfo {author} {\bibfnamefont {D.}~\bibnamefont {{Kasen}}},
  \bibinfo {author} {\bibfnamefont {M.-R.}\ \bibnamefont {{Wu}}}, \ and\
  \bibinfo {author} {\bibfnamefont {G.}~\bibnamefont
  {{Mart{\'{\i}}nez-Pinedo}}},\ }\href {\doibase 10.3847/0004-637X/829/2/110}
  {\bibfield  {journal} {\bibinfo  {journal} {ApJ}\ }\textbf {\bibinfo {volume}
  {829}},\ \bibinfo {eid} {110} (\bibinfo {year} {2016})},\ \Eprint
  {http://arxiv.org/abs/1605.07218} {arXiv:1605.07218 [astro-ph.HE]}
  \BibitemShut {NoStop}%
\bibitem [{\citenamefont {{Zhu}}\ \emph {et~al.}(2018)\citenamefont {{Zhu}},
  \citenamefont {{Wollaeger}}, \citenamefont {{Vassh}}, \citenamefont
  {{Surman}}, \citenamefont {{Sprouse}}, \citenamefont {{Mumpower}},
  \citenamefont {{M{\"o}ller}}, \citenamefont {{McLaughlin}}, \citenamefont
  {{Korobkin}}, \citenamefont {{Kawano}}, \citenamefont {{Jaffke}},
  \citenamefont {{Holmbeck}}, \citenamefont {{Fryer}}, \citenamefont {{Even}},
  \citenamefont {{Couture}},\ and\ \citenamefont {{Barnes}}}]{Cfpaper}%
  \BibitemOpen
  \bibfield  {author} {\bibinfo {author} {\bibfnamefont {Y.}~\bibnamefont
  {{Zhu}}}, \bibinfo {author} {\bibfnamefont {R.~T.}\ \bibnamefont
  {{Wollaeger}}}, \bibinfo {author} {\bibfnamefont {N.}~\bibnamefont
  {{Vassh}}}, \bibinfo {author} {\bibfnamefont {R.}~\bibnamefont {{Surman}}},
  \bibinfo {author} {\bibfnamefont {T.~M.}\ \bibnamefont {{Sprouse}}}, \bibinfo
  {author} {\bibfnamefont {M.~R.}\ \bibnamefont {{Mumpower}}}, \bibinfo
  {author} {\bibfnamefont {P.}~\bibnamefont {{M{\"o}ller}}}, \bibinfo {author}
  {\bibfnamefont {G.~C.}\ \bibnamefont {{McLaughlin}}}, \bibinfo {author}
  {\bibfnamefont {O.}~\bibnamefont {{Korobkin}}}, \bibinfo {author}
  {\bibfnamefont {T.}~\bibnamefont {{Kawano}}}, \bibinfo {author}
  {\bibfnamefont {P.~J.}\ \bibnamefont {{Jaffke}}}, \bibinfo {author}
  {\bibfnamefont {E.~M.}\ \bibnamefont {{Holmbeck}}}, \bibinfo {author}
  {\bibfnamefont {C.~L.}\ \bibnamefont {{Fryer}}}, \bibinfo {author}
  {\bibfnamefont {W.~P.}\ \bibnamefont {{Even}}}, \bibinfo {author}
  {\bibfnamefont {A.~J.}\ \bibnamefont {{Couture}}}, \ and\ \bibinfo {author}
  {\bibfnamefont {J.}~\bibnamefont {{Barnes}}},\ }\href {\doibase
  10.3847/2041-8213/aad5de} {\bibfield  {journal} {\bibinfo  {journal} {ApJL}\
  }\textbf {\bibinfo {volume} {863}},\ \bibinfo {eid} {L23} (\bibinfo {year}
  {2018})},\ \Eprint {http://arxiv.org/abs/1806.09724} {arXiv:1806.09724
  [astro-ph.HE]} \BibitemShut {NoStop}%
\bibitem [{\citenamefont {{Goriely}}\ \emph {et~al.}(2013)\citenamefont
  {{Goriely}}, \citenamefont {{Sida}}, \citenamefont {{Lema{\^i}tre}},
  \citenamefont {{Panebianco}}, \citenamefont {{Dubray}}, \citenamefont
  {{Hilaire}}, \citenamefont {{Bauswein}},\ and\ \citenamefont
  {{Janka}}}]{Goriely+13}%
  \BibitemOpen
  \bibfield  {author} {\bibinfo {author} {\bibfnamefont {S.}~\bibnamefont
  {{Goriely}}}, \bibinfo {author} {\bibfnamefont {J.-L.}\ \bibnamefont
  {{Sida}}}, \bibinfo {author} {\bibfnamefont {J.-F.}\ \bibnamefont
  {{Lema{\^i}tre}}}, \bibinfo {author} {\bibfnamefont {S.}~\bibnamefont
  {{Panebianco}}}, \bibinfo {author} {\bibfnamefont {N.}~\bibnamefont
  {{Dubray}}}, \bibinfo {author} {\bibfnamefont {S.}~\bibnamefont {{Hilaire}}},
  \bibinfo {author} {\bibfnamefont {A.}~\bibnamefont {{Bauswein}}}, \ and\
  \bibinfo {author} {\bibfnamefont {H.-T.}\ \bibnamefont {{Janka}}},\ }\href
  {\doibase 10.1103/PhysRevLett.111.242502} {\bibfield  {journal} {\bibinfo
  {journal} {\prl}\ }\textbf {\bibinfo {volume} {111}},\ \bibinfo {eid}
  {242502} (\bibinfo {year} {2013})}\BibitemShut {NoStop}%
\bibitem [{\citenamefont {{Goriely}}(2015)}]{GorielyGEF}%
  \BibitemOpen
  \bibfield  {author} {\bibinfo {author} {\bibfnamefont {S.}~\bibnamefont
  {{Goriely}}},\ }\href {\doibase 10.1140/epja/i2015-15022-3} {\bibfield
  {journal} {\bibinfo  {journal} {Eur.\ Phys.\ J.\ A}\ }\textbf {\bibinfo
  {volume} {51}},\ \bibinfo {eid} {22} (\bibinfo {year} {2015})}\BibitemShut
  {NoStop}%
\bibitem [{\citenamefont {{Goriely}}\ and\ \citenamefont {{Mart{\'{\i}}nez
  Pinedo}}(2015)}]{GorielyGMPGEF}%
  \BibitemOpen
  \bibfield  {author} {\bibinfo {author} {\bibfnamefont {S.}~\bibnamefont
  {{Goriely}}}\ and\ \bibinfo {author} {\bibfnamefont {G.}~\bibnamefont
  {{Mart{\'{\i}}nez Pinedo}}},\ }\href {\doibase
  10.1016/j.nuclphysa.2015.07.020} {\bibfield  {journal} {\bibinfo  {journal}
  {Nucl.\ Phys.\ A}\ }\textbf {\bibinfo {volume} {944}},\ \bibinfo {pages}
  {158} (\bibinfo {year} {2015})}\BibitemShut {NoStop}%
\bibitem [{\citenamefont {{Panov}}\ \emph {et~al.}(2008)\citenamefont
  {{Panov}}, \citenamefont {{Korneev}},\ and\ \citenamefont
  {{Thielemann}}}]{Panov}%
  \BibitemOpen
  \bibfield  {author} {\bibinfo {author} {\bibfnamefont {I.~V.}\ \bibnamefont
  {{Panov}}}, \bibinfo {author} {\bibfnamefont {I.~Y.}\ \bibnamefont
  {{Korneev}}}, \ and\ \bibinfo {author} {\bibfnamefont {F.-K.}\ \bibnamefont
  {{Thielemann}}},\ }\href {\doibase 10.1007/s11443-008-3006-1} {\bibfield
  {journal} {\bibinfo  {journal} {Astron.\ Lett.}\ }\textbf {\bibinfo {volume}
  {34}},\ \bibinfo {pages} {189} (\bibinfo {year} {2008})}\BibitemShut
  {NoStop}%
\bibitem [{\citenamefont {{Kodama}}\ and\ \citenamefont
  {{Takahashi}}(1975)}]{Kodama}%
  \BibitemOpen
  \bibfield  {author} {\bibinfo {author} {\bibfnamefont {T.}~\bibnamefont
  {{Kodama}}}\ and\ \bibinfo {author} {\bibfnamefont {K.}~\bibnamefont
  {{Takahashi}}},\ }\href {\doibase 10.1016/0375-9474(75)90381-4} {\bibfield
  {journal} {\bibinfo  {journal} {Nucl.\ Phys.\ A}\ }\textbf {\bibinfo {volume}
  {239}},\ \bibinfo {pages} {489} (\bibinfo {year} {1975})}\BibitemShut
  {NoStop}%
\bibitem [{\citenamefont {{Eichler}}\ \emph {et~al.}(2016)\citenamefont
  {{Eichler}}, \citenamefont {{Arcones}}, \citenamefont {{K{\"a}ppeli}},
  \citenamefont {{Korobkin}}, \citenamefont {{Liebend{\"o}rfer}}, \citenamefont
  {{Martinez-Pinedo}}, \citenamefont {{Panov}}, \citenamefont {{Rauscher}},
  \citenamefont {{Rosswog}}, \citenamefont {{Thielemann}},\ and\ \citenamefont
  {{Winteler}}}]{Eichler16}%
  \BibitemOpen
  \bibfield  {author} {\bibinfo {author} {\bibfnamefont {M.}~\bibnamefont
  {{Eichler}}}, \bibinfo {author} {\bibfnamefont {A.}~\bibnamefont
  {{Arcones}}}, \bibinfo {author} {\bibfnamefont {R.}~\bibnamefont
  {{K{\"a}ppeli}}}, \bibinfo {author} {\bibfnamefont {O.}~\bibnamefont
  {{Korobkin}}}, \bibinfo {author} {\bibfnamefont {M.}~\bibnamefont
  {{Liebend{\"o}rfer}}}, \bibinfo {author} {\bibfnamefont {G.}~\bibnamefont
  {{Martinez-Pinedo}}}, \bibinfo {author} {\bibfnamefont {I.~V.}\ \bibnamefont
  {{Panov}}}, \bibinfo {author} {\bibfnamefont {T.}~\bibnamefont {{Rauscher}}},
  \bibinfo {author} {\bibfnamefont {S.}~\bibnamefont {{Rosswog}}}, \bibinfo
  {author} {\bibfnamefont {F.-K.}\ \bibnamefont {{Thielemann}}}, \ and\
  \bibinfo {author} {\bibfnamefont {C.}~\bibnamefont {{Winteler}}},\ }in\ \href
  {\doibase 10.1088/1742-6596/665/1/012054} {\emph {\bibinfo {booktitle}
  {Journal of Physics Conference Series}}},\ \bibinfo {series} {Journal of
  Physics Conference Series}, Vol.\ \bibinfo {volume} {665}\ (\bibinfo {year}
  {2016})\ p.\ \bibinfo {pages} {012054}\BibitemShut {NoStop}%
\bibitem [{\citenamefont {{Shibagaki}}\ \emph {et~al.}(2016)\citenamefont
  {{Shibagaki}}, \citenamefont {{Kajino}}, \citenamefont {{Mathews}},
  \citenamefont {{Chiba}}, \citenamefont {{Nishimura}},\ and\ \citenamefont
  {{Lorusso}}}]{ShibagakiMathews}%
  \BibitemOpen
  \bibfield  {author} {\bibinfo {author} {\bibfnamefont {S.}~\bibnamefont
  {{Shibagaki}}}, \bibinfo {author} {\bibfnamefont {T.}~\bibnamefont
  {{Kajino}}}, \bibinfo {author} {\bibfnamefont {G.~J.}\ \bibnamefont
  {{Mathews}}}, \bibinfo {author} {\bibfnamefont {S.}~\bibnamefont {{Chiba}}},
  \bibinfo {author} {\bibfnamefont {S.}~\bibnamefont {{Nishimura}}}, \ and\
  \bibinfo {author} {\bibfnamefont {G.}~\bibnamefont {{Lorusso}}},\ }\href
  {\doibase 10.3847/0004-637X/816/2/79} {\bibfield  {journal} {\bibinfo
  {journal} {ApJ}\ }\textbf {\bibinfo {volume} {816}},\ \bibinfo {eid} {79}
  (\bibinfo {year} {2016})},\ \Eprint {http://arxiv.org/abs/1505.02257}
  {arXiv:1505.02257 [astro-ph.SR]} \BibitemShut {NoStop}%
\bibitem [{\citenamefont {Gaimard}\ and\ \citenamefont
  {Schmidt}(1991)}]{ABLA91}%
  \BibitemOpen
  \bibfield  {author} {\bibinfo {author} {\bibfnamefont {J.-J.}\ \bibnamefont
  {Gaimard}}\ and\ \bibinfo {author} {\bibfnamefont {K.-H.}\ \bibnamefont
  {Schmidt}},\ }\href {\doibase https://doi.org/10.1016/0375-9474(91)90748-U}
  {\bibfield  {journal} {\bibinfo  {journal} {Nucl.\ Phys.\ A}\ }\textbf
  {\bibinfo {volume} {531}},\ \bibinfo {pages} {709 } (\bibinfo {year}
  {1991})}\BibitemShut {NoStop}%
\bibitem [{\citenamefont {{Kelic}}\ \emph {et~al.}(2009)\citenamefont
  {{Kelic}}, \citenamefont {{Valentina Ricciardi}},\ and\ \citenamefont
  {{Schmidt}}}]{ABLA07}%
  \BibitemOpen
  \bibfield  {author} {\bibinfo {author} {\bibfnamefont {A.}~\bibnamefont
  {{Kelic}}}, \bibinfo {author} {\bibfnamefont {M.}~\bibnamefont {{Valentina
  Ricciardi}}}, \ and\ \bibinfo {author} {\bibfnamefont {K.-H.}\ \bibnamefont
  {{Schmidt}}},\ }\href@noop {} {\bibfield  {journal} {\bibinfo  {journal}
  {arXiv e-prints}\ } (\bibinfo {year} {2009})},\ \Eprint
  {http://arxiv.org/abs/0906.4193} {arXiv:0906.4193 [nucl-th]} \BibitemShut
  {NoStop}%
\bibitem [{\citenamefont {{Wahl}}(2002)}]{Wahl}%
  \BibitemOpen
  \bibfield  {author} {\bibinfo {author} {\bibfnamefont {A.~C.}\ \bibnamefont
  {{Wahl}}},\ }\href@noop {} {\emph {\bibinfo {title} {Systematics of
  Fission-Product Yields}}},\ \bibinfo {type} {Tech. Rep.}\ (\bibinfo
  {institution} {LANL Technical Report LA-13928},\ \bibinfo {year}
  {2002})\BibitemShut {NoStop}%
\bibitem [{\citenamefont {{Schmidt}}\ \emph {et~al.}(2016)\citenamefont
  {{Schmidt}}, \citenamefont {{Jurado}}, \citenamefont {{Amouroux}},\ and\
  \citenamefont {{Schmitt}}}]{GEF}%
  \BibitemOpen
  \bibfield  {author} {\bibinfo {author} {\bibfnamefont {K.-H.}\ \bibnamefont
  {{Schmidt}}}, \bibinfo {author} {\bibfnamefont {B.}~\bibnamefont {{Jurado}}},
  \bibinfo {author} {\bibfnamefont {C.}~\bibnamefont {{Amouroux}}}, \ and\
  \bibinfo {author} {\bibfnamefont {C.}~\bibnamefont {{Schmitt}}},\ }\href
  {\doibase 10.1016/j.nds.2015.12.009} {\bibfield  {journal} {\bibinfo
  {journal} {Nucl.\ Data Sheets}\ }\textbf {\bibinfo {volume} {131}},\ \bibinfo
  {pages} {107} (\bibinfo {year} {2016})}\BibitemShut {NoStop}%
\bibitem [{\citenamefont {{Mendoza-Temis}}\ \emph {et~al.}(2015)\citenamefont
  {{Mendoza-Temis}}, \citenamefont {{Wu}}, \citenamefont {{Langanke}},
  \citenamefont {{Mart{\'{\i}}nez-Pinedo}}, \citenamefont {{Bauswein}},\ and\
  \citenamefont {{Janka}}}]{Mendoza-Temis+15}%
  \BibitemOpen
  \bibfield  {author} {\bibinfo {author} {\bibfnamefont {J.~d.~J.}\
  \bibnamefont {{Mendoza-Temis}}}, \bibinfo {author} {\bibfnamefont {M.-R.}\
  \bibnamefont {{Wu}}}, \bibinfo {author} {\bibfnamefont {K.}~\bibnamefont
  {{Langanke}}}, \bibinfo {author} {\bibfnamefont {G.}~\bibnamefont
  {{Mart{\'{\i}}nez-Pinedo}}}, \bibinfo {author} {\bibfnamefont
  {A.}~\bibnamefont {{Bauswein}}}, \ and\ \bibinfo {author} {\bibfnamefont
  {H.-T.}\ \bibnamefont {{Janka}}},\ }\href {\doibase
  10.1103/PhysRevC.92.055805} {\bibfield  {journal} {\bibinfo  {journal}
  {\prc}\ }\textbf {\bibinfo {volume} {92}},\ \bibinfo {eid} {055805} (\bibinfo
  {year} {2015})},\ \Eprint {http://arxiv.org/abs/1409.6135} {arXiv:1409.6135
  [astro-ph.HE]} \BibitemShut {NoStop}%
\bibitem [{\citenamefont {{Mendoza-Temis}}\ \emph {et~al.}(2016)\citenamefont
  {{Mendoza-Temis}}, \citenamefont {{Wu}}, \citenamefont
  {{Mart{\'{\i}}nez-Pinedo}}, \citenamefont {{Langanke}}, \citenamefont
  {{Bauswein}}, \citenamefont {{Janka}},\ and\ \citenamefont
  {{Frank}}}]{Mendoza-Temis+16}%
  \BibitemOpen
  \bibfield  {author} {\bibinfo {author} {\bibfnamefont {J.~J.}\ \bibnamefont
  {{Mendoza-Temis}}}, \bibinfo {author} {\bibfnamefont {M.~R.}\ \bibnamefont
  {{Wu}}}, \bibinfo {author} {\bibfnamefont {G.}~\bibnamefont
  {{Mart{\'{\i}}nez-Pinedo}}}, \bibinfo {author} {\bibfnamefont
  {K.}~\bibnamefont {{Langanke}}}, \bibinfo {author} {\bibfnamefont
  {A.}~\bibnamefont {{Bauswein}}}, \bibinfo {author} {\bibfnamefont {H.-T.}\
  \bibnamefont {{Janka}}}, \ and\ \bibinfo {author} {\bibfnamefont
  {A.}~\bibnamefont {{Frank}}},\ }in\ \href {\doibase
  10.1088/1742-6596/730/1/012018} {\emph {\bibinfo {booktitle} {Journal of
  Physics Conference Series}}},\ \bibinfo {series} {Journal of Physics
  Conference Series}, Vol.\ \bibinfo {volume} {730}\ (\bibinfo {year} {2016})\
  p.\ \bibinfo {pages} {012018}\BibitemShut {NoStop}%
\bibitem [{\citenamefont {{Roberts}}\ \emph {et~al.}(2011)\citenamefont
  {{Roberts}}, \citenamefont {{Kasen}}, \citenamefont {{Lee}},\ and\
  \citenamefont {{Ramirez-Ruiz}}}]{RobertsWahl}%
  \BibitemOpen
  \bibfield  {author} {\bibinfo {author} {\bibfnamefont {L.~F.}\ \bibnamefont
  {{Roberts}}}, \bibinfo {author} {\bibfnamefont {D.}~\bibnamefont {{Kasen}}},
  \bibinfo {author} {\bibfnamefont {W.~H.}\ \bibnamefont {{Lee}}}, \ and\
  \bibinfo {author} {\bibfnamefont {E.}~\bibnamefont {{Ramirez-Ruiz}}},\ }\href
  {\doibase 10.1088/2041-8205/736/1/L21} {\bibfield  {journal} {\bibinfo
  {journal} {ApJL}\ }\textbf {\bibinfo {volume} {736}},\ \bibinfo {eid} {L21}
  (\bibinfo {year} {2011})},\ \Eprint {http://arxiv.org/abs/1104.5504}
  {arXiv:1104.5504 [astro-ph.HE]} \BibitemShut {NoStop}%
\bibitem [{\citenamefont {{Panov}}\ \emph {et~al.}(2010)\citenamefont
  {{Panov}}, \citenamefont {{Korneev}}, \citenamefont {{Rauscher}},
  \citenamefont {{Mart{\'{\i}}nez-Pinedo}}, \citenamefont {{Keli{\'c}-Heil}},
  \citenamefont {{Zinner}},\ and\ \citenamefont {{Thielemann}}}]{PanovNIF}%
  \BibitemOpen
  \bibfield  {author} {\bibinfo {author} {\bibfnamefont {I.~V.}\ \bibnamefont
  {{Panov}}}, \bibinfo {author} {\bibfnamefont {I.~Y.}\ \bibnamefont
  {{Korneev}}}, \bibinfo {author} {\bibfnamefont {T.}~\bibnamefont
  {{Rauscher}}}, \bibinfo {author} {\bibfnamefont {G.}~\bibnamefont
  {{Mart{\'{\i}}nez-Pinedo}}}, \bibinfo {author} {\bibfnamefont
  {A.}~\bibnamefont {{Keli{\'c}-Heil}}}, \bibinfo {author} {\bibfnamefont
  {N.~T.}\ \bibnamefont {{Zinner}}}, \ and\ \bibinfo {author} {\bibfnamefont
  {F.-K.}\ \bibnamefont {{Thielemann}}},\ }\href {\doibase
  10.1051/0004-6361/200911967} {\bibfield  {journal} {\bibinfo  {journal}
  {A\&A}\ }\textbf {\bibinfo {volume} {513}},\ \bibinfo {eid} {A61} (\bibinfo
  {year} {2010})},\ \Eprint {http://arxiv.org/abs/0911.2181} {arXiv:0911.2181
  [astro-ph.SR]} \BibitemShut {NoStop}%
\bibitem [{GEF()}]{GEFweb}%
  \BibitemOpen
  \href@noop {} {}\bibinfo {howpublished}
  {http://www.khs-erzhausen.de/GEF-2016-1-2.html}\BibitemShut {NoStop}%
\bibitem [{\citenamefont {{Verbeke}}\ \emph {et~al.}(2015)\citenamefont
  {{Verbeke}}, \citenamefont {{Randrup}},\ and\ \citenamefont {{Vogt}}}]{CPC}%
  \BibitemOpen
  \bibfield  {author} {\bibinfo {author} {\bibfnamefont {J.~M.}\ \bibnamefont
  {{Verbeke}}}, \bibinfo {author} {\bibfnamefont {J.}~\bibnamefont
  {{Randrup}}}, \ and\ \bibinfo {author} {\bibfnamefont {R.}~\bibnamefont
  {{Vogt}}},\ }\href {\doibase 10.1016/j.cpc.2015.02.002} {\bibfield  {journal}
  {\bibinfo  {journal} {Comp.\ Phys.\ Comm.}\ }\textbf {\bibinfo {volume}
  {191}},\ \bibinfo {pages} {178} (\bibinfo {year} {2015})}\BibitemShut
  {NoStop}%
\bibitem [{\citenamefont {{Verbeke}}\ \emph {et~al.}(2018)\citenamefont
  {{Verbeke}}, \citenamefont {{Randrup}},\ and\ \citenamefont
  {{Vogt}}}]{CPC_NVA}%
  \BibitemOpen
  \bibfield  {author} {\bibinfo {author} {\bibfnamefont {J.~M.}\ \bibnamefont
  {{Verbeke}}}, \bibinfo {author} {\bibfnamefont {J.}~\bibnamefont
  {{Randrup}}}, \ and\ \bibinfo {author} {\bibfnamefont {R.}~\bibnamefont
  {{Vogt}}},\ }\href {\doibase 10.1016/j.cpc.2017.09.006} {\bibfield  {journal}
  {\bibinfo  {journal} {Comp.\ Phys.\ Comm.}\ }\textbf {\bibinfo {volume}
  {222}},\ \bibinfo {pages} {263} (\bibinfo {year} {2018})}\BibitemShut
  {NoStop}%
\bibitem [{\citenamefont {{Petermann}}\ \emph {et~al.}(2012)\citenamefont
  {{Petermann}}, \citenamefont {{Langanke}}, \citenamefont
  {{Mart{\'{\i}}nez-Pinedo}}, \citenamefont {{Panov}}, \citenamefont
  {{Reinhard}},\ and\ \citenamefont {{Thielemann}}}]{Petermann}%
  \BibitemOpen
  \bibfield  {author} {\bibinfo {author} {\bibfnamefont {I.}~\bibnamefont
  {{Petermann}}}, \bibinfo {author} {\bibfnamefont {K.}~\bibnamefont
  {{Langanke}}}, \bibinfo {author} {\bibfnamefont {G.}~\bibnamefont
  {{Mart{\'{\i}}nez-Pinedo}}}, \bibinfo {author} {\bibfnamefont {I.~V.}\
  \bibnamefont {{Panov}}}, \bibinfo {author} {\bibfnamefont {P.-G.}\
  \bibnamefont {{Reinhard}}}, \ and\ \bibinfo {author} {\bibfnamefont {F.-K.}\
  \bibnamefont {{Thielemann}}},\ }\href {\doibase 10.1140/epja/i2012-12122-6}
  {\bibfield  {journal} {\bibinfo  {journal} {Eur.\ Phys.\ J.\ A}\ }\textbf
  {\bibinfo {volume} {48}},\ \bibinfo {eid} {122} (\bibinfo {year} {2012})},\
  \Eprint {http://arxiv.org/abs/1207.3432} {arXiv:1207.3432 [nucl-th]}
  \BibitemShut {NoStop}%
\bibitem [{\citenamefont {{Giuliani}}\ \emph {et~al.}(2018)\citenamefont
  {{Giuliani}}, \citenamefont {{Mart{\'{\i}}nez-Pinedo}},\ and\ \citenamefont
  {{Robledo}}}]{Samuel18}%
  \BibitemOpen
  \bibfield  {author} {\bibinfo {author} {\bibfnamefont {S.~A.}\ \bibnamefont
  {{Giuliani}}}, \bibinfo {author} {\bibfnamefont {G.}~\bibnamefont
  {{Mart{\'{\i}}nez-Pinedo}}}, \ and\ \bibinfo {author} {\bibfnamefont {L.~M.}\
  \bibnamefont {{Robledo}}},\ }\href {\doibase 10.1103/PhysRevC.97.034323}
  {\bibfield  {journal} {\bibinfo  {journal} {\prc}\ }\textbf {\bibinfo
  {volume} {97}},\ \bibinfo {eid} {034323} (\bibinfo {year} {2018})},\ \Eprint
  {http://arxiv.org/abs/1704.00554} {arXiv:1704.00554 [nucl-th]} \BibitemShut
  {NoStop}%
\bibitem [{\citenamefont {{Thielemann}}\ \emph {et~al.}(1983)\citenamefont
  {{Thielemann}}, \citenamefont {{Metzinger}},\ and\ \citenamefont
  {{Klapdor}}}]{Thielemann+83}%
  \BibitemOpen
  \bibfield  {author} {\bibinfo {author} {\bibfnamefont {F.-K.}\ \bibnamefont
  {{Thielemann}}}, \bibinfo {author} {\bibfnamefont {J.}~\bibnamefont
  {{Metzinger}}}, \ and\ \bibinfo {author} {\bibfnamefont {H.~V.}\ \bibnamefont
  {{Klapdor}}},\ }\href {\doibase 10.1007/BF01413833} {\bibfield  {journal}
  {\bibinfo  {journal} {Z.\ Phys.\ A}\ }\textbf {\bibinfo {volume} {309}},\
  \bibinfo {pages} {301} (\bibinfo {year} {1983})}\BibitemShut {NoStop}%
\bibitem [{\citenamefont {{Panov}}\ \emph {et~al.}(2013)\citenamefont
  {{Panov}}, \citenamefont {{Korneev}}, \citenamefont {{Martinez-Pinedo}},\
  and\ \citenamefont {{Thielemann}}}]{PanovSF}%
  \BibitemOpen
  \bibfield  {author} {\bibinfo {author} {\bibfnamefont {I.~V.}\ \bibnamefont
  {{Panov}}}, \bibinfo {author} {\bibfnamefont {I.~Y.}\ \bibnamefont
  {{Korneev}}}, \bibinfo {author} {\bibfnamefont {G.}~\bibnamefont
  {{Martinez-Pinedo}}}, \ and\ \bibinfo {author} {\bibfnamefont {F.-K.}\
  \bibnamefont {{Thielemann}}},\ }\href {\doibase 10.1134/S1063773713030043}
  {\bibfield  {journal} {\bibinfo  {journal} {Astron.\ Lett.}\ }\textbf
  {\bibinfo {volume} {39}},\ \bibinfo {pages} {150} (\bibinfo {year}
  {2013})}\BibitemShut {NoStop}%
\bibitem [{\citenamefont {{Mumpower}}\ \emph {et~al.}(2018)\citenamefont
  {{Mumpower}}, \citenamefont {{Kawano}}, \citenamefont {{Sprouse}},
  \citenamefont {{Vassh}}, \citenamefont {{Holmbeck}}, \citenamefont
  {{Surman}},\ and\ \citenamefont {{M{\"o}ller}}}]{BDFrp}%
  \BibitemOpen
  \bibfield  {author} {\bibinfo {author} {\bibfnamefont {M.~R.}\ \bibnamefont
  {{Mumpower}}}, \bibinfo {author} {\bibfnamefont {T.}~\bibnamefont
  {{Kawano}}}, \bibinfo {author} {\bibfnamefont {T.~M.}\ \bibnamefont
  {{Sprouse}}}, \bibinfo {author} {\bibfnamefont {N.}~\bibnamefont {{Vassh}}},
  \bibinfo {author} {\bibfnamefont {E.~M.}\ \bibnamefont {{Holmbeck}}},
  \bibinfo {author} {\bibfnamefont {R.}~\bibnamefont {{Surman}}}, \ and\
  \bibinfo {author} {\bibfnamefont {P.}~\bibnamefont {{M{\"o}ller}}},\ }\href
  {\doibase 10.3847/1538-4357/aaeaca} {\bibfield  {journal} {\bibinfo
  {journal} {ApJ}\ }\textbf {\bibinfo {volume} {869}},\ \bibinfo {eid} {14}
  (\bibinfo {year} {2018})},\ \Eprint {http://arxiv.org/abs/1802.04398}
  {arXiv:1802.04398 [nucl-th]} \BibitemShut {NoStop}%
\bibitem [{\citenamefont {{M{\"o}ller}}\ \emph {et~al.}(2016)\citenamefont
  {{M{\"o}ller}}, \citenamefont {{Sierk}}, \citenamefont {{Ichikawa}},\ and\
  \citenamefont {{Sagawa}}}]{FRDM2012}%
  \BibitemOpen
  \bibfield  {author} {\bibinfo {author} {\bibfnamefont {P.}~\bibnamefont
  {{M{\"o}ller}}}, \bibinfo {author} {\bibfnamefont {A.~J.}\ \bibnamefont
  {{Sierk}}}, \bibinfo {author} {\bibfnamefont {T.}~\bibnamefont {{Ichikawa}}},
  \ and\ \bibinfo {author} {\bibfnamefont {H.}~\bibnamefont {{Sagawa}}},\
  }\href {\doibase 10.1016/j.adt.2015.10.002} {\bibfield  {journal} {\bibinfo
  {journal} {At.\ Data Nucl.\ Data Tables}\ }\textbf {\bibinfo {volume}
  {109}},\ \bibinfo {pages} {1} (\bibinfo {year} {2016})},\ \Eprint
  {http://arxiv.org/abs/1508.06294} {arXiv:1508.06294 [nucl-th]} \BibitemShut
  {NoStop}%
\bibitem [{\citenamefont {{Myers}}\ and\ \citenamefont
  {{Swiatecki}}(1996)}]{TFmass}%
  \BibitemOpen
  \bibfield  {author} {\bibinfo {author} {\bibfnamefont {W.~D.}\ \bibnamefont
  {{Myers}}}\ and\ \bibinfo {author} {\bibfnamefont {W.~J.}\ \bibnamefont
  {{Swiatecki}}},\ }\href {\doibase 10.1016/0375-9474(95)00509-9} {\bibfield
  {journal} {\bibinfo  {journal} {Nucl.\ Phys.\ A}\ }\textbf {\bibinfo {volume}
  {601}},\ \bibinfo {pages} {141} (\bibinfo {year} {1996})}\BibitemShut
  {NoStop}%
\bibitem [{\citenamefont {{Goriely}}\ \emph {et~al.}(2009)\citenamefont
  {{Goriely}}, \citenamefont {{Chamel}},\ and\ \citenamefont
  {{Pearson}}}]{HFB17PRL}%
  \BibitemOpen
  \bibfield  {author} {\bibinfo {author} {\bibfnamefont {S.}~\bibnamefont
  {{Goriely}}}, \bibinfo {author} {\bibfnamefont {N.}~\bibnamefont {{Chamel}}},
  \ and\ \bibinfo {author} {\bibfnamefont {J.~M.}\ \bibnamefont {{Pearson}}},\
  }\href {\doibase 10.1103/PhysRevLett.102.152503} {\bibfield  {journal}
  {\bibinfo  {journal} {\prl}\ }\textbf {\bibinfo {volume} {102}},\ \bibinfo
  {eid} {152503} (\bibinfo {year} {2009})},\ \Eprint
  {http://arxiv.org/abs/0906.2607} {arXiv:0906.2607 [nucl-th]} \BibitemShut
  {NoStop}%
\bibitem [{\citenamefont {{Aboussir}}\ \emph {et~al.}(1992)\citenamefont
  {{Aboussir}}, \citenamefont {{Pearson}}, \citenamefont {{Dutta}},\ and\
  \citenamefont {{Tondeur}}}]{ETFSI1992}%
  \BibitemOpen
  \bibfield  {author} {\bibinfo {author} {\bibfnamefont {Y.}~\bibnamefont
  {{Aboussir}}}, \bibinfo {author} {\bibfnamefont {J.~M.}\ \bibnamefont
  {{Pearson}}}, \bibinfo {author} {\bibfnamefont {A.~K.}\ \bibnamefont
  {{Dutta}}}, \ and\ \bibinfo {author} {\bibfnamefont {F.}~\bibnamefont
  {{Tondeur}}},\ }\href {\doibase 10.1016/0375-9474(92)90038-L} {\bibfield
  {journal} {\bibinfo  {journal} {Nucl.\ Phys.\ A}\ }\textbf {\bibinfo {volume}
  {549}},\ \bibinfo {pages} {155} (\bibinfo {year} {1992})}\BibitemShut
  {NoStop}%
\bibitem [{\citenamefont {{Aboussir}}\ \emph {et~al.}(1995)\citenamefont
  {{Aboussir}}, \citenamefont {{Pearson}}, \citenamefont {{Dutta}},\ and\
  \citenamefont {{Tondeur}}}]{ETFSI1995}%
  \BibitemOpen
  \bibfield  {author} {\bibinfo {author} {\bibfnamefont {Y.}~\bibnamefont
  {{Aboussir}}}, \bibinfo {author} {\bibfnamefont {J.~M.}\ \bibnamefont
  {{Pearson}}}, \bibinfo {author} {\bibfnamefont {A.~K.}\ \bibnamefont
  {{Dutta}}}, \ and\ \bibinfo {author} {\bibfnamefont {F.}~\bibnamefont
  {{Tondeur}}},\ }\href {\doibase 10.1016/S0092-640X(95)90014-4} {\bibfield
  {journal} {\bibinfo  {journal} {At.\ Data Nucl.\ Data Tables}\ }\textbf
  {\bibinfo {volume} {61}},\ \bibinfo {pages} {127} (\bibinfo {year}
  {1995})}\BibitemShut {NoStop}%
\bibitem [{\citenamefont {Wu}\ \emph {et~al.}(2019)\citenamefont {Wu},
  \citenamefont {Barnes}, \citenamefont {Mart\'{\i}nez-Pinedo},\ and\
  \citenamefont {Metzger}}]{WuBGMP}%
  \BibitemOpen
  \bibfield  {author} {\bibinfo {author} {\bibfnamefont {M.-R.}\ \bibnamefont
  {Wu}}, \bibinfo {author} {\bibfnamefont {J.}~\bibnamefont {Barnes}}, \bibinfo
  {author} {\bibfnamefont {G.}~\bibnamefont {Mart\'{\i}nez-Pinedo}}, \ and\
  \bibinfo {author} {\bibfnamefont {B.~D.}\ \bibnamefont {Metzger}},\ }\href
  {\doibase 10.1103/PhysRevLett.122.062701} {\bibfield  {journal} {\bibinfo
  {journal} {\prl}\ }\textbf {\bibinfo {volume} {122}},\ \bibinfo {pages}
  {062701} (\bibinfo {year} {2019})}\BibitemShut {NoStop}%
\bibitem [{\citenamefont {{Koning}}\ \emph {et~al.}(2005)\citenamefont
  {{Koning}}, \citenamefont {{Hilaire}},\ and\ \citenamefont
  {{Duijvestijn}}}]{TALYSRef}%
  \BibitemOpen
  \bibfield  {author} {\bibinfo {author} {\bibfnamefont {A.~J.}\ \bibnamefont
  {{Koning}}}, \bibinfo {author} {\bibfnamefont {S.}~\bibnamefont {{Hilaire}}},
  \ and\ \bibinfo {author} {\bibfnamefont {M.~C.}\ \bibnamefont
  {{Duijvestijn}}},\ }in\ \href {\doibase 10.1063/1.1945212} {\emph {\bibinfo
  {booktitle} {International Conference on Nuclear Data for Science and
  Technology}}},\ \bibinfo {series} {American Institute of Physics Conference
  Series}, Vol.\ \bibinfo {volume} {769},\ \bibinfo {editor} {edited by\
  \bibinfo {editor} {\bibfnamefont {R.~C.}\ \bibnamefont {{Haight}}}, \bibinfo
  {editor} {\bibfnamefont {M.~B.}\ \bibnamefont {{Chadwick}}}, \bibinfo
  {editor} {\bibfnamefont {T.}~\bibnamefont {{Kawano}}}, \ and\ \bibinfo
  {editor} {\bibfnamefont {P.}~\bibnamefont {{Talou}}}}\ (\bibinfo {year}
  {2005})\ pp.\ \bibinfo {pages} {1154--1159}\BibitemShut {NoStop}%
\bibitem [{\citenamefont {{Koning}}\ \emph {et~al.}(2012)\citenamefont
  {{Koning}}, \citenamefont {{Hilaire}},\ and\ \citenamefont
  {{Duijvestijn}}}]{TALYScode}%
  \BibitemOpen
  \bibfield  {author} {\bibinfo {author} {\bibfnamefont {A.}~\bibnamefont
  {{Koning}}}, \bibinfo {author} {\bibfnamefont {S.}~\bibnamefont {{Hilaire}}},
  \ and\ \bibinfo {author} {\bibfnamefont {M.}~\bibnamefont {{Duijvestijn}}},\
  }\href@noop {} {\enquote {\bibinfo {title} {{TALYS: Nuclear Reaction
  Simulator}},}\ }\bibinfo {howpublished} {Astrophysics Source Code Library}
  (\bibinfo {year} {2012}),\ \Eprint {http://arxiv.org/abs/1202.004}
  {ascl.net:1202.004} \BibitemShut {NoStop}%
\bibitem [{\citenamefont {{Litaize}}\ \emph {et~al.}(2015)\citenamefont
  {{Litaize}}, \citenamefont {{Serot}},\ and\ \citenamefont
  {{Berge}}}]{FIFRELIN}%
  \BibitemOpen
  \bibfield  {author} {\bibinfo {author} {\bibfnamefont {O.}~\bibnamefont
  {{Litaize}}}, \bibinfo {author} {\bibfnamefont {O.}~\bibnamefont {{Serot}}},
  \ and\ \bibinfo {author} {\bibfnamefont {L.}~\bibnamefont {{Berge}}},\ }\href
  {\doibase 10.1140/epja/i2015-15177-9} {\bibfield  {journal} {\bibinfo
  {journal} {Eur.\ Phys.\ J.\ A}\ }\textbf {\bibinfo {volume} {51}},\ \bibinfo
  {eid} {177} (\bibinfo {year} {2015})}\BibitemShut {NoStop}%
\bibitem [{\citenamefont {{Talou}}\ \emph {et~al.}(2018)\citenamefont
  {{Talou}}, \citenamefont {{Vogt}}, \citenamefont {{Randrup}}, \citenamefont
  {{Rising}}, \citenamefont {{Pozzi}}, \citenamefont {{Verbeke}}, \citenamefont
  {{Andrews}}, \citenamefont {{Clarke}}, \citenamefont {{Jaffke}},
  \citenamefont {{Jandel}}, \citenamefont {{Kawano}}, \citenamefont
  {{Marcath}}, \citenamefont {{Meierbachtol}}, \citenamefont {{Nakae}},
  \citenamefont {{Rusev}}, \citenamefont {{Sood}}, \citenamefont {{Stetcu}},\
  and\ \citenamefont {{Walker}}}]{Talou:2017qlc}%
  \BibitemOpen
  \bibfield  {author} {\bibinfo {author} {\bibfnamefont {P.}~\bibnamefont
  {{Talou}}}, \bibinfo {author} {\bibfnamefont {R.}~\bibnamefont {{Vogt}}},
  \bibinfo {author} {\bibfnamefont {J.}~\bibnamefont {{Randrup}}}, \bibinfo
  {author} {\bibfnamefont {M.~E.}\ \bibnamefont {{Rising}}}, \bibinfo {author}
  {\bibfnamefont {S.~A.}\ \bibnamefont {{Pozzi}}}, \bibinfo {author}
  {\bibfnamefont {J.}~\bibnamefont {{Verbeke}}}, \bibinfo {author}
  {\bibfnamefont {M.~T.}\ \bibnamefont {{Andrews}}}, \bibinfo {author}
  {\bibfnamefont {S.~D.}\ \bibnamefont {{Clarke}}}, \bibinfo {author}
  {\bibfnamefont {P.}~\bibnamefont {{Jaffke}}}, \bibinfo {author}
  {\bibfnamefont {M.}~\bibnamefont {{Jandel}}}, \bibinfo {author}
  {\bibfnamefont {T.}~\bibnamefont {{Kawano}}}, \bibinfo {author}
  {\bibfnamefont {M.~J.}\ \bibnamefont {{Marcath}}}, \bibinfo {author}
  {\bibfnamefont {K.}~\bibnamefont {{Meierbachtol}}}, \bibinfo {author}
  {\bibfnamefont {L.}~\bibnamefont {{Nakae}}}, \bibinfo {author} {\bibfnamefont
  {G.}~\bibnamefont {{Rusev}}}, \bibinfo {author} {\bibfnamefont
  {A.}~\bibnamefont {{Sood}}}, \bibinfo {author} {\bibfnamefont
  {I.}~\bibnamefont {{Stetcu}}}, \ and\ \bibinfo {author} {\bibfnamefont
  {C.}~\bibnamefont {{Walker}}},\ }\href {\doibase 10.1140/epja/i2018-12455-0}
  {\bibfield  {journal} {\bibinfo  {journal} {Eur.\ Phys.\ J.\ A}\ }\textbf
  {\bibinfo {volume} {54}},\ \bibinfo {eid} {9} (\bibinfo {year} {2018})},\
  \Eprint {http://arxiv.org/abs/1710.00107} {arXiv:1710.00107 [nucl-th]}
  \BibitemShut {NoStop}%
\bibitem [{\citenamefont {{Randrup}}\ and\ \citenamefont
  {{M{\"o}ller}}(2011)}]{RandrupMoller}%
  \BibitemOpen
  \bibfield  {author} {\bibinfo {author} {\bibfnamefont {J.}~\bibnamefont
  {{Randrup}}}\ and\ \bibinfo {author} {\bibfnamefont {P.}~\bibnamefont
  {{M{\"o}ller}}},\ }\href {\doibase 10.1103/PhysRevLett.106.132503} {\bibfield
   {journal} {\bibinfo  {journal} {\prl}\ }\textbf {\bibinfo {volume} {106}},\
  \bibinfo {eid} {132503} (\bibinfo {year} {2011})},\ \Eprint
  {http://arxiv.org/abs/1103.0535} {arXiv:1103.0535 [nucl-th]} \BibitemShut
  {NoStop}%
\bibitem [{\citenamefont {{Jaffke}}\ \emph {et~al.}(2018)\citenamefont
  {{Jaffke}}, \citenamefont {{M{\"o}ller}}, \citenamefont {{Talou}},\ and\
  \citenamefont {{Sierk}}}]{PatrickJetal}%
  \BibitemOpen
  \bibfield  {author} {\bibinfo {author} {\bibfnamefont {P.}~\bibnamefont
  {{Jaffke}}}, \bibinfo {author} {\bibfnamefont {P.}~\bibnamefont
  {{M{\"o}ller}}}, \bibinfo {author} {\bibfnamefont {P.}~\bibnamefont
  {{Talou}}}, \ and\ \bibinfo {author} {\bibfnamefont {A.~J.}\ \bibnamefont
  {{Sierk}}},\ }\href {\doibase 10.1103/PhysRevC.97.034608} {\bibfield
  {journal} {\bibinfo  {journal} {\prc}\ }\textbf {\bibinfo {volume} {97}},\
  \bibinfo {eid} {034608} (\bibinfo {year} {2018})},\ \Eprint
  {http://arxiv.org/abs/1712.05511} {arXiv:1712.05511 [nucl-th]} \BibitemShut
  {NoStop}%
\bibitem [{\citenamefont {Schunck}\ and\ \citenamefont
  {Robledo}(2016)}]{schunck2016}%
  \BibitemOpen
  \bibfield  {author} {\bibinfo {author} {\bibfnamefont {N.}~\bibnamefont
  {Schunck}}\ and\ \bibinfo {author} {\bibfnamefont {L.~M.}\ \bibnamefont
  {Robledo}},\ }\href {\doibase 10.1088/0034-4885/79/11/116301} {\bibfield
  {journal} {\bibinfo  {journal} {Rep.\ Prog.\ Phys.}\ }\textbf {\bibinfo
  {volume} {79}},\ \bibinfo {pages} {116301} (\bibinfo {year}
  {2016})}\BibitemShut {NoStop}%
\bibitem [{\citenamefont {{Myers}}\ and\ \citenamefont
  {{Swiatecki}}(1999)}]{TFBH}%
  \BibitemOpen
  \bibfield  {author} {\bibinfo {author} {\bibfnamefont {W.~D.}\ \bibnamefont
  {{Myers}}}\ and\ \bibinfo {author} {\bibfnamefont {W.~J.}\ \bibnamefont
  {{Swiatecki}}},\ }\href {\doibase 10.1103/PhysRevC.60.014606} {\bibfield
  {journal} {\bibinfo  {journal} {\prc}\ }\textbf {\bibinfo {volume} {60}},\
  \bibinfo {eid} {014606} (\bibinfo {year} {1999})}\BibitemShut {NoStop}%
\bibitem [{\citenamefont {{Brosa}}\ \emph {et~al.}(1990)\citenamefont
  {{Brosa}}, \citenamefont {{Grossmann}},\ and\ \citenamefont
  {{M{\"u}ller}}}]{Brosa}%
  \BibitemOpen
  \bibfield  {author} {\bibinfo {author} {\bibfnamefont {U.}~\bibnamefont
  {{Brosa}}}, \bibinfo {author} {\bibfnamefont {S.}~\bibnamefont
  {{Grossmann}}}, \ and\ \bibinfo {author} {\bibfnamefont {A.}~\bibnamefont
  {{M{\"u}ller}}},\ }\href {\doibase 10.1016/0370-1573(90)90114-H} {\bibfield
  {journal} {\bibinfo  {journal} {Phys.\ Rep.}\ }\textbf {\bibinfo {volume}
  {197}},\ \bibinfo {pages} {167} (\bibinfo {year} {1990})}\BibitemShut
  {NoStop}%
\bibitem [{\citenamefont {{Mulgin}}\ and\ \citenamefont
  {{Zhdanov}}(1999)}]{Mulgin}%
  \BibitemOpen
  \bibfield  {author} {\bibinfo {author} {\bibfnamefont {S.~I.}\ \bibnamefont
  {{Mulgin}}}\ and\ \bibinfo {author} {\bibfnamefont {S.~V.}\ \bibnamefont
  {{Zhdanov}}},\ }\href {\doibase 10.1016/S0370-2693(99)00859-X} {\bibfield
  {journal} {\bibinfo  {journal} {Phys.\ Lett.\ B}\ }\textbf {\bibinfo {volume}
  {462}},\ \bibinfo {pages} {29} (\bibinfo {year} {1999})}\BibitemShut
  {NoStop}%
\bibitem [{\citenamefont {England}\ and\ \citenamefont
  {Rider}(1994)}]{england1994endf}%
  \BibitemOpen
  \bibfield  {author} {\bibinfo {author} {\bibfnamefont {T.}~\bibnamefont
  {England}}\ and\ \bibinfo {author} {\bibfnamefont {B.}~\bibnamefont
  {Rider}},\ }\href@noop {} {\bibfield  {journal} {\bibinfo  {journal} {LANL,
  LA-UR, 94e3106}\ } (\bibinfo {year} {1994})}\BibitemShut {NoStop}%
\bibitem [{\citenamefont {Gooden}\ \emph {et~al.}(2016)\citenamefont {Gooden},
  \citenamefont {Arnold}, \citenamefont {Becker}, \citenamefont {Bhatia},
  \citenamefont {Bhike}, \citenamefont {Bond}, \citenamefont {Bredeweg},
  \citenamefont {Fallin}, \citenamefont {Fowler}, \citenamefont {Howell},
  \citenamefont {Kelley}, \citenamefont {Krishichayan}, \citenamefont {Macri},
  \citenamefont {Rusev}, \citenamefont {Ryan}, \citenamefont {Sheets},
  \citenamefont {Stoyer}, \citenamefont {Tonchev}, \citenamefont {Tornow},
  \citenamefont {Vieira},\ and\ \citenamefont {Wilhelmy}}]{gooden2016}%
  \BibitemOpen
  \bibfield  {author} {\bibinfo {author} {\bibfnamefont {M.~E.}\ \bibnamefont
  {Gooden}}, \bibinfo {author} {\bibfnamefont {C.~W.}\ \bibnamefont {Arnold}},
  \bibinfo {author} {\bibfnamefont {J.~A.}\ \bibnamefont {Becker}}, \bibinfo
  {author} {\bibfnamefont {C.}~\bibnamefont {Bhatia}}, \bibinfo {author}
  {\bibfnamefont {M.}~\bibnamefont {Bhike}}, \bibinfo {author} {\bibfnamefont
  {E.~M.}\ \bibnamefont {Bond}}, \bibinfo {author} {\bibfnamefont {T.~A.}\
  \bibnamefont {Bredeweg}}, \bibinfo {author} {\bibfnamefont {B.}~\bibnamefont
  {Fallin}}, \bibinfo {author} {\bibfnamefont {M.~M.}\ \bibnamefont {Fowler}},
  \bibinfo {author} {\bibfnamefont {C.~R.}\ \bibnamefont {Howell}}, \bibinfo
  {author} {\bibfnamefont {J.~H.}\ \bibnamefont {Kelley}}, \bibinfo {author}
  {\bibnamefont {Krishichayan}}, \bibinfo {author} {\bibfnamefont
  {R.}~\bibnamefont {Macri}}, \bibinfo {author} {\bibfnamefont
  {G.}~\bibnamefont {Rusev}}, \bibinfo {author} {\bibfnamefont
  {C.}~\bibnamefont {Ryan}}, \bibinfo {author} {\bibfnamefont {S.~A.}\
  \bibnamefont {Sheets}}, \bibinfo {author} {\bibfnamefont {M.~A.}\
  \bibnamefont {Stoyer}}, \bibinfo {author} {\bibfnamefont {A.~P.}\
  \bibnamefont {Tonchev}}, \bibinfo {author} {\bibfnamefont {W.}~\bibnamefont
  {Tornow}}, \bibinfo {author} {\bibfnamefont {D.~J.}\ \bibnamefont {Vieira}},
  \ and\ \bibinfo {author} {\bibfnamefont {J.~B.}\ \bibnamefont {Wilhelmy}},\
  }\href {\doibase 10.1016/j.nds.2015.12.006} {\bibfield  {journal} {\bibinfo
  {journal} {Nucl. Data Sheets}\ }\bibinfo {series} {Special Issue on Nuclear
  Reaction Data},\ \textbf {\bibinfo {volume} {131}},\ \bibinfo {pages} {319}
  (\bibinfo {year} {2016})}\BibitemShut {NoStop}%
\bibitem [{\citenamefont {{Rubehn}}\ \emph {et~al.}(1996)\citenamefont
  {{Rubehn}}, \citenamefont {{Jing}}, \citenamefont {{Moretto}}, \citenamefont
  {{Phair}}, \citenamefont {{Tso}},\ and\ \citenamefont
  {{Wozniak}}}]{GEFGammaNref}%
  \BibitemOpen
  \bibfield  {author} {\bibinfo {author} {\bibfnamefont {T.}~\bibnamefont
  {{Rubehn}}}, \bibinfo {author} {\bibfnamefont {K.~X.}\ \bibnamefont
  {{Jing}}}, \bibinfo {author} {\bibfnamefont {L.~G.}\ \bibnamefont
  {{Moretto}}}, \bibinfo {author} {\bibfnamefont {L.}~\bibnamefont {{Phair}}},
  \bibinfo {author} {\bibfnamefont {K.}~\bibnamefont {{Tso}}}, \ and\ \bibinfo
  {author} {\bibfnamefont {G.~J.}\ \bibnamefont {{Wozniak}}},\ }\href {\doibase
  10.1103/PhysRevC.54.3062} {\bibfield  {journal} {\bibinfo  {journal} {\prc}\
  }\textbf {\bibinfo {volume} {54}},\ \bibinfo {pages} {3062} (\bibinfo {year}
  {1996})},\ \Eprint {http://arxiv.org/abs/nucl-ex/9607005} {nucl-ex/9607005}
  \BibitemShut {NoStop}%
\bibitem [{\citenamefont {Vogt}\ and\ \citenamefont
  {Randrup}(2017)}]{RVJR_gamma2}%
  \BibitemOpen
  \bibfield  {author} {\bibinfo {author} {\bibfnamefont {R.}~\bibnamefont
  {Vogt}}\ and\ \bibinfo {author} {\bibfnamefont {J.}~\bibnamefont {Randrup}},\
  }\href {\doibase 10.1103/PhysRevC.96.064620} {\bibfield  {journal} {\bibinfo
  {journal} {\prc}\ }\textbf {\bibinfo {volume} {96}},\ \bibinfo {pages}
  {064620} (\bibinfo {year} {2017})}\BibitemShut {NoStop}%
\bibitem [{\citenamefont {Lemaire}\ \emph {et~al.}(2005)\citenamefont
  {Lemaire}, \citenamefont {Talou}, \citenamefont {Kawano}, \citenamefont
  {Chadwick},\ and\ \citenamefont {Madland}}]{Lemaire}%
  \BibitemOpen
  \bibfield  {author} {\bibinfo {author} {\bibfnamefont {S.}~\bibnamefont
  {Lemaire}}, \bibinfo {author} {\bibfnamefont {P.}~\bibnamefont {Talou}},
  \bibinfo {author} {\bibfnamefont {T.}~\bibnamefont {Kawano}}, \bibinfo
  {author} {\bibfnamefont {M.~B.}\ \bibnamefont {Chadwick}}, \ and\ \bibinfo
  {author} {\bibfnamefont {D.~G.}\ \bibnamefont {Madland}},\ }\href {\doibase
  10.1103/PhysRevC.72.024601} {\bibfield  {journal} {\bibinfo  {journal}
  {\prc}\ }\textbf {\bibinfo {volume} {72}},\ \bibinfo {pages} {024601}
  (\bibinfo {year} {2005})}\BibitemShut {NoStop}%
\bibitem [{\citenamefont {{M{\"o}ller}}\ \emph {et~al.}(1997)\citenamefont
  {{M{\"o}ller}}, \citenamefont {{Nix}},\ and\ \citenamefont
  {{Kratz}}}]{MollerSd0}%
  \BibitemOpen
  \bibfield  {author} {\bibinfo {author} {\bibfnamefont {P.}~\bibnamefont
  {{M{\"o}ller}}}, \bibinfo {author} {\bibfnamefont {J.~R.}\ \bibnamefont
  {{Nix}}}, \ and\ \bibinfo {author} {\bibfnamefont {K.-L.}\ \bibnamefont
  {{Kratz}}},\ }\href {\doibase 10.1006/adnd.1997.0746} {\bibfield  {journal}
  {\bibinfo  {journal} {Atomic Data and Nuclear Data Tables}\ }\textbf
  {\bibinfo {volume} {66}},\ \bibinfo {pages} {131} (\bibinfo {year}
  {1997})}\BibitemShut {NoStop}%
\bibitem [{\citenamefont {{Wang}}\ \emph {et~al.}(2017)\citenamefont {{Wang}},
  \citenamefont {{Audi}}, \citenamefont {{Kondev}}, \citenamefont {{Huang}},
  \citenamefont {{Naimi}},\ and\ \citenamefont {{Xu}}}]{AME2016}%
  \BibitemOpen
  \bibfield  {author} {\bibinfo {author} {\bibfnamefont {M.}~\bibnamefont
  {{Wang}}}, \bibinfo {author} {\bibfnamefont {G.}~\bibnamefont {{Audi}}},
  \bibinfo {author} {\bibfnamefont {F.~G.}\ \bibnamefont {{Kondev}}}, \bibinfo
  {author} {\bibfnamefont {W.~J.}\ \bibnamefont {{Huang}}}, \bibinfo {author}
  {\bibfnamefont {S.}~\bibnamefont {{Naimi}}}, \ and\ \bibinfo {author}
  {\bibfnamefont {X.}~\bibnamefont {{Xu}}},\ }\href {\doibase
  10.1088/1674-1137/41/3/030003} {\bibfield  {journal} {\bibinfo  {journal}
  {Chin.\ Phys. C}\ }\textbf {\bibinfo {volume} {41}},\ \bibinfo {eid} {030003}
  (\bibinfo {year} {2017})}\BibitemShut {NoStop}%
\bibitem [{\citenamefont {Audi}\ \emph {et~al.}(2017)\citenamefont {Audi},
  \citenamefont {Kondev}, \citenamefont {Wang}, \citenamefont {Huang},\ and\
  \citenamefont {Naimi}}]{NUBASE2016}%
  \BibitemOpen
  \bibfield  {author} {\bibinfo {author} {\bibfnamefont {G.}~\bibnamefont
  {Audi}}, \bibinfo {author} {\bibfnamefont {F.}~\bibnamefont {Kondev}},
  \bibinfo {author} {\bibfnamefont {M.}~\bibnamefont {Wang}}, \bibinfo {author}
  {\bibfnamefont {W.}~\bibnamefont {Huang}}, \ and\ \bibinfo {author}
  {\bibfnamefont {S.}~\bibnamefont {Naimi}},\ }\href
  {http://stacks.iop.org/1674-1137/41/i=3/a=030001} {\bibfield  {journal}
  {\bibinfo  {journal} {Chin.\ Phys.\ C}\ }\textbf {\bibinfo {volume} {41}},\
  \bibinfo {pages} {030001} (\bibinfo {year} {2017})}\BibitemShut {NoStop}%
\bibitem [{\citenamefont {{Viola, Jr.}}\ and\ \citenamefont
  {{Seaborg}}(1966)}]{VS1966}%
  \BibitemOpen
  \bibfield  {author} {\bibinfo {author} {\bibfnamefont {V.~E.}\ \bibnamefont
  {{Viola, Jr.}}}\ and\ \bibinfo {author} {\bibfnamefont {G.~T.}\ \bibnamefont
  {{Seaborg}}},\ }\href@noop {} {\bibfield  {journal} {\bibinfo  {journal} {J.
  \ Inorg. \ Nucl. \ Chem.}\ }\textbf {\bibinfo {volume} {28}},\ \bibinfo
  {pages} {741} (\bibinfo {year} {1966})}\BibitemShut {NoStop}%
\bibitem [{\citenamefont {{Kawano}}\ \emph {et~al.}(2017)\citenamefont
  {{Kawano}}, \citenamefont {{Mumpower}},\ and\ \citenamefont
  {{Ullmann}}}]{Kawano+17}%
  \BibitemOpen
  \bibfield  {author} {\bibinfo {author} {\bibfnamefont {T.}~\bibnamefont
  {{Kawano}}}, \bibinfo {author} {\bibfnamefont {M.~R.}\ \bibnamefont
  {{Mumpower}}}, \ and\ \bibinfo {author} {\bibfnamefont {J.~L.}\ \bibnamefont
  {{Ullmann}}},\ }in\ \href {\doibase 10.7566/JPSCP.14.011003} {\emph {\bibinfo
  {booktitle} {14th International Symposium on Nuclei in the Cosmos
  (NIC2016)}}},\ \bibinfo {editor} {edited by\ \bibinfo {editor} {\bibfnamefont
  {S.}~\bibnamefont {{Kubono}}}, \bibinfo {editor} {\bibfnamefont
  {T.}~\bibnamefont {{Kajino}}}, \bibinfo {editor} {\bibfnamefont
  {S.}~\bibnamefont {{Nishimura}}}, \bibinfo {editor} {\bibfnamefont
  {T.}~\bibnamefont {{Isobe}}}, \bibinfo {editor} {\bibfnamefont
  {S.}~\bibnamefont {{Nagataki}}}, \bibinfo {editor} {\bibfnamefont
  {T.}~\bibnamefont {{Shima}}}, \ and\ \bibinfo {editor} {\bibfnamefont
  {Y.}~\bibnamefont {{Takeda}}}}\ (\bibinfo {year} {2017})\ p.\ \bibinfo
  {pages} {011003}\BibitemShut {NoStop}%
\bibitem [{\citenamefont {{Mumpower}}\ \emph {et~al.}(2014)\citenamefont
  {{Mumpower}}, \citenamefont {{Kawano}},\ and\ \citenamefont
  {{M{\"o}ller}}}]{Mumpower+14}%
  \BibitemOpen
  \bibfield  {author} {\bibinfo {author} {\bibfnamefont {M.~R.}\ \bibnamefont
  {{Mumpower}}}, \bibinfo {author} {\bibfnamefont {T.}~\bibnamefont
  {{Kawano}}}, \ and\ \bibinfo {author} {\bibfnamefont {P.}~\bibnamefont
  {{M{\"o}ller}}},\ }in\ \href@noop {} {\emph {\bibinfo {booktitle} {APS
  Division of Nuclear Physics Meeting Abstracts}}}\ (\bibinfo {year} {2014})\
  p.\ \bibinfo {pages} {KD.010}\BibitemShut {NoStop}%
\bibitem [{\citenamefont {{Mumpower}}\ \emph {et~al.}(2016)\citenamefont
  {{Mumpower}}, \citenamefont {{Kawano}},\ and\ \citenamefont
  {{M{\"o}ller}}}]{Mumpower+16}%
  \BibitemOpen
  \bibfield  {author} {\bibinfo {author} {\bibfnamefont {M.~R.}\ \bibnamefont
  {{Mumpower}}}, \bibinfo {author} {\bibfnamefont {T.}~\bibnamefont
  {{Kawano}}}, \ and\ \bibinfo {author} {\bibfnamefont {P.}~\bibnamefont
  {{M{\"o}ller}}},\ }\href {\doibase 10.1103/PhysRevC.94.064317} {\bibfield
  {journal} {\bibinfo  {journal} {\prc}\ }\textbf {\bibinfo {volume} {94}},\
  \bibinfo {eid} {064317} (\bibinfo {year} {2016})},\ \Eprint
  {http://arxiv.org/abs/1608.01956} {arXiv:1608.01956 [nucl-th]} \BibitemShut
  {NoStop}%
\bibitem [{\citenamefont {{M{\"o}ller}}\ \emph {et~al.}(2019)\citenamefont
  {{M{\"o}ller}}, \citenamefont {{Mumpower}}, \citenamefont {{Kawano}},\ and\
  \citenamefont {{Myers}}}]{MollerQRPA}%
  \BibitemOpen
  \bibfield  {author} {\bibinfo {author} {\bibfnamefont {P.}~\bibnamefont
  {{M{\"o}ller}}}, \bibinfo {author} {\bibfnamefont {M.~R.}\ \bibnamefont
  {{Mumpower}}}, \bibinfo {author} {\bibfnamefont {T.}~\bibnamefont
  {{Kawano}}}, \ and\ \bibinfo {author} {\bibfnamefont {W.~D.}\ \bibnamefont
  {{Myers}}},\ }\href {\doibase https://doi.org/10.1016/j.adt.2018.03.003}
  {\bibfield  {journal} {\bibinfo  {journal} {At.\ Data Nucl.\ Data Tables}\
  }\textbf {\bibinfo {volume} {125}},\ \bibinfo {pages} {1} (\bibinfo {year}
  {2019})}\BibitemShut {NoStop}%
\bibitem [{\citenamefont {{Mumpower}}\ \emph {et~al.}(2015)\citenamefont
  {{Mumpower}}, \citenamefont {{Surman}}, \citenamefont {{Fang}}, \citenamefont
  {{Beard}}, \citenamefont {{M{\"o}ller}}, \citenamefont {{Kawano}},\ and\
  \citenamefont {{Aprahamian}}}]{Mumpower+15}%
  \BibitemOpen
  \bibfield  {author} {\bibinfo {author} {\bibfnamefont {M.~R.}\ \bibnamefont
  {{Mumpower}}}, \bibinfo {author} {\bibfnamefont {R.}~\bibnamefont
  {{Surman}}}, \bibinfo {author} {\bibfnamefont {D.-L.}\ \bibnamefont
  {{Fang}}}, \bibinfo {author} {\bibfnamefont {M.}~\bibnamefont {{Beard}}},
  \bibinfo {author} {\bibfnamefont {P.}~\bibnamefont {{M{\"o}ller}}}, \bibinfo
  {author} {\bibfnamefont {T.}~\bibnamefont {{Kawano}}}, \ and\ \bibinfo
  {author} {\bibfnamefont {A.}~\bibnamefont {{Aprahamian}}},\ }\href {\doibase
  10.1103/PhysRevC.92.035807} {\bibfield  {journal} {\bibinfo  {journal}
  {\prc}\ }\textbf {\bibinfo {volume} {92}},\ \bibinfo {eid} {035807} (\bibinfo
  {year} {2015})},\ \Eprint {http://arxiv.org/abs/1505.07789} {arXiv:1505.07789
  [nucl-th]} \BibitemShut {NoStop}%
\bibitem [{\citenamefont {{Karpov}}\ \emph {et~al.}(2012)\citenamefont
  {{Karpov}}, \citenamefont {{Zagrebaev}}, \citenamefont {{Martinez
  Palenzuela}}, \citenamefont {{Felipe Ruiz}},\ and\ \citenamefont
  {{Greiner}}}]{Karpov}%
  \BibitemOpen
  \bibfield  {author} {\bibinfo {author} {\bibfnamefont {A.~V.}\ \bibnamefont
  {{Karpov}}}, \bibinfo {author} {\bibfnamefont {V.~I.}\ \bibnamefont
  {{Zagrebaev}}}, \bibinfo {author} {\bibfnamefont {Y.}~\bibnamefont {{Martinez
  Palenzuela}}}, \bibinfo {author} {\bibfnamefont {L.}~\bibnamefont {{Felipe
  Ruiz}}}, \ and\ \bibinfo {author} {\bibfnamefont {W.}~\bibnamefont
  {{Greiner}}},\ }\href {\doibase 10.1142/S0218301312500139} {\bibfield
  {journal} {\bibinfo  {journal} {Int.\ J.\ Mod.\ Phys.\ E}\ }\textbf {\bibinfo
  {volume} {21}},\ \bibinfo {eid} {1250013-1-1250013-20} (\bibinfo {year}
  {2012})}\BibitemShut {NoStop}%
\bibitem [{\citenamefont {{Zagrebaev}}\ \emph {et~al.}(2011)\citenamefont
  {{Zagrebaev}}, \citenamefont {{Karpov}}, \citenamefont {{Mishustin}},\ and\
  \citenamefont {{Greiner}}}]{Zagrebaev}%
  \BibitemOpen
  \bibfield  {author} {\bibinfo {author} {\bibfnamefont {V.~I.}\ \bibnamefont
  {{Zagrebaev}}}, \bibinfo {author} {\bibfnamefont {A.~V.}\ \bibnamefont
  {{Karpov}}}, \bibinfo {author} {\bibfnamefont {I.~N.}\ \bibnamefont
  {{Mishustin}}}, \ and\ \bibinfo {author} {\bibfnamefont {W.}~\bibnamefont
  {{Greiner}}},\ }\href {\doibase 10.1103/PhysRevC.84.044617} {\bibfield
  {journal} {\bibinfo  {journal} {\prc}\ }\textbf {\bibinfo {volume} {84}},\
  \bibinfo {eid} {044617} (\bibinfo {year} {2011})}\BibitemShut {NoStop}%
\bibitem [{\citenamefont {{Sneden}}\ \emph {et~al.}(2008)\citenamefont
  {{Sneden}}, \citenamefont {{Cowan}},\ and\ \citenamefont
  {{Gallino}}}]{Sneden}%
  \BibitemOpen
  \bibfield  {author} {\bibinfo {author} {\bibfnamefont {C.}~\bibnamefont
  {{Sneden}}}, \bibinfo {author} {\bibfnamefont {J.~J.}\ \bibnamefont
  {{Cowan}}}, \ and\ \bibinfo {author} {\bibfnamefont {R.}~\bibnamefont
  {{Gallino}}},\ }\href {\doibase 10.1146/annurev.astro.46.060407.145207}
  {\bibfield  {journal} {\bibinfo  {journal} {ARA\&A}\ }\textbf {\bibinfo
  {volume} {46}},\ \bibinfo {pages} {241} (\bibinfo {year} {2008})}\BibitemShut
  {NoStop}%
\bibitem [{\citenamefont {{Rosswog}}\ \emph {et~al.}(2013)\citenamefont
  {{Rosswog}}, \citenamefont {{Piran}},\ and\ \citenamefont
  {{Nakar}}}]{Rosswog}%
  \BibitemOpen
  \bibfield  {author} {\bibinfo {author} {\bibfnamefont {S.}~\bibnamefont
  {{Rosswog}}}, \bibinfo {author} {\bibfnamefont {T.}~\bibnamefont {{Piran}}},
  \ and\ \bibinfo {author} {\bibfnamefont {E.}~\bibnamefont {{Nakar}}},\ }\href
  {\doibase 10.1093/mnras/sts708} {\bibfield  {journal} {\bibinfo  {journal}
  {Mon.\ Not.\ R.\ Astron.\ Soc.}\ }\textbf {\bibinfo {volume} {430}},\
  \bibinfo {pages} {2585} (\bibinfo {year} {2013})},\ \Eprint
  {http://arxiv.org/abs/1204.6240} {arXiv:1204.6240 [astro-ph.HE]} \BibitemShut
  {NoStop}%
\bibitem [{\citenamefont {{Piran}}\ \emph {et~al.}(2013)\citenamefont
  {{Piran}}, \citenamefont {{Nakar}},\ and\ \citenamefont
  {{Rosswog}}}]{PiranRoss}%
  \BibitemOpen
  \bibfield  {author} {\bibinfo {author} {\bibfnamefont {T.}~\bibnamefont
  {{Piran}}}, \bibinfo {author} {\bibfnamefont {E.}~\bibnamefont {{Nakar}}}, \
  and\ \bibinfo {author} {\bibfnamefont {S.}~\bibnamefont {{Rosswog}}},\ }\href
  {\doibase 10.1093/mnras/stt037} {\bibfield  {journal} {\bibinfo  {journal}
  {Mon.\ Not.\ R.\ Astron.\ Soc.}\ }\textbf {\bibinfo {volume} {430}},\
  \bibinfo {pages} {2121} (\bibinfo {year} {2013})},\ \Eprint
  {http://arxiv.org/abs/1204.6242} {arXiv:1204.6242 [astro-ph.HE]} \BibitemShut
  {NoStop}%
\bibitem [{\citenamefont {{Korobkin}}\ \emph
  {et~al.}(2012{\natexlab{b}})\citenamefont {{Korobkin}}, \citenamefont
  {{Rosswog}}, \citenamefont {{Arcones}},\ and\ \citenamefont
  {{Winteler}}}]{Oleg}%
  \BibitemOpen
  \bibfield  {author} {\bibinfo {author} {\bibfnamefont {O.}~\bibnamefont
  {{Korobkin}}}, \bibinfo {author} {\bibfnamefont {S.}~\bibnamefont
  {{Rosswog}}}, \bibinfo {author} {\bibfnamefont {A.}~\bibnamefont
  {{Arcones}}}, \ and\ \bibinfo {author} {\bibfnamefont {C.}~\bibnamefont
  {{Winteler}}},\ }\href {\doibase 10.1111/j.1365-2966.2012.21859.x} {\bibfield
   {journal} {\bibinfo  {journal} {Mon.\ Not.\ R.\ Astron.\ Soc.}\ }\textbf
  {\bibinfo {volume} {426}},\ \bibinfo {pages} {1940} (\bibinfo {year}
  {2012}{\natexlab{b}})},\ \Eprint {http://arxiv.org/abs/1206.2379}
  {arXiv:1206.2379 [astro-ph.SR]} \BibitemShut {NoStop}%
\bibitem [{\citenamefont {{Metzger}}\ \emph {et~al.}(2010)\citenamefont
  {{Metzger}}, \citenamefont {{Mart{\'{\i}}nez-Pinedo}}, \citenamefont
  {{Darbha}}, \citenamefont {{Quataert}}, \citenamefont {{Arcones}},
  \citenamefont {{Kasen}}, \citenamefont {{Thomas}}, \citenamefont {{Nugent}},
  \citenamefont {{Panov}},\ and\ \citenamefont {{Zinner}}}]{Metzger2010}%
  \BibitemOpen
  \bibfield  {author} {\bibinfo {author} {\bibfnamefont {B.~D.}\ \bibnamefont
  {{Metzger}}}, \bibinfo {author} {\bibfnamefont {G.}~\bibnamefont
  {{Mart{\'{\i}}nez-Pinedo}}}, \bibinfo {author} {\bibfnamefont
  {S.}~\bibnamefont {{Darbha}}}, \bibinfo {author} {\bibfnamefont
  {E.}~\bibnamefont {{Quataert}}}, \bibinfo {author} {\bibfnamefont
  {A.}~\bibnamefont {{Arcones}}}, \bibinfo {author} {\bibfnamefont
  {D.}~\bibnamefont {{Kasen}}}, \bibinfo {author} {\bibfnamefont
  {R.}~\bibnamefont {{Thomas}}}, \bibinfo {author} {\bibfnamefont
  {P.}~\bibnamefont {{Nugent}}}, \bibinfo {author} {\bibfnamefont {I.~V.}\
  \bibnamefont {{Panov}}}, \ and\ \bibinfo {author} {\bibfnamefont {N.~T.}\
  \bibnamefont {{Zinner}}},\ }\href {\doibase 10.1111/j.1365-2966.2010.16864.x}
  {\bibfield  {journal} {\bibinfo  {journal} {Mon.\ Not.\ R.\ Astron.\ Soc.}\
  }\textbf {\bibinfo {volume} {406}},\ \bibinfo {pages} {2650} (\bibinfo {year}
  {2010})},\ \Eprint {http://arxiv.org/abs/1001.5029} {arXiv:1001.5029
  [astro-ph.HE]} \BibitemShut {NoStop}%
\bibitem [{\citenamefont {{Lippuner}}\ and\ \citenamefont
  {{Roberts}}(2015)}]{Jonas}%
  \BibitemOpen
  \bibfield  {author} {\bibinfo {author} {\bibfnamefont {J.}~\bibnamefont
  {{Lippuner}}}\ and\ \bibinfo {author} {\bibfnamefont {L.~F.}\ \bibnamefont
  {{Roberts}}},\ }\href {\doibase 10.1088/0004-637X/815/2/82} {\bibfield
  {journal} {\bibinfo  {journal} {\apj}\ }\textbf {\bibinfo {volume} {815}},\
  \bibinfo {eid} {82} (\bibinfo {year} {2015})},\ \Eprint
  {http://arxiv.org/abs/1508.03133} {arXiv:1508.03133 [astro-ph.HE]}
  \BibitemShut {NoStop}%
\bibitem [{\citenamefont {{Steiner}}\ \emph {et~al.}(2013)\citenamefont
  {{Steiner}}, \citenamefont {{Hempel}},\ and\ \citenamefont
  {{Fischer}}}]{SFHoSteiner}%
  \BibitemOpen
  \bibfield  {author} {\bibinfo {author} {\bibfnamefont {A.~W.}\ \bibnamefont
  {{Steiner}}}, \bibinfo {author} {\bibfnamefont {M.}~\bibnamefont {{Hempel}}},
  \ and\ \bibinfo {author} {\bibfnamefont {T.}~\bibnamefont {{Fischer}}},\
  }\href {\doibase 10.1088/0004-637X/774/1/17} {\bibfield  {journal} {\bibinfo
  {journal} {ApJ}\ }\textbf {\bibinfo {volume} {774}},\ \bibinfo {eid} {17}
  (\bibinfo {year} {2013})},\ \Eprint {http://arxiv.org/abs/1207.2184}
  {arXiv:1207.2184 [astro-ph.SR]} \BibitemShut {NoStop}%
\bibitem [{\citenamefont {{M{\"o}ller}}\ \emph {et~al.}(2009)\citenamefont
  {{M{\"o}ller}}, \citenamefont {{Sierk}}, \citenamefont {{Ichikawa}},
  \citenamefont {{Iwamoto}}, \citenamefont {{Bengtsson}}, \citenamefont
  {{Uhrenholt}},\ and\ \citenamefont {{{\AA}berg}}}]{MollerBH1}%
  \BibitemOpen
  \bibfield  {author} {\bibinfo {author} {\bibfnamefont {P.}~\bibnamefont
  {{M{\"o}ller}}}, \bibinfo {author} {\bibfnamefont {A.~J.}\ \bibnamefont
  {{Sierk}}}, \bibinfo {author} {\bibfnamefont {T.}~\bibnamefont {{Ichikawa}}},
  \bibinfo {author} {\bibfnamefont {A.}~\bibnamefont {{Iwamoto}}}, \bibinfo
  {author} {\bibfnamefont {R.}~\bibnamefont {{Bengtsson}}}, \bibinfo {author}
  {\bibfnamefont {H.}~\bibnamefont {{Uhrenholt}}}, \ and\ \bibinfo {author}
  {\bibfnamefont {S.}~\bibnamefont {{{\AA}berg}}},\ }\href {\doibase
  10.1103/PhysRevC.79.064304} {\bibfield  {journal} {\bibinfo  {journal}
  {\prc}\ }\textbf {\bibinfo {volume} {79}},\ \bibinfo {eid} {064304} (\bibinfo
  {year} {2009})}\BibitemShut {NoStop}%
\bibitem [{\citenamefont {{M{\"o}ller}}\ \emph {et~al.}(2015)\citenamefont
  {{M{\"o}ller}}, \citenamefont {{Sierk}}, \citenamefont {{Ichikawa}},
  \citenamefont {{Iwamoto}},\ and\ \citenamefont {{Mumpower}}}]{MollerBH2}%
  \BibitemOpen
  \bibfield  {author} {\bibinfo {author} {\bibfnamefont {P.}~\bibnamefont
  {{M{\"o}ller}}}, \bibinfo {author} {\bibfnamefont {A.~J.}\ \bibnamefont
  {{Sierk}}}, \bibinfo {author} {\bibfnamefont {T.}~\bibnamefont {{Ichikawa}}},
  \bibinfo {author} {\bibfnamefont {A.}~\bibnamefont {{Iwamoto}}}, \ and\
  \bibinfo {author} {\bibfnamefont {M.~R.}\ \bibnamefont {{Mumpower}}},\ }\href
  {\doibase 10.1103/PhysRevC.91.024310} {\bibfield  {journal} {\bibinfo
  {journal} {\prc}\ }\textbf {\bibinfo {volume} {91}},\ \bibinfo {eid} {024310}
  (\bibinfo {year} {2015})}\BibitemShut {NoStop}%
\bibitem [{\citenamefont {{Mamdouh}}\ \emph {et~al.}(1998)\citenamefont
  {{Mamdouh}}, \citenamefont {{Pearson}}, \citenamefont {{Rayet}},\ and\
  \citenamefont {{Tondeur}}}]{Mamdouh98}%
  \BibitemOpen
  \bibfield  {author} {\bibinfo {author} {\bibfnamefont {A.}~\bibnamefont
  {{Mamdouh}}}, \bibinfo {author} {\bibfnamefont {J.~M.}\ \bibnamefont
  {{Pearson}}}, \bibinfo {author} {\bibfnamefont {M.}~\bibnamefont {{Rayet}}},
  \ and\ \bibinfo {author} {\bibfnamefont {F.}~\bibnamefont {{Tondeur}}},\
  }\href {\doibase 10.1016/S0375-9474(98)00576-4} {\bibfield  {journal}
  {\bibinfo  {journal} {Nucl.\ Phys.\ A}\ }\textbf {\bibinfo {volume} {644}},\
  \bibinfo {pages} {389} (\bibinfo {year} {1998})}\BibitemShut {NoStop}%
\bibitem [{\citenamefont {{Mamdouh}}\ \emph {et~al.}(2001)\citenamefont
  {{Mamdouh}}, \citenamefont {{Pearson}}, \citenamefont {{Rayet}},\ and\
  \citenamefont {{Tondeur}}}]{Mamdouh01}%
  \BibitemOpen
  \bibfield  {author} {\bibinfo {author} {\bibfnamefont {A.}~\bibnamefont
  {{Mamdouh}}}, \bibinfo {author} {\bibfnamefont {J.~M.}\ \bibnamefont
  {{Pearson}}}, \bibinfo {author} {\bibfnamefont {M.}~\bibnamefont {{Rayet}}},
  \ and\ \bibinfo {author} {\bibfnamefont {F.}~\bibnamefont {{Tondeur}}},\
  }\href {\doibase 10.1016/S0375-9474(00)00358-4} {\bibfield  {journal}
  {\bibinfo  {journal} {Nucl.\ Phys.\ A}\ }\textbf {\bibinfo {volume} {679}},\
  \bibinfo {pages} {337} (\bibinfo {year} {2001})},\ \Eprint
  {http://arxiv.org/abs/nucl-th/0010093} {nucl-th/0010093} \BibitemShut
  {NoStop}%
\bibitem [{\citenamefont {{Goriely}}(2004)}]{BRUSLIB}%
  \BibitemOpen
  \bibfield  {author} {\bibinfo {author} {\bibfnamefont {S.}~\bibnamefont
  {{Goriely}}},\ }in\ \href {\doibase 10.1063/1.1737131} {\emph {\bibinfo
  {booktitle} {Tours Symposium on Nuclear Physics V}}},\ \bibinfo {series}
  {American Institute of Physics Conference Series}, Vol.\ \bibinfo {volume}
  {704},\ \bibinfo {editor} {edited by\ \bibinfo {editor} {\bibfnamefont
  {M.}~\bibnamefont {{Arnould}}}, \bibinfo {editor} {\bibfnamefont
  {M.}~\bibnamefont {{Lewitowicz}}}, \bibinfo {editor} {\bibfnamefont
  {G.}~\bibnamefont {{M{\"u}nzenberg}}}, \bibinfo {editor} {\bibfnamefont
  {H.}~\bibnamefont {{Akimune}}}, \bibinfo {editor} {\bibfnamefont
  {M.}~\bibnamefont {{Ohta}}}, \bibinfo {editor} {\bibfnamefont
  {H.}~\bibnamefont {{Utsunomiya}}}, \bibinfo {editor} {\bibfnamefont
  {T.}~\bibnamefont {{Wada}}}, \ and\ \bibinfo {editor} {\bibfnamefont
  {T.}~\bibnamefont {{Yamagata}}}}\ (\bibinfo {year} {2004})\ pp.\ \bibinfo
  {pages} {375--384}\BibitemShut {NoStop}%
\bibitem [{\citenamefont {{Marketin}}\ \emph {et~al.}(2016)\citenamefont
  {{Marketin}}, \citenamefont {{Huther}},\ and\ \citenamefont
  {{Mart{\'{\i}}nez-Pinedo}}}]{Marketin}%
  \BibitemOpen
  \bibfield  {author} {\bibinfo {author} {\bibfnamefont {T.}~\bibnamefont
  {{Marketin}}}, \bibinfo {author} {\bibfnamefont {L.}~\bibnamefont
  {{Huther}}}, \ and\ \bibinfo {author} {\bibfnamefont {G.}~\bibnamefont
  {{Mart{\'{\i}}nez-Pinedo}}},\ }\href {\doibase 10.1103/PhysRevC.93.025805}
  {\bibfield  {journal} {\bibinfo  {journal} {\prc}\ }\textbf {\bibinfo
  {volume} {93}},\ \bibinfo {eid} {025805} (\bibinfo {year} {2016})},\ \Eprint
  {http://arxiv.org/abs/1507.07442} {arXiv:1507.07442 [nucl-th]} \BibitemShut
  {NoStop}%
\bibitem [{\citenamefont {{Horowitz}}\ \emph {et~al.}(2018)\citenamefont
  {{Horowitz}}, \citenamefont {{Arcones}}, \citenamefont {{C{\^o}t{\'e}}},
  \citenamefont {{Dillmann}}, \citenamefont {{Nazarewicz}}, \citenamefont
  {{Roederer}}, \citenamefont {{Schatz}}, \citenamefont {{Aprahamian}},
  \citenamefont {{Atanasov}}, \citenamefont {{Bauswein}}, \citenamefont
  {{Bliss}}, \citenamefont {{Brodeur}}, \citenamefont {{Clark}}, \citenamefont
  {{Frebel}}, \citenamefont {{Foucart}}, \citenamefont {{Hansen}},
  \citenamefont {{Just}}, \citenamefont {{Kankainen}}, \citenamefont
  {{McLaughlin}}, \citenamefont {{Kelly}}, \citenamefont {{Liddick}},
  \citenamefont {{Lee}}, \citenamefont {{Lippuner}}, \citenamefont {{Martin}},
  \citenamefont {{Mendoza-Temis}}, \citenamefont {{Metzger}}, \citenamefont
  {{Mumpower}}, \citenamefont {{Perdikakis}}, \citenamefont {{Pereira}},
  \citenamefont {{O'Shea}}, \citenamefont {{Reifarth}}, \citenamefont
  {{Rogers}}, \citenamefont {{Siegel}}, \citenamefont {{Spyrou}}, \citenamefont
  {{Surman}}, \citenamefont {{Tang}}, \citenamefont {{Uesaka}},\ and\
  \citenamefont {{Wang}}}]{HorowitzRIB2018}%
  \BibitemOpen
  \bibfield  {author} {\bibinfo {author} {\bibfnamefont {C.~J.}\ \bibnamefont
  {{Horowitz}}}, \bibinfo {author} {\bibfnamefont {A.}~\bibnamefont
  {{Arcones}}}, \bibinfo {author} {\bibfnamefont {B.}~\bibnamefont
  {{C{\^o}t{\'e}}}}, \bibinfo {author} {\bibfnamefont {I.}~\bibnamefont
  {{Dillmann}}}, \bibinfo {author} {\bibfnamefont {W.}~\bibnamefont
  {{Nazarewicz}}}, \bibinfo {author} {\bibfnamefont {I.~U.}\ \bibnamefont
  {{Roederer}}}, \bibinfo {author} {\bibfnamefont {H.}~\bibnamefont
  {{Schatz}}}, \bibinfo {author} {\bibfnamefont {A.}~\bibnamefont
  {{Aprahamian}}}, \bibinfo {author} {\bibfnamefont {D.}~\bibnamefont
  {{Atanasov}}}, \bibinfo {author} {\bibfnamefont {A.}~\bibnamefont
  {{Bauswein}}}, \bibinfo {author} {\bibfnamefont {J.}~\bibnamefont {{Bliss}}},
  \bibinfo {author} {\bibfnamefont {M.}~\bibnamefont {{Brodeur}}}, \bibinfo
  {author} {\bibfnamefont {J.~A.}\ \bibnamefont {{Clark}}}, \bibinfo {author}
  {\bibfnamefont {A.}~\bibnamefont {{Frebel}}}, \bibinfo {author}
  {\bibfnamefont {F.}~\bibnamefont {{Foucart}}}, \bibinfo {author}
  {\bibfnamefont {C.~J.}\ \bibnamefont {{Hansen}}}, \bibinfo {author}
  {\bibfnamefont {O.}~\bibnamefont {{Just}}}, \bibinfo {author} {\bibfnamefont
  {A.}~\bibnamefont {{Kankainen}}}, \bibinfo {author} {\bibfnamefont {G.~C.}\
  \bibnamefont {{McLaughlin}}}, \bibinfo {author} {\bibfnamefont {J.~M.}\
  \bibnamefont {{Kelly}}}, \bibinfo {author} {\bibfnamefont {S.~N.}\
  \bibnamefont {{Liddick}}}, \bibinfo {author} {\bibfnamefont {D.~M.}\
  \bibnamefont {{Lee}}}, \bibinfo {author} {\bibfnamefont {J.}~\bibnamefont
  {{Lippuner}}}, \bibinfo {author} {\bibfnamefont {D.}~\bibnamefont
  {{Martin}}}, \bibinfo {author} {\bibfnamefont {J.}~\bibnamefont
  {{Mendoza-Temis}}}, \bibinfo {author} {\bibfnamefont {B.~D.}\ \bibnamefont
  {{Metzger}}}, \bibinfo {author} {\bibfnamefont {M.~R.}\ \bibnamefont
  {{Mumpower}}}, \bibinfo {author} {\bibfnamefont {G.}~\bibnamefont
  {{Perdikakis}}}, \bibinfo {author} {\bibfnamefont {J.}~\bibnamefont
  {{Pereira}}}, \bibinfo {author} {\bibfnamefont {B.~W.}\ \bibnamefont
  {{O'Shea}}}, \bibinfo {author} {\bibfnamefont {R.}~\bibnamefont
  {{Reifarth}}}, \bibinfo {author} {\bibfnamefont {A.~M.}\ \bibnamefont
  {{Rogers}}}, \bibinfo {author} {\bibfnamefont {D.~M.}\ \bibnamefont
  {{Siegel}}}, \bibinfo {author} {\bibfnamefont {A.}~\bibnamefont {{Spyrou}}},
  \bibinfo {author} {\bibfnamefont {R.}~\bibnamefont {{Surman}}}, \bibinfo
  {author} {\bibfnamefont {X.}~\bibnamefont {{Tang}}}, \bibinfo {author}
  {\bibfnamefont {T.}~\bibnamefont {{Uesaka}}}, \ and\ \bibinfo {author}
  {\bibfnamefont {M.}~\bibnamefont {{Wang}}},\ }\href@noop {} {\bibfield
  {journal} {\bibinfo  {journal} {ArXiv e-prints}\ } (\bibinfo {year}
  {2018})},\ \Eprint {http://arxiv.org/abs/1805.04637} {arXiv:1805.04637
  [astro-ph.SR]} \BibitemShut {NoStop}%
\bibitem [{\citenamefont {{Xu}}\ and\ \citenamefont {{Ren}}(2005)}]{XuRen}%
  \BibitemOpen
  \bibfield  {author} {\bibinfo {author} {\bibfnamefont {C.}~\bibnamefont
  {{Xu}}}\ and\ \bibinfo {author} {\bibfnamefont {Z.}~\bibnamefont {{Ren}}},\
  }\href {\doibase 10.1103/PhysRevC.71.014309} {\bibfield  {journal} {\bibinfo
  {journal} {\prc}\ }\textbf {\bibinfo {volume} {71}},\ \bibinfo {eid} {014309}
  (\bibinfo {year} {2005})}\BibitemShut {NoStop}%
\end{thebibliography}%

\end{document}